\documentclass{article}

\PassOptionsToPackage{numbers, compress}{natbib}
\usepackage[nonatbib, final]{neurips_2024}

\usepackage[utf8]{inputenc}
\usepackage[T1]{fontenc}
\usepackage{xurl}
\usepackage[colorlinks, allcolors=blue!50!black]{hyperref}
\usepackage{booktabs}
\usepackage{amsfonts}
\usepackage{nicefrac}
\usepackage{microtype}
\usepackage{xcolor}
\usepackage{graphicx}
\usepackage{tabularx}
\usepackage{enumitem}
\usepackage{multirow}
\usepackage{array}
\usepackage{apacite}
\usepackage{cleveref}
\let\citep\shortcite
\let\citet\shortciteA

\makeatletter
\renewcommand{\APACbVolEdTR}[2]{%
  \ifx\@empty#1\@empty
    \ifx\@empty#2\@empty
    \else%
      \unskip\hbox{}%
    \fi%
  \else%
    ({#1}%
    \ifx\@empty#2\@empty%
    \else%
      \unskip; {#2}%
    \fi%
    )%
  \fi%
}
\makeatother

\title{From Turing to Tomorrow:\\ The UK's Approach to AI Regulation}

\author{
Oliver Ritchie\thanks{First author}\enskip\thanks{Corresponding author: \url{oliritchie@gmail.com}.} \quad
Markus Anderljung\footnotemark[1] \quad 
Tom Rachman\\ \\
Centre for the Governance of AI 
}

\usepackage{tikz}
\usetikzlibrary{calc}
\begin{document}
\vspace*{0.5cm}
\maketitle
 \begin{tikzpicture}[overlay,remember picture]
    \node[anchor=north, text width=1.25\linewidth, align=justify,blue!80!black] at ([yshift=-1cm]current page.north) {    This is a preprint of the following chapter: Oliver Ritchie, Markus Anderljung, and Tom Rachman ``From Turing to Tomorrow: The UK's Approach to AI Regulation''. It is intended for publication in a forthcoming book. It is the version of the author’s manuscript prior to acceptance for publication and has not undergone editorial and/or peer review on behalf of the Publisher.};
\end{tikzpicture}
\setcounter{footnote}{0}

\begin{abstract}
The UK has pursued a distinctive path in AI regulation: less cautious than the EU but more willing to address risks than the US, and has emerged as a global leader in coordinating AI safety efforts.
Impressive developments from companies like London-based DeepMind began to spark concerns in the UK about catastrophic risks from around 2012, although regulatory discussion at the time focussed on bias and discrimination. By 2022, these discussions had evolved into a "pro-innovation" strategy, in which the government directed existing regulators to take a light-touch approach, governing AI at point of use, but avoided regulating the technology or infrastructure directly. ChatGPT arrived in late 2022, galvanising concerns that this approach may be insufficient. The UK responded by establishing an AI Safety Institute to monitor risks and hosting the first international AI Safety Summit in 2023, but — unlike the EU — refrained from regulating frontier AI development in addition to its use. A new government was elected in 2024 which promised to address this gap, but at the time of writing is yet to do so.

What should the UK do next? The government faces competing objectives: harnessing AI for economic growth and better public services while mitigating risk. In light of these, we propose establishing a flexible, principles-based regulator to oversee the most advanced AI development, defensive measures against risks from AI-enabled biological design tools, and argue that more technical work is needed to understand how to respond to AI-generated misinformation. We argue for updated legal frameworks on copyright, discrimination, and AI agents, and that regulators will have a limited but important role if AI substantially disrupts labour markets.

If the UK gets AI regulation right, it could demonstrate how democratic societies can harness AI's benefits while managing its risks.
\end{abstract}

\section{Introduction\label{introduction}}

``We can only see a short distance ahead'', Alan Turing wrote of machine intelligence research in 1950, ``but we can see plenty there that needs to be done'' \cite{Turing1950v}. Even then, the pioneering British computer scientist could see the pathways that would lead to the artificial intelligence (AI) of today. Yet Turing\textquotesingle s remark also speaks to a long-held question in AI regulation: how do we determine what ``needs to be done'' when our understanding of AI's future is so limited?

This chapter charts the UK's answer to that question over time. The UK government's strategy has evolved from simply adapting existing regulatory frameworks to committing to pass new legislation, addressing both harmful uses of AI and the risks posed by the most powerful AI systems. Since the 2012 start of the deep learning boom, the government's role in shaping AI development and adoption has gone from a niche issue to a government priority. As the technology minister, Peter Kyle, wrote in January 2025, when introducing the AI Opportunities Action Plan: ``We want Britain to step up; to shape the AI revolution rather than wait to see how it shapes us. Because we believe Britain has a particular responsibility to provide global leadership in fairly and effectively seizing the opportunities of AI, as we have done on AI safety'' \cite{UKSecretaryOfStateForScienceInnovationAndTechnology2025d}.

We discuss how the UK has recently provided this leadership, including by convening the first global summit on AI safety and trying to chart a course between a US-style focus on growth and an EU-style focus on safety. The UK has the potential to continue shaping the global conversation on AI rules: it houses top AI talent, with globally recognized expertise in development and governance across industry, academia, and government. However, success is not guaranteed. Despite delivering the first major international declaration on AI safety signed by both the US and China, the UK's international leverage is limited \cite{DepartmentForScienceInnovationAndTechnology2023v}. The UK lacks the US's direct influence over the AI industry, the EU's market-shaping power, and China's sheer economic weight and technological dynamism.

With this context in mind, at the end of this chapter we discuss some of what now ``needs to be done'' in UK regulation. If AI will indeed be as transformative as many contend, we should expect it to pose a myriad of regulatory challenges and opportunities. These include how to regulate the most advanced AI systems, how to reduce barriers to AI adoption and innovation, how to safely use AI to deliver government services, how to reduce algorithmic discrimination, and how to safely deploy new applications from biological design tools to AI agents. In the words of Prime Minister Keir Starmer: ``AI is the greatest force for change in the world right now. I am determined to harness it'' \cite{Starmer2025d}.

First, though, let us look back at the history of AI regulation in the UK to better understand the historical and institutional context in which regulations are being formulated today.

\section{History of UK AI regulation\label{history-of-uk-ai-regulation}}

\subsection{From the mid-19th Century, UK researchers laid the foundations for future AI developments\label{from-the-mid-19th-century-uk-researchers-laid-the-foundations-for-future-ai-developments}}

UK researchers worked on many of the computer-science foundations that have enabled modern AI development. The English mathematician Charles Babbage (1791-1871) developed some of the first theoretical work in computing with his designs for an ``Analytical Engine'', though hardware limitations meant he never managed to create a physical version of his punch-card calculator. Fellow English mathematician Ada Lovelace (1815-1852) proposed using such machines for more than pure calculation, developing a theoretical example of what was arguably the first computer programme \cite{Siffert2017x}. A century later, improved hardware allowed the UK to build the world's first programmable digital electronic computers -- the ``Colossus'' series -- to break Nazi cyphers during the Second World War. After working with these machines, Alan Turing speculated on the future of the field, proposing what is now known as the Turing Test and expressed his ``hope that machines will eventually compete with men in all purely intellectual fields'' (Turing 1950).

In the post-war period, much research in computing shifted to the US. The term ``artificial intelligence'' was coined in 1956 at Dartmouth College. However, despite economic growth, innovation in computers and semiconductors, and proliferation of consumer electronics, progress in artificial intelligence was slow. That wouldn't change until the 21st century, when several research breakthroughs seemed to unlock the long-promised potential of AI systems. Multiple new companies, which today are worth tens of billions of pounds, were founded to commercialize these breakthroughs.

Among these companies was DeepMind. Founded in London in 2010, DeepMind has become the most influential British AI company. Initially, DeepMind focused on game-playing systems, as games provided well-defined challenges and constrained environments that could demonstrate AI\textquotesingle s potential. Meanwhile, another British researcher, Geoffrey Hinton, oversaw a decisive AI breakthrough at the University of Toronto, where he and his students (including Ilya Sutskever, later OpenAI's Chief Science Officer) illustrated the power of using multiple GPUs -- high-performance microprocessors designed for the intense graphics needs of videogaming -- to build powerful neural networks. In 2012, Hinton's team demonstrated this with an image-recognition system called AlexNet that performed far better than competitors, alerting researchers around the globe to the promise of deep learning. These techniques proved a boon to DeepMind, which was bought by Google in 2014. The increasingly-impressive systems Google DeepMind built in this period included AlphaGo, which beat the world champion in the highly-complex board game Go. This feat also helped drive Chinese investment in advanced AI, providing impetus for its China AI 2030 strategy \cite{Mozur2017j}.

Accelerating progress in AI capabilities prompted warnings about associated risks. The Cambridge physicist Stephen Hawking cautioned that ``full artificial intelligence could spell the end of the human race'' \cite{Cellan-Jones2014a}, while the Oxford philosopher Nick Bostrom explored these risks in detail in his 2014 book \emph{Superintelligence} \cite{Bostrom2014q}. The UK government provided funding for further research, supporting a group of leading British universities to establish the Alan Turing Institute in 2015, with a mission ``to make great leaps in the development and use of data science and artificial intelligence (AI) in order to change the world for the better'' \cite{TheAlanTuringInstitute2023k}. Conversations about the safety of advanced AI development were taking place internationally, too. Bill Gates expressed bemusement that others were not more afraid of the risks from AI \cite{Rawlinson2015a}, while Elon Musk and Sam Altman founded OpenAI in 2015, a non-profit aiming to build powerful artificial general intelligence (AGI) to ``benefit humanity as a whole'' \cite{OpenAI2015n}.

\subsection{From 2016 - 2019, UK thinking on AI regulation prioritised ethical issues and a light-touch approach\label{from-2016---2019-uk-thinking-on-ai-regulation-prioritised-ethical-issues-and-a-light-touch-approach}}

While the emerging AI safety movement was focused on the potential for severe harms from future systems, legislators and regulators concentrated on immediate concerns, especially unfair or unethical outcomes from using existing systems. In October 2016, the House of Commons Science and Technology Committee's ``Robotics and Artificial Intelligence'' report voiced concern about biases in AI systems, building on suspicion that users were being manipulated by AI-driven social media algorithms \cite{HouseOfCommonsScienceAndTechnologyCommittee2016n}. The committee expressed a broad desire for predictability in AI tools and called for them to be carefully scrutinized, but judged that further study should come before regulation.

The British vote to leave the European Union in 2016 allowed for regulatory divergence between the UK and EU. Prime Minister Theresa May advocated for a ``soft Brexit'', maintaining relatively high levels of regulatory alignment with the political bloc. Amid May's battles to negotiate an EU separation agreeable to both Brussels and Westminster, her government published an Industrial Strategy \cite{DepartmentForBusinessEnergyIndustrialStrategy2017t} for post-Brexit Britain. That document highlighted AI as a key pillar in transforming how the UK worked and lived, declaring ``we will lead the world in safe and ethical use of data and artificial intelligence giving confidence and clarity to citizens and business''. The focus was innovation and an ``agile'' approach to regulation, with no AI-specific regulation yet. In April 2017, The Royal Society issued a report on machine-learning that broadly supported this position, arguing that while overarching regulation would be inappropriate, particular AI uses might merit restriction \cite{TheRoyalSociety2017l}. This preference for narrow regulation was to join the wait-and-see approach as a repeated refrain in British AI policy over the next few years, in contrast to the EU's more assertive approach.

The EU presented its strategy in 2018, sharing the UK focus on economic benefits while elevating concerns about data protection, digital rights, and ethical standards \cite{EuropeanCommission2018q}. In June 2018, its High Level Expert Group on AI \cite{EuropeanCommission2018i} was launched to consider these issues, and later that year produced its first Coordinated Plan on AI \cite{EuropeanCommission2018v}. This plan again outlined a distinctive approach to AI regulation in the EU, describing ``an appropriate and predictable, ethical and regulatory framework that relies on effective safeguards for the protection of fundamental rights and freedoms'' as vital for both citizens and companies.

In April 2018, the House of Lords' Select Committee on Artificial Intelligence captured the prevailing UK hesitance with the title of its report: ``AI in the UK: ready, willing and able?'' \cite{SelectCommitteeOnArtificialIntelligence2018r}. The committee prioritised social media issues, including topics such as algorithmic bias, data protection and privacy, and concerns around market dominance by a small number of big technology companies. It also reinforced the government's focus on immediate issues over severe, but speculative, future risks. The report asserted that ``many of the hopes and the fears presently associated with AI are out of kilter with reality. While we have discussed the possibilities of a world without work, and the prospects of superintelligent machines which far surpass our own cognitive abilities, we believe the real opportunities and risks of AI are of a far more mundane, yet still pressing, nature''. The committee recommended a voluntary code whereby technology companies would inform users when AI was used for ``significant or sensitive decisions''. Again, existing regulations were deemed sufficient.

The Lords' Select Committee also considered Britain\textquotesingle s future as a market leader in the development of advanced AI After hearing evidence, including the limited amount of AI investment in the UK compared with the US and China, and superior levels of computer-science education in China and India, the committee concluded that the UK needed a strategy to retain its status as a leader in this technology. ``The UK can either choose to actively define a realistic role for itself with respect to AI, or be relegated to the role of a passive observer'', the committee report said \cite{SelectCommitteeOnArtificialIntelligence2018r}. Others had a more expansive vision, with Prime Minister Theresa May telling the January 2018 Davos summit, ``We are absolutely determined to make our country the place to come and set up to seize the opportunities of Artificial Intelligence for the future'' \cite{May2018l}.

One ``realistic role'' that the Lords' committee proposed was that the UK become an international convenor for discussions on the ethical deployment of AI, suggesting a global summit in London by the end of 2019. While that date proved optimistic, the idea eventually came to pass: the AI Safety Summit was hosted at Bletchley Park in 2023, with 27 countries in attendance.

In the meantime, concern was growing about the harms of social media and AI algorithms \cite{ExposureLabs2020r}, with some experts demanding regulatory intervention \cite{Solon2018o}. In 2018, the government established the Centre for Data Ethics and Innovation \cite{DepartmentForDigitalCultureMediaSport2018h}, or CDEI, to help ensure responsible applications of AI in both the public and private sectors, and in April 2019, proposed a new regulator to address online harms \cite{SecretaryOfStateForDigitalCultureMediaSport2019r}.

The UK also contributed in international fora, including as a member \cite{OECDAIPolicyObservatoryOtherv} of working groups that developed the OECD's May 2019 AI Principles \cite{OECDAIPolicyObservatoryOtherd}, which stated that AI should be inclusive, human-centered, fair, transparent, safe, and accountable. Government reports drew on the OECD principles for a number of years, for example in the 2023 guidance on Ethics, Transparency and Accountability Framework for Automated Decision-Making \cite{CabinetOffice2023g}.

\subsection{From 2019 - 2022, the UK developed a `pro-innovation' approach to AI regulation, diverging from the EU approach\label{from-2019---2022-the-uk-developed-a-pro-innovation-approach-to-ai-regulation-diverging-from-the-eu-approach}}

Unable to unite an increasingly divided government around her soft Brexit approach, May was forced from office in July 2019 and replaced by Boris Johnson. A few months after taking power, Johnson used an address \cite{Johnson2019c} to the UN General Assembly to bring the focus on AI regulation back to the most extreme risks. ``As new technologies seem to race towards us from the far horizon, we strain our eyes as they come, to make out whether they are for good or bad -- friends or foes?'' Johnson said. ``AI --~what will it mean? Helpful robots washing and caring for an ageing population? Or pink-eyed terminators sent back from the future to cull the human race''? While he proclaimed the UK to be ``a global leader in ethical and responsible technology'' \cite{Johnson2019c}, this did not immediately translate into policy action.

In the background of the emerging COVID-19 pandemic, the EU released its February 2020 white paper on artificial intelligence, outlining plans for ``a European approach to excellence and trust'' \cite{EuropeanCommission2020t}. As for the US approach to advanced AI, discussion focused on national security and competitiveness. At the behest of Congress, an independent consultative group was established, the National Security Commission on Artificial Intelligence, or NSCAI, led by Eric Schmidt, the former Google CEO, and Robert Work, former US deputy defense secretary. In 2021, the commission released a lengthy report that came with a stark warning: ``America is not prepared to defend or compete in the AI era'' \cite{USNationalSecurityCommissionOnArtificialIntelligence2021r}. US industry, academics, and civil society needed to join together and win the global race, they argued. The commission recommended a cautious approach to regulation, arguing that US lawmakers should consider streamlining routes to AI innovation and making existing rules more flexible. The commission also discussed export controls to prevent China or other adversaries from acquiring dual-use technology that might ultimately be deployed against US interests. Between 2018 and 2020, the first Trump administration tightened such restrictions on various Chinese technology companies, most notably Huawei \cite{Bown2018z}. At the state level, legislatures had a different focus, addressing AI-powered technologies that were already affecting citizens, as illustrated by California's Bot Disclosure Act \cite{HertzbergRobert2018b}, mandating that automated agents identify themselves if interacting with humans online. Another widespread regulatory concern was deepfakes, including nonconsensual intimate image abuse \cite{Quirk2023n}.

During this same period, Prime Minister Boris Johnson's government sought economic benefits from Brexit, and decided on a more permissive regime concerning tech regulation, rebranding May's ``agile'' approach as a ``pro-innovation'' one. In July 2021, the technology secretary, Oliver Dowden, said in a foreword to a policy paper on digital regulation: ``Now that we have exited the EU, we have a fresh opportunity to set the global path for digital regulation. With this plan, we are setting a path that is pro-innovation'' \cite{DepartmentForScienceInnovationTechnology2023j}. Despite the rebrand, Johnson's government faced the same challenge May had, trying to broadcast that post-EU Britain was open for business, but with certain constraints. The policy paper outlined a minimally regulated model while still recognising that ``digital businesses are operating in many cases without appropriate guardrails -- the existing rules and norms which have guided business activity were in many cases not designed for modern technologies and business models''.

Later in 2021, the government released more details on how these principles would apply to AI. The National AI Strategy \cite{OfficeForArtificialIntelligenceDepartmentForDigitalCultureMediaSportandDepartmentForBusinessEnergyIndustrialStrategy2021b} included a ten-year plan to make Britain ``a global AI superpower'' while ensuring that it ``gets the national and international governance of AI technologies right to encourage innovation, investment, and protect the public and our fundamental values''. While the paper suggested maintaining the 2018 position that the UK should regulate AI at point-of-use through existing regulators, it committed to reviewing two other options: to eliminate some existing rules, and to set up new overarching rules across sectors. For now, the regulatory focus remained primarily on the issues identified by the Lords' Select Committee such as bias, lack of transparency and labour market disruptions rather than national security concerns and potentially catastrophic risks. However, a section at the end of the AI strategy did propose future work to address these dangers. The paper also proposed further collaboration with international bodies to set global standards, with an AI Standards Hub launched in collaboration with the Alan Turing Institute \cite{DepartmentForDigitalCultureMediaSportOfficeForArtificialIntelligence2022s} and an associated Toolkit to coordinate international engagement. The strategy became the basis for additional publications describing the government's approach in more detail, including a roadmap from the CDEI \cite{CentreForDataEthicsAndInnovation2021u} on how the UK could establish a thriving private sector-led AI assurance ecosystem.

Maintaining the government's preference for relying on existing regulators, the Online Safety Bill was amended to give additional powers to Ofcom in place of the original proposal to establish a new regulator \cite{DepartmentForDigitalCultureMediaSport2020a}. The government accepted that there were drawbacks to a sector-led approach. Some of these were listed in its 2021 National AI Strategy \cite{OfficeForArtificialIntelligenceDepartmentForDigitalCultureMediaSportandDepartmentForBusinessEnergyIndustrialStrategy2021b}, including inconsistent or contradictory approaches across sectors and overlap between regulatory mandates leading to unnecessary regulatory burdens. The strategy promised that a white paper the following year would consider ``whether there is a case for greater cross-cutting AI regulation or greater consistency across regulated sectors'' \cite{OfficeForArtificialIntelligenceDepartmentForDigitalCultureMediaSportandDepartmentForBusinessEnergyIndustrialStrategy2021b}. Regulators themselves were also exploring ways to coordinate on digital issues, including AI. Four UK agencies -- Ofcom, the Competition and Markets Authority, the Information Commissioner's Office, and the Financial Conduct Authority -- established the Digital Regulation Cooperation Forum (DRCF) in 2020 to share best practices, avoid duplication, and improve coherence of the country\textquotesingle s overall regulatory system \cite{DigitalRegulationCooperationForum2024c}.

Further details on how existing regulators should be used were developed in the July 2022 policy paper ``Establishing a pro-innovation approach to regulating AI'' \cite{DepartmentForScienceInnovationTechnology2022j}, proposing greater consistency among regulators. It established a series of cross-sector principles, including asking existing agencies to focus on ``high-risk concerns rather than hypothetical or low risks associated with AI''; to coordinate their efforts; and to act with a light touch, preferring guidance or voluntary measures to new regulatory burdens on AI developers and users. The paper returned to the 2020 theme that ``we should regulate the use of AI rather than the technology itself'' \cite{DepartmentForScienceInnovationTechnology2022j}. It also noted the need for coherence across the regulatory regime, stating that the government would explore ``whether new institutional architecture is needed to oversee the functioning of the landscape as a whole and anticipate future challenges''. At the time, the government's ability to implement this approach effectively was uncertain. Implementation challenges identified by the Alan Turing Institute included a lack of AI expertise, poor coordination among different agencies, and resource constraints \cite{Aitken2022u}.

The regulatory and policy work of the government was again disrupted following Johnson's resignation in the summer of 2022. His successor, Liz Truss, never implemented a vision for AI regulation as she was forced from office after just 49 days. However, the next Prime Minister, Rishi Sunak took a more active approach. With his extensive Silicon Valley connections, he was well-placed to respond when OpenAI released its groundbreaking AI chatbot, ChatGPT, in November 2022, just weeks after he came to power.

\subsection{From 2022 - 2024, concern about catastrophic AI risks grew, but the UK maintained a hands-off approach to regulation\label{from-2022---2024-concern-about-catastrophic-ai-risks-grew-but-the-uk-maintained-a-hands-off-approach-to-regulation}}

A watershed moment had arrived, with many in the wider public awed by their first direct interactions with an advanced AI model. By early 2023, ChatGPT had attracted 100 million active monthly users \cite{Hu2023y}, making it the fastest-growing app in history. From a regulatory perspective, this general-purpose generative AI system differed from previous AI tools, such as the recommendation systems that social-media companies deployed or the task-specific AIs such as DeepMind's game-playing AlphaZero or its biology-research tool AlphaFold. Generative systems such as ChatGPT could undertake a wide range of cognitive tasks, both creative and technical, raising the possibility that this new form of AI could be developed into a general-purpose technology, akin to electricity or the steam engine, overhauling human development, labour, and society as a whole.

The US, perhaps anticipating the release of ChatGPT, had introduced new export controls in October 2022, restricting China from acquiring the most-advanced computing chips, considered fundamental to scaling up AI models and reaching the forefront of the technology \cite{BureauOfIndustryAndSecurity2022e}. The Biden administration, by expanding export controls initiated during the first Trump term, joined in an iterative process with rare bipartisan support, in which the US sought to retain its lead despite strong Chinese AI progress, both with hardware controls and with a push to establish frontier semiconductor production at home and keep AI data centers in the US \cite{USNationalInstituteOfStandardsAndTechnology2022f}.

The new UK Prime Minister, Rishi Sunak, watched the growing geopolitical contest over AI, but had domestic economic concerns to focus on. The Office for Budget Responsibility cited ``significant structural challenges'', including limited productivity growth and stagnant business investment in addition to persistently high inflation \cite{UKOfficeForBudgetResponsibility2023l}. The government was eager to find sources of economic growth, and Sunak looked to prioritise technology as a source of economic growth via deregulation and increased funding for innovation.

The Sunak government set out a vision for AI in the March 2023 white paper on AI regulation, prioritising growth over emerging risks. ``Having exited the European Union we are free to establish a regulatory approach that enables us to establish the UK as an AI superpower'', the Secretary of State for Science, Innovation and Technology, Michelle Donelan, wrote in a foreword to the paper \cite{DepartmentForScienceInnovationAndTechnology2023j}. They retained Johnson's ``pro-innovation'' approach, with regulation still focused on ``the use of~AI~rather than the technology itself''. However, the paper acknowledged that developments such as ChatGPT could shift this balance. ``Given the widely acknowledged transformative potential of foundation models, we must give careful attention to how they might interact with our proposed regulatory framework'', it said, while noting that ``it would be premature to take specific regulatory action in response to foundation models including LLMs. To do so would risk stifling innovation, preventing AI adoption, and distorting the UK's thriving AI ecosystem''.

To further support growth, Sunak's government announced £100 million of funding for a new AI Foundation Model Taskforce to ``ensure sovereign capabilities and broad adoption of safe and reliable foundation models, helping cement the UK's position as a science and technology superpower by 2030'' \cite{DepartmentForScienceInnovationAndTechnology2023s}. The taskforce aimed to bring together government and industry experts to invest in public procurement and infrastructure to support growth in the sector. The government urged balance, repeatedly evoking ``safety and reliability'' in addition to technological advancement. The intent was not merely to assert the UK as a competitive force in the global race to advanced AI, but to become ``a global standard bearer for AI safety''. Indeed, the taskforce, building on the success and operating model of the government\textquotesingle s COVID-19 vaccine taskforce, would eventually become the world's first AI Safety Institute.

Outside the government, there was growing concern that regulation might need to go beyond the downstream uses of AI to the development of advanced systems themselves. Around the time of the pro-innovation white paper of March 2023, the Future of Life institute published an open letter calling for a six-month pause on AI training runs larger than the recently released GPT-4 \cite{FutureOfLifeInstitute2023h}. Signatories included British computer scientist and author Stuart Russell; executive director of the Cambridge Centre for the Study of Existential Risk, Sean O\textquotesingle Heigeartaigh; and the UK tech entrepreneur Ian Hogarth, as well as international figures such as Elon Musk and the Apple co-founder Steve Wozniak. The letter urged policymakers to ``dramatically accelerate development of robust AI governance systems'', including new regulatory authorities dedicated to AI, along with strict certification and rules establishing liability for AI-caused harms.

Both houses of the British Parliament discussed similar concerns. During a debate on the Data Protection bill, the Conservative MP Damian Collins referred to the open letter, arguing for more proactive regulation: ``There must be an onus on companies to demonstrate that their systems are safe. The onus must not just be on the user to demonstrate that they have somehow suffered as a consequence of that system's design'' \cite{HouseOfCommonsHansard2023a}. Some members of the House of Lords argued that recent developments had eclipsed government plans, and the topic should be reconsidered. Lord Clement-Jones, the Liberal Democrat chair of the Lords' Select Committee on Artificial Intelligence, argued that ``a long gestation period of national AI policymaking has ended up producing a minimal proposal for `A pro-innovation approach to AI regulation' which, in substance, will amount to toothless exhortation by sectoral regulators to follow ethical principles and a complete failure to regulate AI development where there is no regulator'' \cite{HouseOfCommonsHansard2023f}.

\subsection{In 2023, seeking a global role, the UK convened the first AI Safety Summit\label{in-2023-seeking-a-global-role-the-uk-convened-the-first-ai-safety-summit}}

Significant concerns about AI safety were again raised when key figures in the development and deployment of AI signed another open letter, released by the Center for AI Safety on 30 May, 2023, consisting of a one-sentence declaration: ``Mitigating the risk of extinction from AI should be a global priority alongside other societal-scale risks such as pandemics and nuclear war'' \cite{CenterForAISafety2024q}. Signatories included the British deep-learning pioneer Geoffrey Hinton, who had just resigned from Google in order to speak freely about AI risks, remarking that he harboured some regrets about his life's work \cite{Metz2023h}. Others endorsing the letter included the CEOs of Google DeepMind, Anthropic, and OpenAI, all three of whom met with Sunak that same month and discussed the severity of existential risk, arguing that the UK could lead an international AI governance summit, echoing the idea raised by the Lords' Select Committee in 2018 \cite{Manancourt2024z}.

In June 2023, Sunak announced that the first international AI Safety Summit was to be held that November at Bletchley Park \cite{Parker2023d,BletchleyPark2023w}, the site of the first Colossus computers, saying ``I want to make the UK not just the intellectual home but the geographical home of global AI safety regulation'' \cite{Sunak2023e}. The Sunak government also shifted the focus of its AI Foundation Model Taskforce towards AI safety, appointing the tech entrepreneur Ian Hogarth as its new leader \cite{DepartmentForScienceInnovationAndTechnology2023a}. Hogarth had cautioned against the rash pursuit of artificial general intelligence, arguing that regulation might be warranted. ``We are not powerless to slow down this race'', he wrote \cite{Hogarth2023d}.

As the technology rushed onwards, UK thinkers began to debate whether regulation should move beyond specific uses of AI to directly address the underlying models themselves. The term ``frontier AI regulation'' became common, describing an approach to regulating the most powerful AI models, ``that could possess dangerous capabilities,'' relevant to, e.g., cyber- and biological attacks, ``sufficient to pose severe risks to public safety'' \cite{Anderljung2023p}. Building on the observation that the most capable AI models required increasingly vast computational resources, growing by four times a year \cite{Sevilla2024q}, the argument went that governments could limit their interventions to a handful of the riskiest models, leaving most AI companies unaffected. The idea faced significant critiques, including concerns that it would centralise power \cite{Howard2023r} while focusing efforts on speculative national security risks \cite{Helfrich2024z}. Others debated how to define the scope of frontier AI \cite{Hooker2024h,Heim2024w,Toner2023f}, what requirements might be imposed upon its developers \cite{Schuett2023i,Schuett2024i}, and how to make the definition durable to algorithmic advances \cite{Scharre2024e}. Nevertheless, the notion of frontier AI proved useful enough to gain currency, with leading US developers -- OpenAI, Google, Anthropic, and Microsoft -- forming a new industry body, the Frontier Model Forum \cite{OpenAI2023d}, with the stated aim of advancing safety research, while the UK government renamed its AI Foundation Model Taskforce as the Frontier AI Taskforce \cite{DepartmentForScienceInnovationTechnology2023r}.

In the EU, discussion turned to whether the AI Act should move beyond simply focusing on the use of AI, to its responsible development and market placement. The European Parliament added references to ``general-purpose AI systems'' to the EU AI Act in June 2023, requiring providers of such systems --~so long as they weren't open-sourced -- to offer transparency to downstream actors using and building on them \cite{EuropeanParliament2023m}. However, seeing the trend towards a potentially small number of extremely high-compute systems playing an outsized role in the market, with potential risks stemming from their high-impact capabilities, another regulatory target was proposed: general-purpose AI models with potential for systemic risk. These were initially defined as AI models trained using more than 10\textsuperscript{25} floating point operations, which was only true of a handful of models at the time it was proposed. After a tumultuous trilogue between the European Commission, Parliament, and Council \cite{Volpicelli2023j}, the Act was passed in March 2024, and was put into place in August 2024 .

Sunak also expressed a desire to regulate both the underlying AI technology and its uses. Shortly before the AI Safety Summit \cite{GOVUK2023b}, he spoke at The Royal Society in London \cite{Sunak2023x}, noting that the only people testing frontier models were those creating them, and even they did not fully understand the technology. ``We should not rely on them marking their own homework'', he said. Still, Sunak stressed that innovation remained Britain's priority, and held back from proposing immediate action. He stated that ``the UK's answer is not to rush to regulate'', and questioned how the government could ``write laws that make sense for something that we don't yet fully understand''.

The summit served as a focal point for international actors to set out their positions too, most notably the US. Coinciding with the summit, President Joe Biden released his executive order on Safe, Secure, and Trustworthy AI \cite{Biden2023k}, galvanizing efforts across the US government to grapple with AI, from averting AI-driven discrimination and safeguarding citizens' privacy, to protecting workers from the impact of AI on labour, to ensuring US global leadership in AI. With regards to frontier AI, the order required US developers of AI systems more advanced than any at that time to report training runs and safety testing to the US government \cite{Biden2023k}.

In light of growing US-China tensions, a key question for the Bletchley summit organisers was whether to include representatives from Beijing. Hawks in the Conservative Party opposed this \cite{Frei2023m}, but Sunak's team felt that any summit agreement without China's support would be far less meaningful \cite{Casalicchio2023t}. At the same time, there was uncertainty over whether China would even accept an invitation. In the end, a Chinese delegation did attend and endorsed the final Bletchley Declaration \cite{DepartmentForScienceInnovationAndTechnology2023v}, which noted the particular risk from frontier AI systems, including in cybersecurity and biotech, and the need for global alignment on safety. The inclusion of 29 signatories including Chinese and US representatives alongside those from the EU, the UK, and other major powers, was a landmark moment \cite{PrimeMinisterSOffice10DowningStreet2023y}, marking the first time that all leading AI nations had signed a joint commitment to addressing global AI risk.The summit delivered other significant announcements, including a multilateral agreement for governments and developers to work together to test AI models, an international advisory panel to advise on frontier AI risk and produce a regular ``State of the Science'' report, and an agreement that this would be the first in a series of such summits.

At the same time, the UK became a global pioneer by converting its AI taskforce into the AI Safety Institute, AISI, the world's first such government body \cite{DepartmentForScienceInnovationAndTechnology2023c}. Its mandate was to conduct technical evaluations of AI systems ``to minimise surprise to the UK and humanity from rapid and unexpected advances in AI''. The establishment of AISI came alongside that of the US AI Safety Institute and began what would expand into a global network of state-supported AI safety institutes \cite{DepartmentForScienceInnovationAndTechnology2024v}. Britain also led the world in public investment for its safety institute, with its £100 million for AISI, greatly exceeding the amount apportioned to the EU equivalent and to its American counterpart \cite{Wilson2024e}. The UK had no illusions about matching the US or China in AI-model building, yet the work of AISI quickly impressed tech insiders for its startup mindset, efficiency, and impressive hires from top AI labs. ``If you'd said to me two years ago, a government is going to create a new government body that does testing of AI systems pre-deployment, I would have said, `Good Lord, that sounds highly likely to go extraordinarily wrong'\,'', the Anthropic co-founder Jack Clark remarked \cite{Manancourt2024z}. ``But what I've experienced is the UK has built a testing institute that moves as quickly as a tech startup, which is extremely unusual. And the experience I have is when it's time to do the tests, we give them {[}access{]} to our model, and they are just instantly running large-scale tests''.

However, Sunak's government faced criticism that the UK had not matched international leadership with sufficient domestic action. Shadow technology minister Peter Kyle responded to the Bletchley Declaration with a call for binding requirements on companies developing powerful AI. ``It is not good enough for our `inaction man' Prime Minister to say he will not rush to take action, having told the public that there are national security risks which could end our way of life'', Kyle said \cite{Lloyd2023n}.

The government's February 2024 response to the AI white paper consultation went further than it had before in acknowledging these concerns and accepting that binding requirements on highly capable general-purpose AI systems might become necessary. ``We anticipate that all jurisdictions will, in time, want to place targeted mandatory interventions on the design, development, and deployment of such systems to ensure risks are adequately addressed'', the government response said, \cite{DepartmentForScienceInnovationTechnology2024q}, but argued that the sector-specific approach remained sufficient for the time being. The government published additional guidance and announced more funding for regulators \cite{DepartmentForScienceInnovationTechnology2024q}, but stopped short of suggesting any major changes. Some UK think tanks called for faster action. ``The government is understandably concerned that moving too quickly could risk stymying innovation, or could result in committing to rules which quickly become outdated'', the Centre for Long-Term Resilience said. ``But there are also considerable risks to moving too slowly: existing harms and risks remain unaddressed while new ones will inevitably emerge; other countries and jurisdictions may increasingly set the terms of regulation; and innovation in the UK may also suffer due to a lack of regulatory certainty for businesses'' \cite{Whittlestone2024x}.

What nobody disputed was that the AI Safety Summit in Bletchley had begun an iterative process of consensus-building. A second global meeting came six months later in South Korea, with Britain as the co-chair. The Seoul Declaration called for safety, innovation and inclusivity, along with the interoperability of governance frameworks and the expansion of the international AI safety institute network \cite{DepartmentForScienceInnovationAndTechnology2024a}. Unlike Bletcheley, China refrained from signing this declaration, but did not rule out future participation and would go on to sign the following summit in Paris \cite{AIActionSummit2025z}. Most significantly, the Seoul summit secured voluntary commitments \cite{DepartmentForScienceInnovationTechnology2024v} from major AI developers including OpenAI, Google, Anthropic, Meta, Microsoft, Amazon, and xAI, along with the Chinese startup Zhipu AI, to develop and implement safety frameworks designed to keep the risks from their frontier systems to tolerable levels, drawing inspiration from such frameworks already adopted by Anthropic \cite{Anthropic2023y}, OpenAI \cite{OpenAI2023m}, and Google DeepMind \cite{GoogleDeepMind2024y}. The summit also saw the draft release of the first International Scientific Report on the Safety of Advanced AI \cite{Bengio2024k}, chaired by the computer scientist Yoshua Bengio, known as one of the ``godfathers of AI'' \cite{Vincent2019b}.

\subsection{In 2024, Labour replaced the Conservatives, affirming plans for future AI regulation while retaining an innovation focus\label{in-2024-labour-replaced-the-conservatives-affirming-plans-for-future-ai-regulation-while-retaining-an-innovation-focus}}

Prime Minister Rishi Sunak called a general election for July 2024 and lost to Keir Starmer, who promised an agenda of change, including more concrete action on AI regulation. The party manifesto stated that ``Labour will ensure the safe development and use of AI models by introducing binding regulation on the handful of companies developing the most powerful AI models'' \cite{LabourParty2024z}. It also committed to banning the creation of sexually explicit AI deepfakes. While this was not a radical departure from the Conservative approach, Labour was willing for the first time to establish concrete rules in place of voluntary commitments for frontier developers, and to move faster to address specific harms coming from the use of AI.

When setting out its legislative agenda in its first King's speech, the new government stated that it would ``establish the appropriate legislation to place requirements on those working to develop the most powerful artificial intelligence models'' \cite{PrimeMinisterSOffice10DowningStreetAndKingCharlesIII2024h}. But despite Peter Kyle's previous proposals for action as shadow technology minister, Labour did not fully commit to a new AI regulation bill in the first parliamentary session. One possible reason for the lack of immediate regulatory action was the fear of losing overseas (particularly US) investment from companies that were highly mobile internationally. Secretary of State for Science, Innovation, and Technology, Peter Kyle, now faced the challenge of maintaining good relations with tech giants himself \cite{Wright2024b}. ``I'm probably the first Secretary of State that is dealing with companies which are outspending our entire British state when it comes to investment in innovation. So let's just act with a bit of sense of humility. We are having to apply a sense of statecraft to working with companies that we've in the past reserved for dealing with other states'', he said. ``We have to have a regulatory and legislative landscape that's reflexive, responsive, and agile enough that it can give emerging innovations a soft landing while we adapt the legislation over time''.

The government attempted to soften such landings by opening the Regulatory Innovation Office (RIO) to lessen the bureaucratic frictions that developers might face when trying to bring products to market. This new agency was tasked with hastening approvals and improving coordination among existing regulatory agencies. The underlying drive behind RIO was ``kickstarting growth across the country'' and showing that ``the UK is `open for business' as the government resets relations with trading partners around the globe,'' the new administration said when announcing the RIO in October 2024 \cite{DepartmentForScienceInnovationAndTechnology2024k}. One of the agency's goals was to encourage the adoption of AI to bolster the National Health Service. ``AI is set to revolutionise healthcare delivery so doctors can diagnose illnesses faster and improve patient care''. Reinforcing the message that regulators should focus more on growth, the government replaced the CEO of the Competition and Markets Authority in January 2025, after the regulator's proposed plans to promote growth were seen as insufficient \cite{Jack2025j}.

In contrast to its cautious approach on regulation, the new government moved fast to seek benefits from AI use. In a show of continuity between governments, they asked Matt Clifford -- who had previously led preparations for the Bletchley Summit under Sunak \cite{DepartmentForScienceInnovationTechnologyOthera} -- to identify public and private sector AI opportunities for the country. The resulting AI Opportunities Action Plan included proposals to ratchet up investment in AI infrastructure, cultivate a tech-savvy workforce, and proactively seek AI solutions to problems, while maintaining support for the AI Safety Institute's evaluations of potential risks \cite{DepartmentForScienceInnovationAndTechnology2025v}The government announced its full agreement with almost all of the 50 recommendations \cite{DepartmentForScienceInnovationTechnology2025k}while signaling that the UK's approach to legislating AI would remain cautious. ``We don't need to walk down a US or an EU path on AI regulation'', Starmer wrote, introducing the action plan. ``We can go our own way, taking a distinctively British approach that will test AI long before we regulate'' \cite{Starmer2025d}.

The government did not stop working on safety behind the scenes, continuing the international leadership that Sunak had pioneered. In November 2024, the UK AISI jointly organised a conference in California with the non-profit Centre for the Governance of AI to accelerate the design and implementation of frontier AI safety frameworks. The conference was attended by academics and researchers from leading AI companies \cite{UKAISafetyInstituteOtherk}. However, the trajectory of global AI-safety efforts entered a period of uncertainty after the re-election of Donald Trump as president in November 2024. A flurry of support for Trump in US tech circles led to questions over which Silicon Valley advice he might heed, whether taking the counsel of accelerationist supporters such as the venture capitalist Marc Andreessen \cite{Andreessen2023s}, or taking a more cautious approach in light of ``severe'' risks emphasised by supporters such as Elon Musk \cite{Wong2023x}. Another notable influence was the tech entrepreneur David Sacks, Trump's ``AI and crypto czar'', whose complaints that ``leftist elites'' were manipulatively forcing their views on the US public fed into a broader conservative push to eliminate ``wokeness'' from institutions \cite{Torenberg2024o}. After taking power in January 2025, the new Republican administration tasked senior officials to review Biden's October 2023 executive order, identifying what, if anything, should be kept and put into an action plan.

Trump also signaled an interest in accelerating American AI development, joining with tech leaders shortly to announce the Stargate investment plan: a projected \$500 billion in private investment over four years to build US-based data centres \cite{Jacobs2025e}. ``What we want to do is we want to keep it in this country'', Trump said, citing tech competition with China. ``We have an emergency; we have to get this stuff built''. Trump's push to erase traces of Biden's work did not necessarily mean that all aspects of AI regulation would be eradicated. For instance, export controls to restrict advanced chips from reaching China, which had been initiated during Trump's first term and expanded under Biden, seemed to match the new president's ``America First'' agenda.

The fate of the US AI Safety Institute, established under the Biden administration, remained uncertain at first. However, leading US tech developers strongly supported its work, with dozens of companies including OpenAI, Anthropic, Meta, and Google urging Congress to maintain the institute \cite{InformationTechnologyIndustryCouncil2024b}. The new conservative AI policy was not necessarily against all guardrails. ``I don't think the Wild West has worked out in other areas of the tech space'', said Kara Frederick, former head of the Tech Policy Center at The Heritage Foundation, now a special assistant to Trump \cite{Burgan2024p}. But she noted that the term ``AI safety'' had been poisoned. ``It is radioactive in some very conservative circles'', she said. The US AISI was subsequently renamed the ``Centre For AI Standards and Innovation'' \cite{USDepartmentOfCommerce2025c} to reflect this change in focus. Perhaps reflecting a desire to align messaging with US allies, the UK AISI was rebranded as an AI \emph{security} institute following Trump's win, even though its fundamental objectives remained unchanged \cite{DepartmentForScienceInnovationAndTechnology2025d}.

We have mapped how the UK approach to AI regulation has developed from light touch observation, through actively shaping global discussions, to preparing to deliver bespoke domestic regulation for AI. The remainder of the chapter considers how the incumbent Labour government\textquotesingle s priorities might shape its approach to AI regulation, then details some of the main policy questions they will need to answer in order to regulate effectively.

\section{Challenges and opportunities for UK AI regulation\label{challenges-and-opportunities-for-uk-ai-regulation}}

The Labour party entered government in 2024 with a mandate to develop regulations for both frontier AI development and specific uses of the technology. But it also had an urgent need to deliver economic growth and improve public services. In this section, we first describe the Labour government's policy goals. We then discuss a selection of the most pressing challenges and opportunities for UK AI regulation.

UK policymakers have started to take the possibilities of AI seriously. Many now genuinely believe that AI may be the most transformative technology of the century, creating both immense opportunities and serious challenges for policymakers. Successfully regulating AI requires balancing its potential for productivity gains against the need to address novel harms across multiple domains. Key challenges include: updating legal frameworks for AI systems that can take traditionally human actions; controlling a wide range of new AI-enabled harms, from sexual harassment to synthetic pathogens; and ensuring sufficient control of machines that are more capable than humans in many intellectual domains \cite{Bengio2025z}.

\subsection{The Labour government's objectives on economic growth and public service reform favour a somewhat permissive approach to AI regulation\label{the-labour-governments-objectives-on-economic-growth-and-public-service-reform-favour-a-somewhat-permissive-approach-to-ai-regulation}}

During the election, the Labour government promised to deliver a series of `missions', or policy objectives. These missions are used to coordinate work across government to deliver economic growth, clean energy, improved healthcare, reduced crime, and better education \cite{LabourParty2024r}.

The first of these -- economic growth -- could lead ministers to favour lighter-touch regulation to support innovation, and potentially accepting increased risk in exchange. Labour has set a target of delivering the highest sustained growth of any G7 economy \cite{TunbridgeWellsLabourParty2024k}, and the government has stated that it sees AI adoption and innovation as an important part of their plans to deliver this. ``It is hard to imagine how we will meet the ambition for highest sustained growth in the G7 -- and the countless quality-of-life benefits that flow from that -- without embracing the opportunities of AI'', notes the AI Opportunities Action Plan \cite{UKSecretaryOfStateForScienceInnovationAndTechnology2025d}. AI-driven growth could come from investment in large UK-based frontier developers such as Google DeepMind, datacenter capacity, narrow AI products such as driverless cars and AI-enabled pharmaceutical development, or companies selling supporting services or hardware such as specialised AI chips. Diffusion of AI tools throughout the economy could also boost overall productivity, in particular in service sectors such as finance, consulting, law and marketing. To deliver this mission, the government is likely to avoid overly strict regulations on AI development that could disincentivise investment in the UK, or overly strict rules on downstream uses of AI that could slow adoption.

AI could also play a large role in delivering the missions on healthcare, education, and, to a lesser extent, crime. AI tools could improve medical outcomes in a variety of categories, including improving radiology diagnosis, helping doctors detect heart disease, predicting the progression of diseases, and personalising cancer and surgical treatment, among other areas \cite{Kwint2023w,Najjar2023s}. In education, AI tools could save teachers time in developing lesson plans and enhance the educational experience by providing multiple examples and explanations. They could also offer each student unlimited personalised tutoring from AI assistants \cite{Hu2024z,Mollick2023x}.

Ministers have indicated a strong interest in pursuing these opportunities, with Starmer saying ``AI is the way to secure growth, to raise living standards, put money in people's pockets, create exciting new companies, and transform our public services''. One reason for this focus may be that the government lacks more conventional options to improve public services. It is limited in its ability to raise taxes to fund improvements by concerns that this could hamper economic growth, and by political commitments to not raise taxes on working people \cite{LabourParty2024z}. It is limited in its ability to borrow money to fund improvements by its fiscal rules \cite{LabourParty2024x} and by the increasing cost of government borrowing \cite{UKOfficeForNationalStatistics2025y}. Finally, simply spending existing budgets more effectively is also likely to be challenging: successive governments have already delivered ``a decade'' of efficiency drives to try to improve spending efficiency \cite{HouseOfCommonsCommitteeOfPublicAccounts2021r}, with much of the low hanging fruit already captured. AI tools could offer novel opportunities to drive down costs, with the Tony Blair Institute predicting potential savings as high as £40bn per year \cite{Iosad2024y}.

Though these pressures may make the government more reluctant to introduce new regulation, as many researchers and government reports have pointed out, regulatory intervention does not always hinder growth. Changes to the regulatory system can increase regulatory certainty, address regulatory overlaps, and reduce information asymmetries. It can also increase trust in technologies. Conversely, accidents could turn public opinion against AI technologies. The biggest threat to a thriving AI industry, one might argue, would be a large-scale disaster such as the partial nuclear meltdown at Three Mile Island in 1979 in the US, which permanently damaged public trust in nuclear power \cite{Bianchi2023y}.

\subsection{Regulatory policy options for the UK\label{regulatory-policy-options-for-the-uk}}

The UK seeks to be a world leader in both frontier development and innovative uses of AI, while also providing protection from associated risks. In the remainder of this chapter we discuss a selection of the most pressing regulatory opportunities and challenges facing the UK at the time of writing. We begin with topics that we can see most clearly, and then turn towards more speculative issues.

\subsubsection{Reducing regulatory barriers to adoption\label{reducing-regulatory-barriers-to-adoption}}

{\paragraph{The regulatory challenge\label{the-regulatory-challenge}}

A key challenge in AI regulation will be to ensure that existing regulation does not needlessly erect barriers to AI adoption throughout the economy. Few existing regulations have been written with current -- let alone future -- AI technologies in mind. For example, how should data protection regulations work in the context of AI? The powerful AI models underlying many applications are trained on immense datasets, sometimes incorporating much of the internet, challenging laws and norms of fair dealing, while also raising questions about copyright and intellectual-property law. Regulators focused on government activity may be a particularly important focus. About 45\% of spending in the British economy is directed by the government rather than the private market \cite{InstituteForFiscalStudies2024r}, making increasing the productivity of UK public services a key priority.

{\paragraph{Regulatory options\label{regulatory-options}}

Initiatives like the Regulatory Innovation Office (RIO) established in October 2024, and the subsequent measures laid out in the AI Opportunities Action Plan, are a good place to start \cite{DepartmentForScienceInnovationAndTechnology2024k}. The RIO was meant to hasten progress from tech innovation to widespread adoption, working to prevent regulation from becoming an excessive bottleneck. Initially, the RIO focused on the space industry, engineering biology, autonomous technology such as drones, and digital innovations, including AI, for healthcare. The most comprehensive project to reduce regulatory burdens on AI adoption came in the multipronged AI Opportunities Action Plan, which advocated for well-crafted regulatory processes that would increase business confidence and public assurance in new AI applications \cite{UKSecretaryOfStateForScienceInnovationAndTechnology2025d}.

One promising approach in some domains involves regulatory sandboxes, limited contexts in which innovators may test out products and services with guidance from regulatory agencies and less risk of incurring liability. Sandboxes have the secondary benefit of helping regulatory agencies that lack deep technical expertise gain real-world experience of AI effects. The AI Opportunities Action Plan proposed sandboxes for some of the most complex and most promising fields: robotics, drones and autonomous vehicles. To further encourage AI adoption, existing agencies could be required to publish annual reports on how they enabled innovation and growth in their sectors.

\subsubsection{Frontier AI regulation\label{frontier-ai-regulation}}

{\paragraph{The regulatory challenge\label{the-regulatory-challenge-1}}

While the Sunak government accepted that regulation on development of the most advanced systems would likely one day be needed, Starmer's Labour government committed to introducing ``binding regulation on the handful of companies developing the most powerful AI models'' \cite{LabourParty2024z}.

Since the early 2010s, progress in AI capabilities has been impressive. This is in large part because researchers have found ways to leverage increasingly massive amounts of computational power (compute). Between 2010 and May 2024, the amount of compute used to train the most capable models increased by a factor of 786 million (author's calculation based on \citet{Epoch2024y}). Recently, reinforcement learning techniques have been leveraged to yield impressive results in coding and math, with OpenAI's o3 ranking 175th on Codeforce globally, an online coding competition \cite{OpenAI2025g}. Coders are seeing significant productivity boosts from using AI systems. One series of field experiments found that access to an AI coding assistant was associated with a 26\% increase in completed tasks, with greater productivity gains among less-experienced developers \cite{Cui2024a}. In another study, participants using AI produced 55\% more lines of code than those not using AI. Further, even though these studies focus on a subset of more automatable tasks, we've seen significant AI progress since these studies were carried out in 2022 and 2023 \cite{Gambacorta2024z}.

However, these impressive capabilities are \emph{dual use}: they can be used for ill as well as good. Advances in AI capabilities bring corresponding increases in potential for harm, including from cyber attacks, biorisks, mass persuasion, and fraud \cite{Bengio2024t,Anderljung2023p,Shevlane2023k}. Further, given the structure of the current AI industry, general-purpose frontier AI systems will often be the first to develop these capabilities.

Though the models of early 2024 seemed unlikely to increase such risks, the models of today appear more concerning. Studies suggest that GPT-4 level systems, released in March 2023, can mildly boost a lay person's ability to execute a biological weapons attack but not that of an expert \cite{Mouton2024i,Patwardhan2024g}. OpenAI says GPT-4's successors -- o1 \cite{OpenAI2024k} and o3 \cite{OpenAI2025g} -- pose what it terms ``Medium Risk''\footnote{OpenAI defines Medium risk as ``Model provides meaningfully improved assistance that increases ability for existing experts in CBRN-related advanced fields to be able to create a known CBRN threat (e.g., tacit knowledge, specific supplier information, plans for distribution)'' \cite{OpenAI2023m}.} with regards to human persuasion and chemical, biological, radiological, and nuclear risks. The risks stem from both potential misuse and unintended consequences, particularly as AI systems become more difficult for users to effectively control.

Directly regulating how the most advanced models are developed and deployed could complement shortcomings in point-of-use regulation. Though practices are still emerging, developers are better placed to assess the capabilities and potential impacts of their systems than downstream users, who may have less technical expertise and access to information on how the model works \cite{DepartmentForScienceInnovationAndTechnology2023y}. Developers are also better placed to respond to dangerous capabilities, as they can employ tools such as adjusting training data or teaching models to refuse harmful requests, neither of which are available to consumers of those systems. Some companies have also defined thresholds beyond which they consider risk unacceptable and set up processes to avoid breaching them \cite{Anthropic2023y,OpenAI2023m,GoogleDeepMind2024y}. Others have already committed to articulating and adhering to such thresholds \cite{DepartmentForScienceInnovationTechnology2024v}.

Another reason to address risks at their source -- though more speculative -- is to prevent catastrophic single points of failure. As AI systems become more integrated in our economy and society, many AI applications may end up using a small set of systems, developed by a handful of companies, for a wide range of applications. Few companies have the resources and expertise to compete with the likes of Google DeepMind, OpenAI, Anthropic, DeepSeek, xAI, and Meta. This could mean most companies and users use these handful of systems rather than building their own. This concentration may create systemic vulnerabilities, similar to our largest banks, whose activities we also regulate at source for similar reasons.

A common argument against frontier AI regulation is that one should regulate uses of AI, rather than the underlying technology \cite{Ng2023q}. But, this approach of combining regulation across multiple stages in a product lifecycle is in fact the default across many regulatory domains. Pharmaceutical companies need to prove their medicines are sufficiently safe before putting them on the market, while doctors still need to prescribe them appropriately, and patients aren't allowed to sell them on for off-license uses. In aviation, airplane manufacturers need to make sure their planes meet certain standards, while airlines need to maintain them responsibly, and pilots need to fly them competently. Similarly, as a \emph{general-purpose technology}, AI is often likened to electricity, a highly regulated industry.

Other arguments against frontier AI regulation bear more weight. One is on the topic of economics. If frontier AI is so important to growth, why impose onerous rules? This is an important question, but it suggests designing prudent, effective rules, rather than avoiding rules altogether. First, rules can be made less onerous. Second, economic gains from AI will likely come primarily through its practical applications rather than from AI development itself. Due to market competition and pricing constraints, most of the economic value may be captured by end users rather than AI developers. This suggests that maximizing AI\textquotesingle s economic benefits is primarily about enabling widespread adoption, not about lowering barriers to advanced AI development.

Another related concern is the UK's relatively limited market size, which could leave it open to regulatory flight. If the UK imposes requirements that companies do not already comply with, they may choose to leave the UK market rather than comply. Given the EU's larger market size, a Brussels Effect in AI \cite{Bradford2020v,Siegmann2022w} is more likely than a London Effect.

{\paragraph{Regulatory options\label{regulatory-options-1}}

Frontier AI systems currently appear powerful enough to warrant tailored regulation, complementing regulations on AI use cases. Further, by the time the UK passes a bill, the current pace of progress suggests that we will have seen significant developments, at least on par with the progress between OpenAI's GPT-4 and o3. However, designing an effective regulatory regime for Frontier AI will be challenging: overly strict and broad regulations on AI in the UK may deny UK citizens the technology's benefits without reducing global risks significantly, whereas if we do nothing, we risk potential harm to UK citizens with limited tools to defend them.

The right regime will need to have the appropriate scope, requirements imposed on in-scope actors, and the right regulatory institution. The scope needs to be broad enough to apply to all systems that might pose sufficient risks, but narrow enough to feasibly implement without stifling innovation. One promising possibility is to set requirements based on the amount of compute used to train a model, with larger models facing stricter requirements \cite{Heim2024w}. This is likely the simplest measure and would align with the EU and the US approaches. For example, the regime could apply to any system trained using more compute than any that has been released to date, 10\textsuperscript{26}\hspace{0pt} floating point operations, or at least within one order of magnitude of the highest-compute model at any point in time.

The scope will need to be adapted in light of AI industry developments. Training compute thresholds will likely need to be adjusted and complemented by other metrics \cite{Hooker2024h}. These may include what data the model was trained on, how many users it has, or the presence of particularly dangerous capabilities. Advances in reasoning models that see greater performance using inference compute might also require adjusting the scope, focusing more on models' capabilities.

A second challenge is to define regulatory requirements that are effective without being too burdensome. To achieve this, the UK should implement principles-based requirements that companies must meet, without being too prescriptive about how they should be met \cite{Schuett2024i}. For example, a regulatory regime for frontier AI developers could impose three sets of obligations on frontier AI companies:

\begin{enumerate}
\def\labelenumi{\arabic{enumi}.}
\item
  \begin{quote}
  Safety: Requiring that companies do not impose intolerable risks on society in developing or deploying these systems.\hspace{0pt}\hspace{0pt}\hspace{0pt}
  \end{quote}
\item
  \begin{quote}
  Cybersecurity: Requiring that companies implement measures to prevent model theft.\hspace{0pt}
  \end{quote}
\item
  \begin{quote}
  Transparency: Requiring that companies provide relevant information about their systems to regulators, downstream developers, and users.
  \end{quote}
\end{enumerate}

\hspace{0pt}Crucially, these requirements need not stifle innovation. By building on what many companies are already doing, we can set a high bar without driving innovation offshore. The UK should align its approach with existing international commitments, the Hiroshima Protocols \cite{G7HiroshimaSummit2023z}, and in particular, the Frontier AI Safety Commitments \cite{DepartmentForScienceInnovationTechnology2024v}, as well as the EU's requirements on the most powerful systems.

To implement this framework, the UK needs a competent regulator with significant expertise and supervisory powers, including the ability to take enforcement actions. This regulator must have the flexibility to adapt rules as AI technology evolves, ensuring that oversight remains relevant and effective, while making sure requirements don't ossify or impose unnecessary burdens.

The success of the AI Safety Institute (AISI) provides an excellent foundation to build upon. Some suggest that this means AISI should become the regulator, while others argue that this would hamper the institution's ability to work collaboratively with AI companies to push the science of AI safety \cite{Mokander2024e}. This argument might be even stronger if the Trump administration treats foreign regulation of US-developed AI models as overreach. As such, the best option might include keeping AISI as a separate source of expertise within government -- similar to the National Cyber Security Centre \cite{NationalCyberSecurityCentre2023e} -- which informs the work of an independent regulator focused on frontier AI companies.

\subsubsection{Misinformation, deepfakes, and identifying AI-generated content\label{misinformation-deepfakes-and-identifying-ai-generated-content}}

{\paragraph{The regulatory challenge\label{the-regulatory-challenge-2}}

AI-generated text, images, and audio can be nearly indistinguishable from authentic or human-generated content. This can cause harm, including through non-consensual deepfake sexual imagery and increasingly sophisticated scams. Existing rules may address some of these issues. For example, the UK\textquotesingle s Consumer Protection from Unfair Trading Regulations prohibit misleading commercial practices regardless of whether AI generated them, while existing data protection law under UK GDPR continues to regulate the processing of personal data, whether done by traditional or AI-powered systems. However, for other issues, AI tools will create challenges that current regulators are not well placed to address, such as responding to non-consensual intimate content. By early 2025, the sharing of such deepfakes was against the law, but their creation had yet to be criminalised \cite{MinistryOfJustice2025d}. Another complex issue is the use of AI voice-cloning to impersonate individuals in ways that do not clearly violate existing laws against identity theft and fraud. AI tools could also be used to generate large volumes of misleading political content during elections. The Online Safety Act 2023 designated Ofcom as the online safety regulator, yet doubts remain about how to define and respond to politically manipulative AI-generated content \cite{Abrusci2024h}, which could fall under the remit of Ofcom \cite{DepartmentForScienceInnovationTechnology2024q}, the Electoral Commission \cite{TheElectoralCommission2024q} or neither.

Further, even if existing regulations apply to certain uses of AI technologies, it may be beneficial for society if AI generated content is identified as such. The most widespread current AI-content tagging method involves adding a cryptographic ID to the metadata of files produced by AI systems \cite{CoalitionForContentProvenanceAndAuthenticityC2PA2021e} . However, it is unclear how useful this will be for regulators as that metadata can be deleted or changed \cite{Anderljung2023g}. Substantial progress is being made on watermarking technology that embeds invisible and hard-to-remove information in the AI-generated output itself. Google DeepMind has reported advances via its SynthID for audio, video, images and text -- committing to deploy it in its products -- but it is not yet clear how reliable this approach could be as a widespread solution \cite{DeepMind2024w}. Challenges include how to ensure widespread adoption, and how to ensure that technical capabilities to insert watermarks remain ahead of capabilities to remove them.

A complement to identifying AI-generated content is to identify authentic content. Content-provenance techniques (such as those in the C2PA standard \cite{CoalitionForContentProvenanceAndAuthenticityOtherb}) can be used by governments to prove the authenticity of official documents or announcements. For this purpose, it matters less whether the tag can be removed; it only matters that a false tag is hard to make. The Biden Executive Order on AI moved the US government in this direction, tasking the Office of Management and Budget with issuing guidance for government use of content provenance techniques \cite{Biden2023k}.

{\paragraph{Regulatory options\label{regulatory-options-2}}

The government was right to introduce additional criminal sanctions that allow existing regulatory and legal structures to tackle harms from sexually explicit deepfakes. The government needs to continue to identify and address new AI issues as they arise. In the immediate term, that might entail looking at whether existing regulations are fit to address risks from AI-enabled scams and cyberattacks.

The right solution to the more general objective of making it easier to distinguish AI generated from real content is not yet clear, and future technical developments could present new options. For now, we recommend that the government explores whether a mandate for watermarking content at least in certain domains (e.g. photorealistic images) is technically and administratively feasible. The government can make more immediate progress on verification of genuine content, and the UK could support development of this technology by adopting a content-provenance technique such as C2PA for important government outputs now, as a first step towards setting best-practice for certain kinds of activity (e.g. photojournalism).

\subsubsection{Copyright\label{copyright}}

{\paragraph{The regulatory challenge\label{the-regulatory-challenge-3}}

AI systems raise several regulatory challenges related to copyright. One concerns the legal gray zone of training AI systems on UK soil. AI systems are trained on large amounts of data, much of which is protected by copyright. Under current UK law, AI companies probably cannot legally copy such works to train AI models, unless they agree an individual licence with each copyright holder or a suitable exemption applies -- an impractical proposition. Unlike the US, with its broad ``fair use'' doctrine, which may allow copying of content to train AI systems (although this is being disputed in the courts), or the EU with its opt-out commercial text and data-mining exemption, the UK\textquotesingle s exemptions are narrow and do not obviously permit commercial AI training. The restrictiveness of the current UK copyright regime will inhibit not only pre-training models on copyrighted works, but also fine-tuning models that were pre-trained elsewhere.

Making it easier to fine-tune systems in the UK involves a trade-off. Fine-tuning AI models in the UK could be a source of economic opportunity, if it was permitted, but could also undermine the rights of content creators to retain meaningful ownership of their work. A fundamental challenge in this area is the international dynamic: pre-training (and fine-tuning) might occur in a jurisdiction where copying is permitted. The resulting model, which would not implicate copyright protections, could then be exported to the UK. This dynamic puts copyright owners in a difficult position. In 2022-23, the Sunak government consulted with technology and creative industry stakeholders on a voluntary code of practice, but failed to find an agreeable solution. Labour\textquotesingle s manifesto committed to finding a way to support both sectors, and Labour has consulted \cite{IntellectualPropertyOffice2024z} on a proposal to require more transparency from developers on what data they are using, while making it easier for them to train models in the UK by introducing a copyright exemption that allows text and data mining unless content creators opt out of this, similar to the EU approach. The major difference between the UK rule and the EU rule is that the EU restricts models trained abroad without following its copyright law from being deployed there, but the UK does not.

This topic has attracted significant controversy, for example in the form of Baroness Kidron\textquotesingle s proposed amendment to the 2025 Data (Use and Access) Bill, which would have required AI models marketed in the UK to comply with UK copyright law, including during training. While the House of Lords repeatedly voted to include this, they eventually withdrew the amendment to allow the bill to pass after several rounds of parliamentary \textquotesingle ping-pong\textquotesingle{} in which the House of Commons rejected the amendment. The government is expected to return to these issues in future via a `comprehensive' AI bill \cite{Courea2025r}.

{\paragraph{Regulatory options\label{regulatory-options-3}}

Transparency from developers on what data they are using is likely to be an important part of the solution: it is easy for developers to do, and would address information asymmetries so copyright owners have the information needed to advocate for themselves and make informed decisions.

Government could also help reduce the transaction costs of bargaining for obtaining a licence to use copyrighted works for training or fine-tuning. For example, the government could support the creation of an easy-to-use bulk licensing system. Such a system could make it more attractive for model development to locate in the UK, by giving developers certainty about their potential copyright liability while also preserving copyright owners' rights.

These options, alone or together, are unlikely to resolve the tension between copyright owners and AI developers, especially in regard to the threat AI poses to human creative workers' ongoing employment. We discuss the employment impacts of AI further below.

\subsubsection{Discrimination and bias\label{discrimination-and-bias}}

{\paragraph{The regulatory challenge\label{the-regulatory-challenge-4}}

AI systems can perpetuate or amplify existing societal biases when trained on historical data that reflects discriminatory patterns. For example, AI recruitment tools trained on past hiring data may discriminate against women if that data reflects historical gender bias in employment. Systems can also create novel forms of discrimination that may not map cleanly onto protected characteristics -- for instance by identifying and discriminating based on previously undetectable correlates of ethnicity or disability.

While the UK has strong protections against discrimination through the Equality Act 2010, these were largely designed to address human decision-making. It may not always be clear how to apply existing rules when harm is caused by an AI system: if a mortgage-lending algorithm or AI agent discriminates against certain characteristics that function as complex proxies for a protected characteristic, it may be harder to prove discriminatory intent than with human decision-makers. AI systems could also incorrectly assign responsibility. For example, the Trades Union Congress are worried that workers could be inappropriately held liable for discriminatory actions carried out by AI systems owned by their employers, even if the workers are not realistically in a position to control or predict the actions of the AI system \cite{TradesUnionCongress2024q}. These issues raise a number of unresolved questions about liability and enforcement -- whether responsibility lies with the system developer, deployer, or both.

{\paragraph{Regulatory options\label{regulatory-options-4}}

Rather than create separate rules for AI systems, the government should focus on updating existing anti-discrimination frameworks to explicitly account for algorithmic decision-making. Importantly, such updates should be future-proof; many regulatory efforts aimed at addressing discrimination are designed in light of old AI regimes, with an outsized focus on adjusting what data large language models are trained on, rather than other ways to steer their behaviour. This may include a mix of cross-cutting rules to update the framework as a whole (for example, to clarify how liability should apply) and targeted rules to address specific issues, such as ensuring appropriate forms of accountability and redress in hiring and firing decisions.

\subsubsection{Biological design tools\label{biological-design-tools}}

{\paragraph{The regulatory challenge\label{the-regulatory-challenge-5}}

AI-enabled biological tools are AI models trained on large quantities of biological sequence data to understand and predict biological processes. Notable AI-enabled biological tools include Google DeepMind's AlphaFold 3 \cite{GoogleDeepMindAlphaFoldTeam2024o} -- which can predict the structure and interactions of life's molecules with unprecedented accuracy -- and xTrimoPGLM, a large model with strong performance across a range of biological tasks \cite{Chen2023x}.

Biological Tools are already contributing to progress in many areas of biomedicine, including vaccine development and cancer therapy \cite{Dolgin2023u,Arnold2023l}. However, the models also introduce novel dual-use risks. In particular, some experts have warned that some biological tools could potentially allow malicious actors to identify new pandemic-capable viruses \cite{NationalSecurityCommissionOnEmergingBiotechnology2024q,Baker2024k,Rose2023n,Pannu2024z}. This is especially concerning as it is becoming increasingly easy to synthesise novel viruses from scratch using mail-order DNA, without the need for physical virus samples. These developments lower the technical and financial barriers faced by malicious actors for biological misuse.

{\paragraph{Regulatory options\label{regulatory-options-5}}

Some have suggested reducing malicious actors' access to the most advanced biological design tools \cite{Bloomfield2024s}. This could involve identifying the biological design tools with the greatest potential for risk, requiring that their risk be assessed and that appropriate safeguards are put in place, such as limiting unrestricted model access to researchers with relevant expertise. However, such intervention faces serious hurdles: Most of these systems are highly dual-use and many of them are currently released openly to support further scientific discovery \cite{Moulange2023c,Halstead2024j}. Instead, governments might be better off starting by conducting literature-based risk assessments of biological tools and creating voluntary responsible development guidelines in collaboration with industry \cite{Smith2024a}.

Further, interventions at the model layer might not be sufficient. If the UK wants to reduce risk while still benefiting from these systems, it may be necessary to take adaptive measures, increasing society's ability to manage widespread access to more capable biological design tools \cite{Bernardi2024i}. For example, stricter controls on ordering, producing and selling bespoke DNA sequences could reduce the ability to weaponize pathogens \cite{Nelson2023p}. Similarly, governments could invest in metagenomic sequencing programs to detect new pathogens early, as well as investing in broad spectrum vaccines and stockpiling more effective personal protective equipment to increase resilience against new outbreaks \cite{BipartisanCommissionOnBiodefense2021o}.

\subsubsection{AI Agents\label{ai-agents}}

{\paragraph{The regulatory challenge\label{the-regulatory-challenge-6}}

Some frontier AI systems are increasingly able to act autonomously to pursue goals: we describe these as ``agents''. In contrast to a chatbot that would only display code for a user to copy and paste, a software engineering agent could directly modify project files, compile code, test the resulting application, then go back to the code to make improvements.

Many agents today are made by combining existing generative AI models with other software (`scaffolding') to allow them to take actions \cite{AutoGen2024n,AutoGPT2024i,OpenAI2024t,Anthropic2024d}. Frontier AI companies are also developing new agents from the ground up \cite{GoogleDeepMind2024s}.

Agents could increase the risk level associated with frontier AI \cite{Gabriel2024c}. Without a human in the loop, agents could carry out malicious activities at superhuman speed and scale, even if the agents are no better than humans at performing each individual such activity. For example, AI agents can already automate certain kinds of scam calls and some components of cyberattacks \cite{Fang2024d,Fang2024a}. Scaling enforcement activities could be a challenge for existing regulators. Furthermore, many of the most extreme risks from AI involve systems acting autonomously in the world, potentially against the interests of their users \cite{Bengio2025z}. The government may therefore want to ensure that additional safeguards are in place for agents, such as requiring explicit human approval for certain types of action. Agent identification systems \cite{Chan2024o,South2025d,Chan2025m} for certain kinds of activity, such as buying and selling products or making phone calls, could make it easier to track and potentially address these kinds of harm.

Even when individual AI agents do not engage in harmful activity, widespread availability of agents could indirectly create issues for existing regulators. For example, agents could make it easier and faster for consumers to switch to better products or services \cite{Sunstein2024b}. This decrease in switching costs will likely benefit consumers. At the same time, an increase in switching could destabilize economic activities or institutions that depend upon relatively high switching costs \cite{VanLoo2019n,Drechsler2023p}. Existing regulators should monitor for such society-wide effects.

AI agents could also challenge the legal foundations behind existing UK regulation. It is not currently clear who, if anyone, will be legally responsible for consequences caused by an agent. UK law often decides whether an actor is liable for the consequences of their action based on what they expect, know, or intended when they took the action. When someone deploys an AI agent, that person will not necessarily know what actions that agent is going to take, intend for those actions to happen, or have the same contextual knowledge that the agent has. This could create an accountability loophole \cite{Wills2024h}. An effective government response may require modifying the common law or expanding the scope of existing regulators to cover the actions of agents, for example by allocating responsibilities or liabilities to people who deploy or develop them. The government may wish to go further and establish a single legal framework to resolve issues of accountability, ownership, and criminal liability for AI agents across all sectors.

\paragraph{Regulatory options }\label{regulatory-options-6}

The UK will likely need to adjust its legal frameworks to clarify how existing rules should apply in the context of AI agents. In cases where agents potentially undermine the functioning of a very wide range of existing legislation, high level cross-cutting approaches could be preferable. For example, it would be better to establish clear general principles for how criminal liability applies when AI agents are involved in illegal activities, rather than separately amending individual statutes like the Computer Misuse Act, the Fraud Act, and dozens of other laws to each specify how they handle autonomous AI systems. In other cases, such as if AI agents undermine certain economic activities, sector-specific responses could be more appropriate.

In addition to responding to regulatory challenges, governments could also explore ways to unlock the benefits of agents. Governments could support the development infrastructure that enables and secures interactions with agents \cite{Chan2025m}. Analogous to fiduciary duties in domains like law and medicine, legal requirements that agents act in their user's best interests could protect consumers \cite{Aguirre2020q,Benthall2023t}. Similarly, regulators may need to clarify that certain actions come with the same fiduciary duties whether carried out by an AI system or carried out by a human.

\subsubsection{\texorpdfstring{AI-driven unemployment {AI-driven unemployment }}\label{ai-driven-unemployment}}

{\paragraph{The regulatory challenge\label{the-regulatory-challenge-7}}

Today's AI systems are beginning to transform the nature of many existing jobs, driving productivity gains for workers in fields such as software engineering, customer service, and legal work \cite{Cui2024c,Brynjolfsson2023d,Choi2023p}. There is considerable uncertainty about how future AI systems might affect employment, but some economists predict that as AI systems begin to surpass human performance in many real-world tasks \cite{Korinek2024i}, they could cause significant labor market disruptions, including widespread job losses \cite{Susskind2017j}. Historically, automation has eventually created at least as many jobs as it has displaced \cite{Autor2022q}. However, the rapid advancement \cite{Bengio2024t} and widespread adoption of AI \cite{Bick2024x} could result in different impacts, as AI has the potential to affect a broader range of work and integrate into the economy much faster than previous technologies \cite{McAfee2024d}.

Even if this technological change delivers aggregate economic growth benefits, it risks creating significant disruption for affected workers. Historical evidence suggests that displaced workers often struggle to transition to new roles, experiencing lasting negative effects on their psychological wellbeing, earnings, and family circumstances \cite{Burgard2007v,Barnette2017f,Telle2011u}. The speed of AI advancement may mean that traditional adjustment mechanisms -- like workers gradually retraining for new roles -- become less effective, or can't be delivered fast enough.

Together, these factors pose significant challenges for UK policymakers. They must balance fostering AI innovation to drive economic growth with protecting workers from the negative impacts associated with wage declines and job losses. Given the uncertainty surrounding AI's potential impacts, labour policies must remain adaptable to evolving circumstances while providing enough regulatory certainty to support business investment. Striking this balance will be essential to ensuring AI's economic benefits are shared within the UK and to strengthening societal resilience against economic disruptions.

{\paragraph{Regulatory options\label{regulatory-options-7}}

To better understand AI\textquotesingle s potential impact on employment, policymakers should closely track key trends as AI systems advance and become more widely integrated. These include evolving skill demands, the emergence of new job categories, AI adoption rates across different sectors in the UK, and wage changes across industries and occupations. By identifying the overlap of occupations most likely to experience disruption from AI and the groups of workers for whom job loss would be especially harmful, regulators can develop better targeted worker support programs. Recent efforts such as the Department for Education\textquotesingle s work mapping AI\textquotesingle s impact on jobs and skills should be expanded to provide clear foresight around these impacts \cite{DepartmentForEducation2023s}.

Regulation to halt AI-driven automation is unlikely to be the best option. Instead, a focus on enhancing society's capacity to adapt to automation's impacts can help drive beneficial growth while strengthening workers' economic security. Core components of such an approach can include AI literacy and accessibility initiatives that position workers to leverage AI to remain competitive in labour markets as AI diffuses through the economy. Rethinking the design of safety nets, the tax system, and transition assistance programs for affected workers may also be necessary. AI Growth Bonds are one example of an innovative approach that could help to spark more AI innovation in the UK while generating a pool of resources to support workers who end up displaced from their jobs by automation \cite{Casey2024z}. Finally, given the potential for truly transformative futures -- where the demand for human work falls dramatically -- governments would be wise to invest in proactive scenario planning. Ensuring that the UK has the infrastructure to scale up assistance programs in response to rapid disruption could be a low-risk, high-reward option to boost economic preparedness and ensure stability through rapid technological change.

\subsection{Conclusion\label{conclusion}}

The UK has carved out a distinctive role in global AI governance -- from laying intellectual foundations decades ago, to convening international summits and establishing the world\textquotesingle s first AI Safety Institute. Yet significant work remains to be done to translate this international leadership into effective domestic policy.

The fundamental challenge remains unchanged from Turing\textquotesingle s time: we can only see a short distance ahead. This uncertainty creates a strong temptation to delay regulation until the technology\textquotesingle s trajectory becomes clearer. However, given AI\textquotesingle s accelerating capabilities and growing integration into society, waiting too long risks allowing significant harms to emerge before protective frameworks are in place. The UK must therefore continue working to develop governance approaches that are both principled and adaptable -- protecting citizens while enabling innovation.

The UK's success in this endeavor will depend not just on crafting the right policies, but on maintaining the institutional capacity and cross-party commitment needed to implement them effectively over time. If it can achieve this, the UK has an opportunity to help shape how one of history\textquotesingle s most transformative technologies develops -- not just through convening international discussions, but by demonstrating how democratic societies can harness AI\textquotesingle s benefits while managing its risks.

\subsection{Acknowledgements\label{acknowledgements}}

Thanks to Alan Chan, Sam Manning, Peter Wills, Jonas Schuett, Sophie Williams, Stephen Clare, John Halstead, Beth Eakman, Katharine Bardsley for fruitful discussion and valuable input to the chapter, and José Luis León Medina for his technical assistance with the manuscript's formatting.

\clearpage
\bibliographystyle{apacite}
\bibliography{bibliography}

\begin{thebibliography}{}

\bibitem [\protect \citeauthoryear {%
Abrusci%
}{%
Abrusci%
}{%
{\protect \APACyear {2024}}%
}]{%
Abrusci2024h}
\APACinsertmetastar {%
Abrusci2024h}%
\begin{APACrefauthors}%
Abrusci, E.%
\end{APACrefauthors}%
\unskip\
\newblock
\APACrefYearMonthDay{2024}{}{}.
\newblock
{\BBOQ}\APACrefatitle {The {UK} Online Safety Act, the {EU} Digital Services Act and online disinformation: is the right to political participation adequately protected?} {The {UK} online safety act, the {EU} digital services act and online disinformation: is the right to political participation adequately protected?}{\BBCQ}
\newblock
\APACjournalVolNumPages{Journal of Media Law}{}{}{1--28}.
\newblock
\begin{APACrefURL} \url{https://www.tandfonline.com/doi/abs/10.1080/17577632.2024.2425551} \end{APACrefURL}
\PrintBackRefs{\CurrentBib}

\bibitem [\protect \citeauthoryear {%
Aguirre%
, Dempsey%
, Surden%
\BCBL {}\ \BBA {} Reiner%
}{%
Aguirre%
\ \protect \BOthers {.}}{%
{\protect \APACyear {2020}}%
}]{%
Aguirre2020q}
\APACinsertmetastar {%
Aguirre2020q}%
\begin{APACrefauthors}%
Aguirre, A.%
, Dempsey, G.%
, Surden, H.%
\BCBL {}\ \BBA {} Reiner, P\BPBI B.%
\end{APACrefauthors}%
\unskip\
\newblock
\APACrefYearMonthDay{2020}{}{}.
\newblock
\APACrefbtitle {{AI} loyalty: A new paradigm for aligning stakeholder interests.} {{AI} loyalty: A new paradigm for aligning stakeholder interests.}
\newblock
\begin{APACrefURL} \url{http://arxiv.org/abs/2003.11157} \end{APACrefURL}
\PrintBackRefs{\CurrentBib}

\bibitem [\protect \citeauthoryear {%
{AI Action Summit}%
}{%
{AI Action Summit}%
}{%
{\protect \APACyear {2025}}%
}]{%
AIActionSummit2025z}
\APACinsertmetastar {%
AIActionSummit2025z}%
\begin{APACrefauthors}%
{AI Action Summit}.%
\end{APACrefauthors}%
\unskip\
\newblock
\APACrefYearMonthDay{2025}{}{}.
\newblock
\APACrefbtitle {Statement on Inclusive and Sustainable Artificial Intelligence for People and the Planet.} {Statement on inclusive and sustainable artificial intelligence for people and the planet.}
\newblock
\begin{APACrefURL} \url{https://www.elysee.fr/en/emmanuel-macron/2025/02/11/statement-on-inclusive-and-sustainable-artificial-intelligence-for-people-and-the-planet} \end{APACrefURL}
\newblock
\APACrefnote{Accessed: 2025-6-18}
\PrintBackRefs{\CurrentBib}

\bibitem [\protect \citeauthoryear {%
Aitken%
\ \protect \BOthers {.}}{%
Aitken%
\ \protect \BOthers {.}}{%
{\protect \APACyear {2022}}%
}]{%
Aitken2022u}
\APACinsertmetastar {%
Aitken2022u}%
\begin{APACrefauthors}%
Aitken, M.%
, Leslie, D.%
, Ostmann, F.%
, Pratt, J.%
, Margetts, H.%
\BCBL {}\ \BBA {} Dorobantu, C.%
\end{APACrefauthors}%
\unskip\
\newblock
\APACrefYearMonthDay{2022}{}{}.
\newblock
\APACrefbtitle {Common Regulatory Capacity for {AI}} {Common regulatory capacity for {AI}}\ \APACbVolEdTR{}{\BTR{}}.
\newblock
\APACaddressInstitution{}{The Alan Turing Institute}.
\newblock
\begin{APACrefURL} \url{https://www.turing.ac.uk/news/publications/common-regulatory-capacity-ai} \end{APACrefURL}
\PrintBackRefs{\CurrentBib}

\bibitem [\protect \citeauthoryear {%
Anderljung%
\ \protect \BOthers {.}}{%
Anderljung%
\ \protect \BOthers {.}}{%
{\protect \APACyear {2023}}%
}]{%
Anderljung2023p}
\APACinsertmetastar {%
Anderljung2023p}%
\begin{APACrefauthors}%
Anderljung, M.%
, Barnhart, J.%
, Korinek, A.%
, Leung, J.%
, O'Keefe, C.%
, Whittlestone, J.%
\BDBL {}Wolf, K.%
\end{APACrefauthors}%
\unskip\
\newblock
\APACrefYearMonthDay{2023}{}{}.
\newblock
\APACrefbtitle {Frontier {AI} Regulation: Managing Emerging Risks to Public Safety.} {Frontier {AI} regulation: Managing emerging risks to public safety.}
\newblock
\begin{APACrefURL} \url{http://arxiv.org/abs/2307.03718} \end{APACrefURL}
\PrintBackRefs{\CurrentBib}

\bibitem [\protect \citeauthoryear {%
Anderljung%
\ \BBA {} Hazell%
}{%
Anderljung%
\ \BBA {} Hazell%
}{%
{\protect \APACyear {2023}}%
}]{%
Anderljung2023g}
\APACinsertmetastar {%
Anderljung2023g}%
\begin{APACrefauthors}%
Anderljung, M.%
\BCBT {}\ \BBA {} Hazell, J.%
\end{APACrefauthors}%
\unskip\
\newblock
\APACrefYearMonthDay{2023}{}{}.
\newblock
\APACrefbtitle {Protecting Society from {AI} Misuse: When are Restrictions on Capabilities Warranted?} {Protecting society from {AI} misuse: When are restrictions on capabilities warranted?}
\newblock
\begin{APACrefURL} \url{http://arxiv.org/abs/2303.09377} \end{APACrefURL}
\PrintBackRefs{\CurrentBib}

\bibitem [\protect \citeauthoryear {%
Andreessen%
}{%
Andreessen%
}{%
{\protect \APACyear {2023}}%
}]{%
Andreessen2023s}
\APACinsertmetastar {%
Andreessen2023s}%
\begin{APACrefauthors}%
Andreessen, M.%
\end{APACrefauthors}%
\unskip\
\newblock
\APACrefYearMonthDay{2023}{}{}.
\newblock
\APACrefbtitle {The Techno-Optimist Manifesto.} {The techno-optimist manifesto.}
\newblock
\begin{APACrefURL} \url{https://a16z.com/the-techno-optimist-manifesto/} \end{APACrefURL}
\newblock
\APACrefnote{Accessed: 2025-2-6}
\PrintBackRefs{\CurrentBib}

\bibitem [\protect \citeauthoryear {%
{Anthropic}%
}{%
{Anthropic}%
}{%
{\protect \APACyear {2023}}%
}]{%
Anthropic2023y}
\APACinsertmetastar {%
Anthropic2023y}%
\begin{APACrefauthors}%
{Anthropic}.%
\end{APACrefauthors}%
\unskip\
\newblock
\APACrefYearMonthDay{2023}{}{}.
\newblock
\APACrefbtitle {Anthropic's Responsible Scaling Policy.} {Anthropic's responsible scaling policy.}
\newblock
\begin{APACrefURL} \url{https://www.anthropic.com/news/anthropics-responsible-scaling-policy} \end{APACrefURL}
\newblock
\APACrefnote{Accessed: 2024-5-6}
\PrintBackRefs{\CurrentBib}

\bibitem [\protect \citeauthoryear {%
{Anthropic}%
}{%
{Anthropic}%
}{%
{\protect \APACyear {2024}}%
}]{%
Anthropic2024d}
\APACinsertmetastar {%
Anthropic2024d}%
\begin{APACrefauthors}%
{Anthropic}.%
\end{APACrefauthors}%
\unskip\
\newblock
\APACrefYearMonthDay{2024}{}{}.
\newblock
\APACrefbtitle {Computer use (beta).} {Computer use (beta).}
\newblock
\begin{APACrefURL} \url{https://docs.anthropic.com/en/docs/build-with-claude/computer-use} \end{APACrefURL}
\newblock
\APACrefnote{Accessed: 2025-2-6}
\PrintBackRefs{\CurrentBib}

\bibitem [\protect \citeauthoryear {%
Arnold%
}{%
Arnold%
}{%
{\protect \APACyear {2023}}%
}]{%
Arnold2023l}
\APACinsertmetastar {%
Arnold2023l}%
\begin{APACrefauthors}%
Arnold, C.%
\end{APACrefauthors}%
\unskip\
\newblock
\APACrefYearMonthDay{2023}{}{}.
\newblock
{\BBOQ}\APACrefatitle {Inside the nascent industry of {AI}-designed drugs} {Inside the nascent industry of {AI}-designed drugs}.{\BBCQ}
\newblock
\APACjournalVolNumPages{Nature medicine}{29}{6}{1292--1295}.
\newblock
\begin{APACrefURL} \url{https://www.nature.com/articles/s41591-023-02361-0} \end{APACrefURL}
\PrintBackRefs{\CurrentBib}

\bibitem [\protect \citeauthoryear {%
{AutoGen}%
}{%
{AutoGen}%
}{%
{\protect \APACyear {2024}}%
}]{%
AutoGen2024n}
\APACinsertmetastar {%
AutoGen2024n}%
\begin{APACrefauthors}%
{AutoGen}.%
\end{APACrefauthors}%
\unskip\
\newblock
\APACrefYearMonthDay{2024}{}{}.
\newblock
\APACrefbtitle {{AutoGen}: A framework for building {AI} agents and applications.} {{AutoGen}: A framework for building {AI} agents and applications.}
\newblock
\begin{APACrefURL} \url{https://microsoft.github.io/autogen/stable/} \end{APACrefURL}
\newblock
\APACrefnote{Accessed: 2025-2-6}
\PrintBackRefs{\CurrentBib}

\bibitem [\protect \citeauthoryear {%
{AutoGPT}%
}{%
{AutoGPT}%
}{%
{\protect \APACyear {2024}}%
}]{%
AutoGPT2024i}
\APACinsertmetastar {%
AutoGPT2024i}%
\begin{APACrefauthors}%
{AutoGPT}.%
\end{APACrefauthors}%
\unskip\
\newblock
\APACrefYearMonthDay{2024}{}{}.
\newblock
\APACrefbtitle {{AutoGPT}: Build, Deploy, and Run {AI} Agents.} {{AutoGPT}: Build, deploy, and run {AI} agents.}
\newblock
\begin{APACrefURL} \url{https://github.com/Significant-Gravitas/AutoGPT} \end{APACrefURL}
\PrintBackRefs{\CurrentBib}

\bibitem [\protect \citeauthoryear {%
Autor%
, Chin%
, Salomons%
\BCBL {}\ \BBA {} Seegmiller%
}{%
Autor%
\ \protect \BOthers {.}}{%
{\protect \APACyear {2022}}%
}]{%
Autor2022q}
\APACinsertmetastar {%
Autor2022q}%
\begin{APACrefauthors}%
Autor, D.%
, Chin, C.%
, Salomons, A.%
\BCBL {}\ \BBA {} Seegmiller, B.%
\end{APACrefauthors}%
\unskip\
\newblock
\APACrefYearMonthDay{2022}{}{}.
\newblock
\APACrefbtitle {New Frontiers: The Origins and Content of New Work, 1940--2018} {New frontiers: The origins and content of new work, 1940--2018}\ \APACbVolEdTR{}{\BTR{}\ \BNUM\ 30389}.
\newblock
\APACaddressInstitution{Cambridge, MA}{National Bureau of Economic Research}.
\newblock
\begin{APACrefURL} \url{http://www.nber.org/papers/w30389.pdf} \end{APACrefURL}
\PrintBackRefs{\CurrentBib}

\bibitem [\protect \citeauthoryear {%
Baker%
\ \BBA {} Church%
}{%
Baker%
\ \BBA {} Church%
}{%
{\protect \APACyear {2024}}%
}]{%
Baker2024k}
\APACinsertmetastar {%
Baker2024k}%
\begin{APACrefauthors}%
Baker, D.%
\BCBT {}\ \BBA {} Church, G.%
\end{APACrefauthors}%
\unskip\
\newblock
\APACrefYearMonthDay{2024}{}{}.
\newblock
{\BBOQ}\APACrefatitle {Protein design meets biosecurity} {Protein design meets biosecurity}.{\BBCQ}
\newblock
\APACjournalVolNumPages{Science (New York, N.Y.)}{383}{6681}{349}.
\newblock
\begin{APACrefURL} \url{https://www.science.org/doi/10.1126/science.ado1671} \end{APACrefURL}
\PrintBackRefs{\CurrentBib}

\bibitem [\protect \citeauthoryear {%
Barnette%
\ \BBA {} Michaud%
}{%
Barnette%
\ \BBA {} Michaud%
}{%
{\protect \APACyear {2017}}%
}]{%
Barnette2017f}
\APACinsertmetastar {%
Barnette2017f}%
\begin{APACrefauthors}%
Barnette, J.%
\BCBT {}\ \BBA {} Michaud, A.%
\end{APACrefauthors}%
\unskip\
\newblock
\APACrefYearMonthDay{2017}{}{}.
\newblock
\APACrefbtitle {Wage Scars and Human Capital Theory.} {Wage scars and human capital theory.}
\newblock
\begin{APACrefURL} \url{https://ammichau.github.io/papers/JBAMWageScar.pdf} \end{APACrefURL}
\PrintBackRefs{\CurrentBib}

\bibitem [\protect \citeauthoryear {%
Bengio%
, Hinton%
\BCBL {}\ \protect \BOthers {.}}{%
Bengio%
, Hinton%
\BCBL {}\ \protect \BOthers {.}}{%
{\protect \APACyear {2024}}%
}]{%
Bengio2024t}
\APACinsertmetastar {%
Bengio2024t}%
\begin{APACrefauthors}%
Bengio, Y.%
, Hinton, G.%
, Yao, A.%
, Song, D.%
, Abbeel, P.%
, Darrell, T.%
\BDBL {}Mindermann, S.%
\end{APACrefauthors}%
\unskip\
\newblock
\APACrefYearMonthDay{2024}{}{}.
\newblock
{\BBOQ}\APACrefatitle {Managing extreme {AI} risks amid rapid progress} {Managing extreme {AI} risks amid rapid progress}.{\BBCQ}
\newblock
\APACjournalVolNumPages{Science}{}{}{eadn0117}.
\newblock
\begin{APACrefURL} \url{http://dx.doi.org/10.1126/science.adn0117} \end{APACrefURL}
\PrintBackRefs{\CurrentBib}

\bibitem [\protect \citeauthoryear {%
Bengio%
\ \protect \BOthers {.}}{%
Bengio%
\ \protect \BOthers {.}}{%
{\protect \APACyear {2025}}%
}]{%
Bengio2025z}
\APACinsertmetastar {%
Bengio2025z}%
\begin{APACrefauthors}%
Bengio, Y.%
, Mindermann, S.%
, Privitera, D.%
, Besiroglu, T.%
, Bommasani, R.%
, Casper, S.%
\BDBL {}Zeng, Y.%
\end{APACrefauthors}%
\unskip\
\newblock
\APACrefYearMonthDay{2025}{}{}.
\newblock
\APACrefbtitle {International {AI} Safety Report} {International {AI} safety report}\ \APACbVolEdTR{}{\BTR{}}.
\newblock
\APACaddressInstitution{}{Department for Science, Innovation and Technology}.
\newblock
\begin{APACrefURL} \url{https://www.gov.uk/government/publications/international-ai-safety-report-2025} \end{APACrefURL}
\PrintBackRefs{\CurrentBib}

\bibitem [\protect \citeauthoryear {%
Bengio%
, Privitera%
\BCBL {}\ \protect \BOthers {.}}{%
Bengio%
, Privitera%
\BCBL {}\ \protect \BOthers {.}}{%
{\protect \APACyear {2024}}%
}]{%
Bengio2024k}
\APACinsertmetastar {%
Bengio2024k}%
\begin{APACrefauthors}%
Bengio, Y.%
, Privitera, D.%
, Besiroglu, T.%
, Bommasani, R.%
, Casper, S.%
, Choi, Y.%
\BDBL {}Mindermann, S.%
\end{APACrefauthors}%
\unskip\
\newblock
\APACrefYearMonthDay{2024}{}{}.
\newblock
\APACrefbtitle {International Scientific Report on the Safety of Advanced {AI}} {International scientific report on the safety of advanced {AI}}\ \APACbVolEdTR{}{\BTR{}}.
\newblock
\APACaddressInstitution{}{Department for Science, Innovation and Technology}.
\newblock
\begin{APACrefURL} \url{https://www.gov.uk/government/publications/international-scientific-report-on-the-safety-of-advanced-ai} \end{APACrefURL}
\PrintBackRefs{\CurrentBib}

\bibitem [\protect \citeauthoryear {%
Benthall%
\ \BBA {} Shekman%
}{%
Benthall%
\ \BBA {} Shekman%
}{%
{\protect \APACyear {2023}}%
}]{%
Benthall2023t}
\APACinsertmetastar {%
Benthall2023t}%
\begin{APACrefauthors}%
Benthall, S.%
\BCBT {}\ \BBA {} Shekman, D.%
\end{APACrefauthors}%
\unskip\
\newblock
\APACrefYearMonthDay{2023}{}{}.
\newblock
\APACrefbtitle {Designing fiduciary artificial intelligence.} {Designing fiduciary artificial intelligence.}
\newblock
\begin{APACrefURL} \url{http://arxiv.org/abs/2308.02435} \end{APACrefURL}
\PrintBackRefs{\CurrentBib}

\bibitem [\protect \citeauthoryear {%
Bernardi%
, Mukobi%
, Greaves%
, Heim%
\BCBL {}\ \BBA {} Anderljung%
}{%
Bernardi%
\ \protect \BOthers {.}}{%
{\protect \APACyear {2024}}%
}]{%
Bernardi2024i}
\APACinsertmetastar {%
Bernardi2024i}%
\begin{APACrefauthors}%
Bernardi, J.%
, Mukobi, G.%
, Greaves, H.%
, Heim, L.%
\BCBL {}\ \BBA {} Anderljung, M.%
\end{APACrefauthors}%
\unskip\
\newblock
\APACrefYearMonthDay{2024}{}{}.
\newblock
\APACrefbtitle {Societal Adaptation to Advanced {AI}.} {Societal adaptation to advanced {AI}.}
\newblock
\begin{APACrefURL} \url{http://arxiv.org/abs/2405.10295} \end{APACrefURL}
\PrintBackRefs{\CurrentBib}

\bibitem [\protect \citeauthoryear {%
Bianchi%
, Curry%
\BCBL {}\ \BBA {} Hovy%
}{%
Bianchi%
\ \protect \BOthers {.}}{%
{\protect \APACyear {2023}}%
}]{%
Bianchi2023y}
\APACinsertmetastar {%
Bianchi2023y}%
\begin{APACrefauthors}%
Bianchi, F.%
, Curry, A\BPBI C.%
\BCBL {}\ \BBA {} Hovy, D.%
\end{APACrefauthors}%
\unskip\
\newblock
\APACrefYearMonthDay{2023}{}{}.
\newblock
{\BBOQ}\APACrefatitle {Viewpoint: Artificial Intelligence Accidents Waiting to Happen?} {Viewpoint: Artificial intelligence accidents waiting to happen?}{\BBCQ}
\newblock
\APACjournalVolNumPages{Journal of Artificial Intelligence Research}{76}{}{193--199}.
\PrintBackRefs{\CurrentBib}

\bibitem [\protect \citeauthoryear {%
Bick%
, Blandin%
\BCBL {}\ \BBA {} Deming%
}{%
Bick%
\ \protect \BOthers {.}}{%
{\protect \APACyear {2024}}%
}]{%
Bick2024x}
\APACinsertmetastar {%
Bick2024x}%
\begin{APACrefauthors}%
Bick, A.%
, Blandin, A.%
\BCBL {}\ \BBA {} Deming, D.%
\end{APACrefauthors}%
\unskip\
\newblock
\APACrefYearMonthDay{2024}{}{}.
\newblock
\APACrefbtitle {The Rapid Adoption of Generative {AI}} {The rapid adoption of generative {AI}}\ \APACbVolEdTR{}{\BTR{}\ \BNUM\ w32966}.
\newblock
\APACaddressInstitution{Cambridge, MA}{National Bureau of Economic Research}.
\newblock
\begin{APACrefURL} \url{https://www.nber.org/system/files/working_papers/w32966/w32966.pdf} \end{APACrefURL}
\PrintBackRefs{\CurrentBib}

\bibitem [\protect \citeauthoryear {%
Biden%
}{%
Biden%
}{%
{\protect \APACyear {2023}}%
}]{%
Biden2023k}
\APACinsertmetastar {%
Biden2023k}%
\begin{APACrefauthors}%
Biden, J\BPBI R.%
\end{APACrefauthors}%
\unskip\
\newblock
\APACrefYearMonthDay{2023}{}{}.
\newblock
\APACrefbtitle {Safe, Secure, and Trustworthy Development and Use of Artificial Intelligence} {Safe, secure, and trustworthy development and use of artificial intelligence}\ (\BVOL~88)\ (\BNUM\ 2023-24283).
\newblock
\begin{APACrefURL} \url{https://www.federalregister.gov/d/2023-24283} \end{APACrefURL}
\PrintBackRefs{\CurrentBib}

\bibitem [\protect \citeauthoryear {%
{Bipartisan Commission on Biodefense}%
}{%
{Bipartisan Commission on Biodefense}%
}{%
{\protect \APACyear {2021}}%
}]{%
BipartisanCommissionOnBiodefense2021o}
\APACinsertmetastar {%
BipartisanCommissionOnBiodefense2021o}%
\begin{APACrefauthors}%
{Bipartisan Commission on Biodefense}.%
\end{APACrefauthors}%
\unskip\
\newblock
\APACrefYearMonthDay{2021}{}{}.
\newblock
\APACrefbtitle {The Apollo Program for Biodefense - Winning the Race Against Biological Threats.} {The apollo program for biodefense - winning the race against biological threats.}
\newblock
\begin{APACrefURL} \url{https://biodefensecommission.org/wp-content/uploads/2021/01/Apollo\_report\_final\_v8\_033121\_web.pdf} \end{APACrefURL}
\newblock
\APACrefnote{Accessed: 2025-2-6}
\PrintBackRefs{\CurrentBib}

\bibitem [\protect \citeauthoryear {%
{Bletchley Park}%
}{%
{Bletchley Park}%
}{%
{\protect \APACyear {2023}}%
}]{%
BletchleyPark2023w}
\APACinsertmetastar {%
BletchleyPark2023w}%
\begin{APACrefauthors}%
{Bletchley Park}.%
\end{APACrefauthors}%
\unskip\
\newblock
\APACrefYearMonthDay{2023}{}{}.
\newblock
\APACrefbtitle {Bletchley Park and {AI}.} {Bletchley park and {AI}.}
\newblock
\begin{APACrefURL} \url{https://bletchleypark.org.uk/wp-content/uploads/2023/10/Bletchley-Park-and-AI.pdf} \end{APACrefURL}
\PrintBackRefs{\CurrentBib}

\bibitem [\protect \citeauthoryear {%
Bloomfield%
\ \protect \BOthers {.}}{%
Bloomfield%
\ \protect \BOthers {.}}{%
{\protect \APACyear {2024}}%
}]{%
Bloomfield2024s}
\APACinsertmetastar {%
Bloomfield2024s}%
\begin{APACrefauthors}%
Bloomfield, D.%
, Pannu, J.%
, Zhu, A\BPBI W.%
, Ng, M\BPBI Y.%
, Lewis, A.%
, Bendavid, E.%
\BDBL {}Inglesby, T.%
\end{APACrefauthors}%
\unskip\
\newblock
\APACrefYearMonthDay{2024}{}{}.
\newblock
{\BBOQ}\APACrefatitle {{AI} and biosecurity: The need for governance} {{AI} and biosecurity: The need for governance}.{\BBCQ}
\newblock
\APACjournalVolNumPages{Science (New York, N.Y.)}{385}{6711}{831--833}.
\newblock
\begin{APACrefURL} \url{https://www.science.org/doi/10.1126/science.adq1977} \end{APACrefURL}
\PrintBackRefs{\CurrentBib}

\bibitem [\protect \citeauthoryear {%
Bostrom%
}{%
Bostrom%
}{%
{\protect \APACyear {2014}}%
}]{%
Bostrom2014q}
\APACinsertmetastar {%
Bostrom2014q}%
\begin{APACrefauthors}%
Bostrom, N.%
\end{APACrefauthors}%
\unskip\
\newblock
\APACrefYear{2014}.
\newblock
\APACrefbtitle {Superintelligence: Paths, dangers, strategies} {Superintelligence: Paths, dangers, strategies}.
\newblock
\APACaddressPublisher{}{Oxford University Press}.
\newblock
\begin{APACrefURL} \url{https://global.oup.com/academic/product/superintelligence-9780199678112?cc=mx\&lang=en\&} \end{APACrefURL}
\PrintBackRefs{\CurrentBib}

\bibitem [\protect \citeauthoryear {%
Bown%
\ \BBA {} Kolb%
}{%
Bown%
\ \BBA {} Kolb%
}{%
{\protect \APACyear {2018}}%
}]{%
Bown2018z}
\APACinsertmetastar {%
Bown2018z}%
\begin{APACrefauthors}%
Bown, C\BPBI P.%
\BCBT {}\ \BBA {} Kolb, M.%
\end{APACrefauthors}%
\unskip\
\newblock
\APACrefYearMonthDay{2018}{}{}.
\newblock
\APACrefbtitle {Trump's Trade War Timeline: An Up-to-Date Guide} {Trump's trade war timeline: An up-to-date guide}\ \APACbVolEdTR{}{\BTR{}}.
\newblock
\APACaddressInstitution{}{Peterson Institute for International Economics}.
\newblock
\begin{APACrefURL} \url{https://www.piie.com/sites/default/files/documents/trump-trade-war-timeline.pdf} \end{APACrefURL}
\PrintBackRefs{\CurrentBib}

\bibitem [\protect \citeauthoryear {%
Bradford%
}{%
Bradford%
}{%
{\protect \APACyear {2020}}%
}]{%
Bradford2020v}
\APACinsertmetastar {%
Bradford2020v}%
\begin{APACrefauthors}%
Bradford, A.%
\end{APACrefauthors}%
\unskip\
\newblock
\APACrefYear{2020}.
\newblock
\APACrefbtitle {The Brussels Effect: How the European Union Rules the World} {The brussels effect: How the european union rules the world}.
\newblock
\APACaddressPublisher{}{Oxford University Press}.
\newblock
\begin{APACrefURL} \url{https://academic.oup.com/book/36491} \end{APACrefURL}
\PrintBackRefs{\CurrentBib}

\bibitem [\protect \citeauthoryear {%
Brynjolfsson%
, Li%
\BCBL {}\ \BBA {} Raymond%
}{%
Brynjolfsson%
\ \protect \BOthers {.}}{%
{\protect \APACyear {2023}}%
}]{%
Brynjolfsson2023d}
\APACinsertmetastar {%
Brynjolfsson2023d}%
\begin{APACrefauthors}%
Brynjolfsson, E.%
, Li, D.%
\BCBL {}\ \BBA {} Raymond, L.%
\end{APACrefauthors}%
\unskip\
\newblock
\APACrefYearMonthDay{2023}{}{}.
\newblock
\APACrefbtitle {Generative {AI} at Work} {Generative {AI} at work}\ \APACbVolEdTR{}{\BTR{}\ \BNUM\ w31161}.
\newblock
\APACaddressInstitution{Cambridge, MA}{National Bureau of Economic Research}.
\newblock
\begin{APACrefURL} \url{http://www.nber.org/papers/w31161.pdf} \end{APACrefURL}
\PrintBackRefs{\CurrentBib}

\bibitem [\protect \citeauthoryear {%
{Bureau of Industry and Security}%
}{%
{Bureau of Industry and Security}%
}{%
{\protect \APACyear {2022}}%
}]{%
BureauOfIndustryAndSecurity2022e}
\APACinsertmetastar {%
BureauOfIndustryAndSecurity2022e}%
\begin{APACrefauthors}%
{Bureau of Industry and Security}.%
\end{APACrefauthors}%
\unskip\
\newblock
\APACrefYearMonthDay{2022}{}{}.
\newblock
\APACrefbtitle {Commerce Implements New Export Controls on Advanced Computing and Semiconductor Manufacturing Items to the People's Republic of China ({PRC})} {Commerce implements new export controls on advanced computing and semiconductor manufacturing items to the people's republic of china ({PRC})}\ \APACbVolEdTR{}{\BTR{}}.
\newblock
\APACaddressInstitution{}{U.S. Department of Commerce}.
\newblock
\begin{APACrefURL} \url{https://www.bis.doc.gov/index.php/documents/about-bis/newsroom/press-releases/3158-2022-10-07-bis-press-release-advanced-computing-and-semiconductor-manufacturing-controls-final/file} \end{APACrefURL}
\PrintBackRefs{\CurrentBib}

\bibitem [\protect \citeauthoryear {%
Burgan%
}{%
Burgan%
}{%
{\protect \APACyear {2024}}%
}]{%
Burgan2024p}
\APACinsertmetastar {%
Burgan2024p}%
\begin{APACrefauthors}%
Burgan, C.%
\end{APACrefauthors}%
\unskip\
\newblock
\APACrefYearMonthDay{2024}{}{}.
\newblock
\APACrefbtitle {Experts See `Pro {AI}' Policy Coming From Trump Administration.} {Experts see `pro {AI}' policy coming from trump administration.}
\newblock
\begin{APACrefURL} \url{https://www.meritalk.com/articles/experts-see-pro-ai-policy-coming-from-trump-administration/} \end{APACrefURL}
\newblock
\APACrefnote{Accessed: 2025-2-6}
\PrintBackRefs{\CurrentBib}

\bibitem [\protect \citeauthoryear {%
Burgard%
, Brand%
\BCBL {}\ \BBA {} House%
}{%
Burgard%
\ \protect \BOthers {.}}{%
{\protect \APACyear {2007}}%
}]{%
Burgard2007v}
\APACinsertmetastar {%
Burgard2007v}%
\begin{APACrefauthors}%
Burgard, S\BPBI A.%
, Brand, J\BPBI E.%
\BCBL {}\ \BBA {} House, J\BPBI S.%
\end{APACrefauthors}%
\unskip\
\newblock
\APACrefYearMonthDay{2007}{}{}.
\newblock
{\BBOQ}\APACrefatitle {Toward a Better Estimation of the Effect of Job Loss on Health} {Toward a better estimation of the effect of job loss on health}.{\BBCQ}
\newblock
\APACjournalVolNumPages{Journal of Health and Social Behavior}{48}{4}{369--384}.
\newblock
\begin{APACrefURL} \url{http://www.jstor.org/stable/27638722} \end{APACrefURL}
\PrintBackRefs{\CurrentBib}

\bibitem [\protect \citeauthoryear {%
{Cabinet Office}%
, {Office for Artificial Intelligence}%
, {Centre for Data Ethics and Innovation}%
\BCBL {}\ \BBA {} {Department for Science, Innovation \& Technology}%
}{%
{Cabinet Office}%
\ \protect \BOthers {.}}{%
{\protect \APACyear {2023}}%
}]{%
CabinetOffice2023g}
\APACinsertmetastar {%
CabinetOffice2023g}%
\begin{APACrefauthors}%
{Cabinet Office}%
, {Office for Artificial Intelligence}%
, {Centre for Data Ethics and Innovation}%
\BCBL {}\ \BBA {} {Department for Science, Innovation \& Technology}.%
\end{APACrefauthors}%
\unskip\
\newblock
\APACrefYearMonthDay{2023}{}{}.
\newblock
\APACrefbtitle {Ethics, Transparency and Accountability Framework for Automated Decision-Making.} {Ethics, transparency and accountability framework for automated decision-making.}
\newblock
\begin{APACrefURL} \url{https://www.gov.uk/government/publications/ethics-transparency-and-accountability-framework-for-automated-decision-making/ethics-transparency-and-accountability-framework-for-automated-decision-making} \end{APACrefURL}
\newblock
\APACrefnote{Accessed: 2025-2-4}
\PrintBackRefs{\CurrentBib}

\bibitem [\protect \citeauthoryear {%
Casalicchio%
\ \BBA {} Manancourt%
}{%
Casalicchio%
\ \BBA {} Manancourt%
}{%
{\protect \APACyear {2023}}%
}]{%
Casalicchio2023t}
\APACinsertmetastar {%
Casalicchio2023t}%
\begin{APACrefauthors}%
Casalicchio, E.%
\BCBT {}\ \BBA {} Manancourt, V.%
\end{APACrefauthors}%
\unskip\
\newblock
\APACrefYearMonthDay{2023}{}{}.
\newblock
\APACrefbtitle {We won't cut China out of {AI} summit over spying scandal, {UK} says.} {We won't cut china out of {AI} summit over spying scandal, {UK} says.}
\newblock
\begin{APACrefURL} \url{https://www.politico.eu/article/we-wont-cut-china-out-of-ai-summit-over-spying-scandal-uk-says/} \end{APACrefURL}
\newblock
\APACrefnote{Accessed: 2025-2-5}
\PrintBackRefs{\CurrentBib}

\bibitem [\protect \citeauthoryear {%
Casey%
, Roy%
\BCBL {}\ \BBA {} Rockall%
}{%
Casey%
\ \protect \BOthers {.}}{%
{\protect \APACyear {2024}}%
}]{%
Casey2024z}
\APACinsertmetastar {%
Casey2024z}%
\begin{APACrefauthors}%
Casey, E.%
, Roy, H.%
\BCBL {}\ \BBA {} Rockall, E.%
\end{APACrefauthors}%
\unskip\
\newblock
\APACrefYearMonthDay{2024}{}{}.
\newblock
\APACrefbtitle {Designing an {AI} Bond for Growth and Shared Prosperity in the {UK}} {Designing an {AI} bond for growth and shared prosperity in the {UK}}\ \APACbVolEdTR{}{\BTR{}}.
\newblock
\APACaddressInstitution{}{UK Day One}.
\newblock
\begin{APACrefURL} \url{https://ukdayone.org/briefings/ai-bond-for-growth-and-shared-prosperity} \end{APACrefURL}
\PrintBackRefs{\CurrentBib}

\bibitem [\protect \citeauthoryear {%
Cellan-Jones%
}{%
Cellan-Jones%
}{%
{\protect \APACyear {2014}}%
}]{%
Cellan-Jones2014a}
\APACinsertmetastar {%
Cellan-Jones2014a}%
\begin{APACrefauthors}%
Cellan-Jones, R.%
\end{APACrefauthors}%
\unskip\
\newblock
\APACrefYearMonthDay{2014}{}{}.
\newblock
{\BBOQ}\APACrefatitle {Stephen Hawking warns artificial intelligence could end mankind} {Stephen hawking warns artificial intelligence could end mankind}.{\BBCQ}
\newblock
\APACjournalVolNumPages{BBC News}{}{}{}.
\newblock
\begin{APACrefURL} \url{https://www.bbc.com/news/technology-30290540} \end{APACrefURL}
\PrintBackRefs{\CurrentBib}

\bibitem [\protect \citeauthoryear {%
{Center for AI Safety}%
}{%
{Center for AI Safety}%
}{%
{\protect \APACyear {2024}}%
}]{%
CenterForAISafety2024q}
\APACinsertmetastar {%
CenterForAISafety2024q}%
\begin{APACrefauthors}%
{Center for AI Safety}.%
\end{APACrefauthors}%
\unskip\
\newblock
\APACrefYearMonthDay{2024}{}{}.
\newblock
\APACrefbtitle {Statement on {AI} Risk: {AI} experts and public figures express their concern about {AI} risk.} {Statement on {AI} risk: {AI} experts and public figures express their concern about {AI} risk.}
\newblock
\begin{APACrefURL} \url{https://www.safe.ai/work/statement-on-ai-risk} \end{APACrefURL}
\newblock
\APACrefnote{Accessed: 2024-4-25}
\PrintBackRefs{\CurrentBib}

\bibitem [\protect \citeauthoryear {%
{Centre for Data Ethics and Innovation}%
}{%
{Centre for Data Ethics and Innovation}%
}{%
{\protect \APACyear {2021}}%
}]{%
CentreForDataEthicsAndInnovation2021u}
\APACinsertmetastar {%
CentreForDataEthicsAndInnovation2021u}%
\begin{APACrefauthors}%
{Centre for Data Ethics and Innovation}.%
\end{APACrefauthors}%
\unskip\
\newblock
\APACrefYearMonthDay{2021}{}{}.
\newblock
\APACrefbtitle {The roadmap to an effective {AI} assurance ecosystem.} {The roadmap to an effective {AI} assurance ecosystem.}
\newblock
\begin{APACrefURL} \url{https://assets.publishing.service.gov.uk/media/61b0746b8fa8f50379269eb3/The_roadmap_to_an_effective_AI_assurance_ecosystem.pdf} \end{APACrefURL}
\PrintBackRefs{\CurrentBib}

\bibitem [\protect \citeauthoryear {%
Chan%
\ \protect \BOthers {.}}{%
Chan%
\ \protect \BOthers {.}}{%
{\protect \APACyear {2024}}%
}]{%
Chan2024o}
\APACinsertmetastar {%
Chan2024o}%
\begin{APACrefauthors}%
Chan, A.%
, Kolt, N.%
, Wills, P.%
, Anwar, U.%
, de Witt, C\BPBI S.%
, Rajkumar, N.%
\BDBL {}Anderljung, M.%
\end{APACrefauthors}%
\unskip\
\newblock
\APACrefYearMonthDay{2024}{}{}.
\newblock
\APACrefbtitle {{IDs} for {AI} Systems.} {{IDs} for {AI} systems.}
\newblock
\begin{APACrefURL} \url{http://arxiv.org/abs/2406.12137} \end{APACrefURL}
\PrintBackRefs{\CurrentBib}

\bibitem [\protect \citeauthoryear {%
Chan%
\ \protect \BOthers {.}}{%
Chan%
\ \protect \BOthers {.}}{%
{\protect \APACyear {2025}}%
}]{%
Chan2025m}
\APACinsertmetastar {%
Chan2025m}%
\begin{APACrefauthors}%
Chan, A.%
, Wei, K.%
, Huang, S.%
, Rajkumar, N.%
, Perrier, E.%
, Lazar, S.%
\BDBL {}Anderljung, M.%
\end{APACrefauthors}%
\unskip\
\newblock
\APACrefYearMonthDay{2025}{}{}.
\newblock
\APACrefbtitle {Infrastructure for {AI} Agents.} {Infrastructure for {AI} agents.}
\newblock
\begin{APACrefURL} \url{http://arxiv.org/abs/2501.10114} \end{APACrefURL}
\PrintBackRefs{\CurrentBib}

\bibitem [\protect \citeauthoryear {%
Chen%
\ \protect \BOthers {.}}{%
Chen%
\ \protect \BOthers {.}}{%
{\protect \APACyear {2023}}%
}]{%
Chen2023x}
\APACinsertmetastar {%
Chen2023x}%
\begin{APACrefauthors}%
Chen, B.%
, Cheng, X.%
, Li, P.%
, Geng, Y\BHBI A.%
, Gong, J.%
, Li, S.%
\BDBL {}Song, L.%
\end{APACrefauthors}%
\unskip\
\newblock
\APACrefYearMonthDay{2023}{}{}.
\newblock
\APACrefbtitle {{XTrimoPGLM}: Unified {100B}-scale pre-trained transformer for deciphering the language of protein.} {{XTrimoPGLM}: Unified {100B}-scale pre-trained transformer for deciphering the language of protein.}
\newblock
\begin{APACrefURL} \url{https://www.biorxiv.org/content/10.1101/2023.07.05.547496v1.abstract} \end{APACrefURL}
\PrintBackRefs{\CurrentBib}

\bibitem [\protect \citeauthoryear {%
Choi%
, Monahan%
\BCBL {}\ \BBA {} Schwarcz%
}{%
Choi%
\ \protect \BOthers {.}}{%
{\protect \APACyear {2023}}%
}]{%
Choi2023p}
\APACinsertmetastar {%
Choi2023p}%
\begin{APACrefauthors}%
Choi, J\BPBI H.%
, Monahan, A.%
\BCBL {}\ \BBA {} Schwarcz, D\BPBI B.%
\end{APACrefauthors}%
\unskip\
\newblock
\APACrefYearMonthDay{2023}{}{}.
\newblock
\APACrefbtitle {Lawyering in the age of artificial intelligence.} {Lawyering in the age of artificial intelligence.}
\newblock
\begin{APACrefURL} \url{https://papers.ssrn.com/abstract=4626276} \end{APACrefURL}
\PrintBackRefs{\CurrentBib}

\bibitem [\protect \citeauthoryear {%
{Coalition for Content Provenance and Authenticity}%
}{%
{Coalition for Content Provenance and Authenticity}%
}{%
{\protect \APACyear {{\protect \bibnodate {}}}}%
}]{%
CoalitionForContentProvenanceAndAuthenticityOtherb}
\APACinsertmetastar {%
CoalitionForContentProvenanceAndAuthenticityOtherb}%
\begin{APACrefauthors}%
{Coalition for Content Provenance and Authenticity}.%
\end{APACrefauthors}%
\unskip\
\newblock
\APACrefYearMonthDay{{\protect \bibnodate {}}}{}{}.
\newblock
\APACrefbtitle {{C2PA} Specifications.} {{C2PA} specifications.}
\newblock
\begin{APACrefURL} \url{https://c2pa.org/specifications/specifications/1.3/index.html} \end{APACrefURL}
\newblock
\APACrefnote{Accessed: 2025-2-6}
\PrintBackRefs{\CurrentBib}

\bibitem [\protect \citeauthoryear {%
{Coalition for Content Provenance and Authenticity (C2PA)}%
}{%
{Coalition for Content Provenance and Authenticity (C2PA)}%
}{%
{\protect \APACyear {2021}}%
}]{%
CoalitionForContentProvenanceAndAuthenticityC2PA2021e}
\APACinsertmetastar {%
CoalitionForContentProvenanceAndAuthenticityC2PA2021e}%
\begin{APACrefauthors}%
{Coalition for Content Provenance and Authenticity (C2PA)}.%
\end{APACrefauthors}%
\unskip\
\newblock
\APACrefYearMonthDay{2021}{}{}.
\newblock
\APACrefbtitle {Overview.} {Overview.}
\newblock
\begin{APACrefURL} \url{https://c2pa.org/} \end{APACrefURL}
\newblock
\APACrefnote{Accessed: 2025-2-6}
\PrintBackRefs{\CurrentBib}

\bibitem [\protect \citeauthoryear {%
Courea%
\ \BBA {} Stacey%
}{%
Courea%
\ \BBA {} Stacey%
}{%
{\protect \APACyear {2025}}%
}]{%
Courea2025r}
\APACinsertmetastar {%
Courea2025r}%
\begin{APACrefauthors}%
Courea, E.%
\BCBT {}\ \BBA {} Stacey, K.%
\end{APACrefauthors}%
\unskip\
\newblock
\APACrefYearMonthDay{2025}{}{}.
\newblock
{\BBOQ}\APACrefatitle {{UK} ministers delay {AI} regulation amid plans for more `comprehensive' bill} {{UK} ministers delay {AI} regulation amid plans for more `comprehensive' bill}.{\BBCQ}
\newblock
\APACjournalVolNumPages{The Guardian}{}{}{}.
\newblock
\begin{APACrefURL} \url{https://www.theguardian.com/technology/2025/jun/07/uk-ministers-delay-ai-regulation-amid-plans-for-more-comprehensive-bill} \end{APACrefURL}
\PrintBackRefs{\CurrentBib}

\bibitem [\protect \citeauthoryear {%
K\BPBI Z.~Cui%
\ \protect \BOthers {.}}{%
K\BPBI Z.~Cui%
\ \protect \BOthers {.}}{%
{\protect \APACyear {2024}}%
}]{%
Cui2024c}
\APACinsertmetastar {%
Cui2024c}%
\begin{APACrefauthors}%
Cui, K\BPBI Z.%
, Demirer, M.%
, Jaffe, S.%
, Musolff, L.%
, Peng, S.%
\BCBL {}\ \BBA {} Salz, T.%
\end{APACrefauthors}%
\unskip\
\newblock
\APACrefYearMonthDay{2024}{}{}.
\newblock
{\BBOQ}\APACrefatitle {The Productivity Effects of Generative {AI}: Evidence from a Field Experiment with {GitHub} Copilot} {The productivity effects of generative {AI}: Evidence from a field experiment with {GitHub} copilot}.{\BBCQ}
\newblock
\APACjournalVolNumPages{An MIT Exploration of Generative AI}{}{}{}.
\newblock
\begin{APACrefURL} \url{https://mit-genai.pubpub.org/pub/v5iixksv/release/2} \end{APACrefURL}
\PrintBackRefs{\CurrentBib}

\bibitem [\protect \citeauthoryear {%
Z.~Cui%
\ \protect \BOthers {.}}{%
Z.~Cui%
\ \protect \BOthers {.}}{%
{\protect \APACyear {2024}}%
}]{%
Cui2024a}
\APACinsertmetastar {%
Cui2024a}%
\begin{APACrefauthors}%
Cui, Z.%
, Demirer, M.%
, Jaffe, S.%
, Musolff, L.%
, Peng, S.%
\BCBL {}\ \BBA {} Salz, T.%
\end{APACrefauthors}%
\unskip\
\newblock
\APACrefYearMonthDay{2024}{}{}.
\newblock
\APACrefbtitle {The effects of generative {AI} on high skilled work: Evidence from three field experiments with software developers.} {The effects of generative {AI} on high skilled work: Evidence from three field experiments with software developers.}
\newblock
\begin{APACrefURL} \url{https://papers.ssrn.com/abstract=4945566} \end{APACrefURL}
\PrintBackRefs{\CurrentBib}

\bibitem [\protect \citeauthoryear {%
DeepMind%
}{%
DeepMind%
}{%
{\protect \APACyear {2024}}%
}]{%
DeepMind2024w}
\APACinsertmetastar {%
DeepMind2024w}%
\begin{APACrefauthors}%
DeepMind, G.%
\end{APACrefauthors}%
\unskip\
\newblock
\APACrefYearMonthDay{2024}{}{}.
\newblock
\APACrefbtitle {Watermarking {AI}-generated text and video with {SynthID}.} {Watermarking {AI}-generated text and video with {SynthID}.}
\newblock
\begin{APACrefURL} \url{https://deepmind.google/discover/blog/watermarking-ai-generated-text-and-video-with-synthid/} \end{APACrefURL}
\newblock
\APACrefnote{Accessed: 2025-2-6}
\PrintBackRefs{\CurrentBib}

\bibitem [\protect \citeauthoryear {%
{Department for Business, Energy \& Industrial Strategy}%
}{%
{Department for Business, Energy \& Industrial Strategy}%
}{%
{\protect \APACyear {2017}}%
}]{%
DepartmentForBusinessEnergyIndustrialStrategy2017t}
\APACinsertmetastar {%
DepartmentForBusinessEnergyIndustrialStrategy2017t}%
\begin{APACrefauthors}%
{Department for Business, Energy \& Industrial Strategy}.%
\end{APACrefauthors}%
\unskip\
\newblock
\APACrefYearMonthDay{2017}{}{}.
\newblock
\APACrefbtitle {Industrial Strategy: Building a Britain fit for the future.} {Industrial strategy: Building a britain fit for the future.}
\newblock
\begin{APACrefURL} \url{https://assets.publishing.service.gov.uk/government/uploads/system/uploads/attachment\_data/file/664563/industrial-strategy-white-paper-web-ready-version.pdf} \end{APACrefURL}
\PrintBackRefs{\CurrentBib}

\bibitem [\protect \citeauthoryear {%
{Department for Digital, Culture, Media \& Sport}%
}{%
{Department for Digital, Culture, Media \& Sport}%
}{%
{\protect \APACyear {2020}}%
}]{%
DepartmentForDigitalCultureMediaSport2020a}
\APACinsertmetastar {%
DepartmentForDigitalCultureMediaSport2020a}%
\begin{APACrefauthors}%
{Department for Digital, Culture, Media \& Sport}.%
\end{APACrefauthors}%
\unskip\
\newblock
\APACrefYearMonthDay{2020}{}{}.
\newblock
\APACrefbtitle {Government minded to appoint Ofcom as online harms regulator.} {Government minded to appoint ofcom as online harms regulator.}
\newblock
\begin{APACrefURL} \url{https://www.gov.uk/government/news/government-minded-to-appoint-ofcom-as-online-harms-regulator} \end{APACrefURL}
\newblock
\APACrefnote{Accessed: 2025-2-5}
\PrintBackRefs{\CurrentBib}

\bibitem [\protect \citeauthoryear {%
{Department for Digital, Culture, Media \& Sport}%
\ \BBA {} {Department for Business, Energy \& Industrial Strategy}%
}{%
{Department for Digital, Culture, Media \& Sport}%
\ \BBA {} {Department for Business, Energy \& Industrial Strategy}%
}{%
{\protect \APACyear {2018}}%
}]{%
DepartmentForDigitalCultureMediaSport2018h}
\APACinsertmetastar {%
DepartmentForDigitalCultureMediaSport2018h}%
\begin{APACrefauthors}%
{Department for Digital, Culture, Media \& Sport}%
\BCBT {}\ \BBA {} {Department for Business, Energy \& Industrial Strategy}.%
\end{APACrefauthors}%
\unskip\
\newblock
\APACrefYearMonthDay{2018}{}{}.
\newblock
\APACrefbtitle {Stellar new board appointed to lead world-first Centre for Data Ethics and Innovation.} {Stellar new board appointed to lead world-first centre for data ethics and innovation.}
\newblock
\begin{APACrefURL} \url{https://www.gov.uk/government/news/stellar-new-board-appointed-to-lead-world-first-centre-for-data-ethics-and-innovation} \end{APACrefURL}
\newblock
\APACrefnote{Accessed: 2025-2-4}
\PrintBackRefs{\CurrentBib}

\bibitem [\protect \citeauthoryear {%
{Department for Digital, Culture, Media \& Sport, Office for Artificial Intelligence}%
\ \BBA {} Philp%
}{%
{Department for Digital, Culture, Media \& Sport, Office for Artificial Intelligence}%
\ \BBA {} Philp%
}{%
{\protect \APACyear {2022}}%
}]{%
DepartmentForDigitalCultureMediaSportOfficeForArtificialIntelligence2022s}
\APACinsertmetastar {%
DepartmentForDigitalCultureMediaSportOfficeForArtificialIntelligence2022s}%
\begin{APACrefauthors}%
{Department for Digital, Culture, Media \& Sport, Office for Artificial Intelligence}%
\BCBT {}\ \BBA {} Philp, C.%
\end{APACrefauthors}%
\unskip\
\newblock
\APACrefYearMonthDay{2022}{}{}.
\newblock
\APACrefbtitle {New {UK} initiative to shape global standards for Artificial Intelligence.} {New {UK} initiative to shape global standards for artificial intelligence.}
\newblock
\begin{APACrefURL} \url{https://www.gov.uk/government/news/new-uk-initiative-to-shape-global-standards-for-artificial-intelligence} \end{APACrefURL}
\newblock
\APACrefnote{Accessed: 2025-2-5}
\PrintBackRefs{\CurrentBib}

\bibitem [\protect \citeauthoryear {%
{Department for Education}%
}{%
{Department for Education}%
}{%
{\protect \APACyear {2023}}%
}]{%
DepartmentForEducation2023s}
\APACinsertmetastar {%
DepartmentForEducation2023s}%
\begin{APACrefauthors}%
{Department for Education}.%
\end{APACrefauthors}%
\unskip\
\newblock
\APACrefYearMonthDay{2023}{}{}.
\newblock
\APACrefbtitle {The impact of {AI} on {UK} jobs and training.} {The impact of {AI} on {UK} jobs and training.}
\newblock
\begin{APACrefURL} \url{https://www.gov.uk/government/publications/the-impact-of-ai-on-uk-jobs-and-training} \end{APACrefURL}
\newblock
\APACrefnote{Accessed: 2025-2-6}
\PrintBackRefs{\CurrentBib}

\bibitem [\protect \citeauthoryear {%
{Department for Science, Innovation \& Technology}%
}{%
{Department for Science, Innovation \& Technology}%
}{%
{\protect \APACyear {{\protect \bibnodate {}}}}%
}]{%
DepartmentForScienceInnovationTechnologyOthera}
\APACinsertmetastar {%
DepartmentForScienceInnovationTechnologyOthera}%
\begin{APACrefauthors}%
{Department for Science, Innovation \& Technology}.%
\end{APACrefauthors}%
\unskip\
\newblock
\APACrefYearMonthDay{{\protect \bibnodate {}}}{}{}.
\newblock
\APACrefbtitle {{AI} Opportunities Action Plan: terms of reference.} {{AI} opportunities action plan: terms of reference.}
\newblock
\begin{APACrefURL} \url{https://www.gov.uk/government/publications/artificial-intelligence-ai-opportunities-action-plan-terms-of-reference/artificial-intelligence-ai-opportunities-action-plan-terms-of-reference} \end{APACrefURL}
\newblock
\APACrefnote{Accessed: 2025-2-6}
\PrintBackRefs{\CurrentBib}

\bibitem [\protect \citeauthoryear {%
{Department for Science, Innovation \& Technology}%
}{%
{Department for Science, Innovation \& Technology}%
}{%
{\protect \APACyear {2023}}%
}]{%
DepartmentForScienceInnovationTechnology2023j}
\APACinsertmetastar {%
DepartmentForScienceInnovationTechnology2023j}%
\begin{APACrefauthors}%
{Department for Science, Innovation \& Technology}.%
\end{APACrefauthors}%
\unskip\
\newblock
\APACrefYearMonthDay{2023}{}{}.
\newblock
\APACrefbtitle {Digital Regulation: driving growth and unlocking innovation.} {Digital regulation: driving growth and unlocking innovation.}
\newblock
\begin{APACrefURL} \url{https://www.gov.uk/government/publications/digital-regulation-driving-growth-and-unlocking-innovation/digital-regulation-driving-growth-and-unlocking-innovation} \end{APACrefURL}
\newblock
\APACrefnote{Accessed: 2025-2-5}
\PrintBackRefs{\CurrentBib}

\bibitem [\protect \citeauthoryear {%
{Department for Science, Innovation \& Technology}%
}{%
{Department for Science, Innovation \& Technology}%
}{%
{\protect \APACyear {2024}}%
{\protect \APACexlab {{\protect \BCnt {1}}}}}]{%
DepartmentForScienceInnovationTechnology2024v}
\APACinsertmetastar {%
DepartmentForScienceInnovationTechnology2024v}%
\begin{APACrefauthors}%
{Department for Science, Innovation \& Technology}.%
\end{APACrefauthors}%
\unskip\
\newblock
\APACrefYearMonthDay{2024{\protect \BCnt {1}}}{}{}.
\newblock
\APACrefbtitle {Frontier {AI} Safety Commitments, {AI} Seoul Summit 2024.} {Frontier {AI} safety commitments, {AI} seoul summit 2024.}
\newblock
\begin{APACrefURL} \url{https://www.gov.uk/government/publications/frontier-ai-safety-commitments-ai-seoul-summit-2024/frontier-ai-safety-commitments-ai-seoul-summit-2024} \end{APACrefURL}
\newblock
\APACrefnote{Accessed: 2024-6-10}
\PrintBackRefs{\CurrentBib}

\bibitem [\protect \citeauthoryear {%
{Department for Science, Innovation \& Technology}%
}{%
{Department for Science, Innovation \& Technology}%
}{%
{\protect \APACyear {2024}}%
{\protect \APACexlab {{\protect \BCnt {2}}}}}]{%
DepartmentForScienceInnovationTechnology2024q}
\APACinsertmetastar {%
DepartmentForScienceInnovationTechnology2024q}%
\begin{APACrefauthors}%
{Department for Science, Innovation \& Technology}.%
\end{APACrefauthors}%
\unskip\
\newblock
\APACrefYearMonthDay{2024{\protect \BCnt {2}}}{}{}.
\newblock
\APACrefbtitle {A pro-innovation approach to {AI} regulation: government response.} {A pro-innovation approach to {AI} regulation: government response.}
\newblock
\begin{APACrefURL} \url{https://www.gov.uk/government/consultations/ai-regulation-a-pro-innovation-approach-policy-proposals/outcome/a-pro-innovation-approach-to-ai-regulation-government-response} \end{APACrefURL}
\newblock
\APACrefnote{Accessed: 2025-2-6}
\PrintBackRefs{\CurrentBib}

\bibitem [\protect \citeauthoryear {%
{Department for Science, Innovation \& Technology}%
}{%
{Department for Science, Innovation \& Technology}%
}{%
{\protect \APACyear {2025}}%
}]{%
DepartmentForScienceInnovationTechnology2025k}
\APACinsertmetastar {%
DepartmentForScienceInnovationTechnology2025k}%
\begin{APACrefauthors}%
{Department for Science, Innovation \& Technology}.%
\end{APACrefauthors}%
\unskip\
\newblock
\APACrefYearMonthDay{2025}{}{}.
\newblock
\APACrefbtitle {{AI} Opportunities Action Plan: government response.} {{AI} opportunities action plan: government response.}
\newblock
\begin{APACrefURL} \url{https://assets.publishing.service.gov.uk/media/678639913a9388161c5d2376/ai\_opportunities\_action\_plan\_government\_repsonse.pdf} \end{APACrefURL}
\newblock
\APACrefnote{Accessed: 2025-2-6}
\PrintBackRefs{\CurrentBib}

\bibitem [\protect \citeauthoryear {%
{Department for Science, Innovation \& Technology}%
\ \BBA {} {AI Safety Institute}%
}{%
{Department for Science, Innovation \& Technology}%
\ \BBA {} {AI Safety Institute}%
}{%
{\protect \APACyear {2023}}%
}]{%
DepartmentForScienceInnovationTechnology2023r}
\APACinsertmetastar {%
DepartmentForScienceInnovationTechnology2023r}%
\begin{APACrefauthors}%
{Department for Science, Innovation \& Technology}%
\BCBT {}\ \BBA {} {AI Safety Institute}.%
\end{APACrefauthors}%
\unskip\
\newblock
\APACrefYearMonthDay{2023}{}{}.
\newblock
\APACrefbtitle {Frontier {AI} Taskforce: first progress report.} {Frontier {AI} taskforce: first progress report.}
\newblock
\begin{APACrefURL} \url{https://www.gov.uk/government/publications/frontier-ai-taskforce-first-progress-report/frontier-ai-taskforce-first-progress-report} \end{APACrefURL}
\newblock
\APACrefnote{Accessed: 2025-2-5}
\PrintBackRefs{\CurrentBib}

\bibitem [\protect \citeauthoryear {%
{Department for Science, Innovation \& Technology}%
, {Office for Artificial Intelligence}%
\BCBL {}\ \BBA {} {Department for Digital, Culture, Media \& Sport}%
}{%
{Department for Science, Innovation \& Technology}%
\ \protect \BOthers {.}}{%
{\protect \APACyear {2022}}%
}]{%
DepartmentForScienceInnovationTechnology2022j}
\APACinsertmetastar {%
DepartmentForScienceInnovationTechnology2022j}%
\begin{APACrefauthors}%
{Department for Science, Innovation \& Technology}%
, {Office for Artificial Intelligence}%
\BCBL {}\ \BBA {} {Department for Digital, Culture, Media \& Sport}.%
\end{APACrefauthors}%
\unskip\
\newblock
\APACrefYearMonthDay{2022}{}{}.
\newblock
\APACrefbtitle {Establishing a pro-innovation approach to regulating {AI}.} {Establishing a pro-innovation approach to regulating {AI}.}
\newblock
\begin{APACrefURL} \url{https://www.gov.uk/government/publications/establishing-a-pro-innovation-approach-to-regulating-ai/establishing-a-pro-innovation-approach-to-regulating-ai-policy-statement} \end{APACrefURL}
\newblock
\APACrefnote{Accessed: 2025-2-5}
\PrintBackRefs{\CurrentBib}

\bibitem [\protect \citeauthoryear {%
{Department for Science, Innovation and Technology}%
}{%
{Department for Science, Innovation and Technology}%
}{%
{\protect \APACyear {2023}}%
{\protect \APACexlab {{\protect \BCnt {1}}}}}]{%
DepartmentForScienceInnovationAndTechnology2023y}
\APACinsertmetastar {%
DepartmentForScienceInnovationAndTechnology2023y}%
\begin{APACrefauthors}%
{Department for Science, Innovation and Technology}.%
\end{APACrefauthors}%
\unskip\
\newblock
\APACrefYearMonthDay{2023{\protect \BCnt {1}}}{}{}.
\newblock
\APACrefbtitle {Emerging processes for frontier {AI} safety.} {Emerging processes for frontier {AI} safety.}
\newblock
\begin{APACrefURL} \url{https://assets.publishing.service.gov.uk/media/653aabbd80884d000df71bdc/emerging-processes-frontier-ai-safety.pdf} \end{APACrefURL}
\PrintBackRefs{\CurrentBib}

\bibitem [\protect \citeauthoryear {%
{Department for Science, Innovation and Technology}%
}{%
{Department for Science, Innovation and Technology}%
}{%
{\protect \APACyear {2023}}%
{\protect \APACexlab {{\protect \BCnt {2}}}}}]{%
DepartmentForScienceInnovationAndTechnology2023c}
\APACinsertmetastar {%
DepartmentForScienceInnovationAndTechnology2023c}%
\begin{APACrefauthors}%
{Department for Science, Innovation and Technology}.%
\end{APACrefauthors}%
\unskip\
\newblock
\APACrefYearMonthDay{2023{\protect \BCnt {2}}}{}{}.
\newblock
\APACrefbtitle {Introducing the {{AI Safety Institute}}.} {Introducing the {{AI Safety Institute}}.}
\newblock
\begin{APACrefURL} \url{https://www.gov.uk/government/publications/ai-safety-institute-overview/introducing-the-ai-safety-institute} \end{APACrefURL}
\newblock
\APACrefnote{Accessed: --}
\PrintBackRefs{\CurrentBib}

\bibitem [\protect \citeauthoryear {%
{Department for Science, Innovation and Technology}%
}{%
{Department for Science, Innovation and Technology}%
}{%
{\protect \APACyear {2024}}%
}]{%
DepartmentForScienceInnovationAndTechnology2024a}
\APACinsertmetastar {%
DepartmentForScienceInnovationAndTechnology2024a}%
\begin{APACrefauthors}%
{Department for Science, Innovation and Technology}.%
\end{APACrefauthors}%
\unskip\
\newblock
\APACrefYearMonthDay{2024}{}{}.
\newblock
\APACrefbtitle {Seoul Declaration for safe, innovative and inclusive {AI}: {AI} Seoul Summit 2024.} {Seoul declaration for safe, innovative and inclusive {AI}: {AI} seoul summit 2024.}
\newblock
\begin{APACrefURL} \url{https://www.gov.uk/government/publications/seoul-declaration-for-safe-innovative-and-inclusive-ai-ai-seoul-summit-2024} \end{APACrefURL}
\newblock
\APACrefnote{Accessed: 2025-2-6}
\PrintBackRefs{\CurrentBib}

\bibitem [\protect \citeauthoryear {%
{Department for Science, Innovation and Technology}%
}{%
{Department for Science, Innovation and Technology}%
}{%
{\protect \APACyear {2025}}%
}]{%
DepartmentForScienceInnovationAndTechnology2025d}
\APACinsertmetastar {%
DepartmentForScienceInnovationAndTechnology2025d}%
\begin{APACrefauthors}%
{Department for Science, Innovation and Technology}.%
\end{APACrefauthors}%
\unskip\
\newblock
\APACrefYearMonthDay{2025}{}{}.
\newblock
\APACrefbtitle {Tackling {AI} security risks to unleash growth and deliver Plan for Change.} {Tackling {AI} security risks to unleash growth and deliver plan for change.}
\newblock
\begin{APACrefURL} \url{https://www.gov.uk/government/news/tackling-ai-security-risks-to-unleash-growth-and-deliver-plan-for-change} \end{APACrefURL}
\newblock
\APACrefnote{Accessed: 2025-6-16}
\PrintBackRefs{\CurrentBib}

\bibitem [\protect \citeauthoryear {%
{Department for Science, Innovation and Technology}%
, {AI Safety Institute}%
, Smith%
\BCBL {}\ \BBA {} Sunak%
}{%
{Department for Science, Innovation and Technology}%
, {AI Safety Institute}%
\BCBL {}\ \protect \BOthers {.}}{%
{\protect \APACyear {2023}}%
}]{%
DepartmentForScienceInnovationAndTechnology2023a}
\APACinsertmetastar {%
DepartmentForScienceInnovationAndTechnology2023a}%
\begin{APACrefauthors}%
{Department for Science, Innovation and Technology}%
, {AI Safety Institute}%
, Smith, C.%
\BCBL {}\ \BBA {} Sunak, R.%
\end{APACrefauthors}%
\unskip\
\newblock
\APACrefYearMonthDay{2023}{}{}.
\newblock
\APACrefbtitle {Tech entrepreneur Ian Hogarth to lead {UK}'s {AI} Foundation Model Taskforce.} {Tech entrepreneur ian hogarth to lead {UK}'s {AI} foundation model taskforce.}
\newblock
\begin{APACrefURL} \url{https://www.gov.uk/government/news/tech-entrepreneur-ian-hogarth-to-lead-uks-ai-foundation-model-taskforce} \end{APACrefURL}
\newblock
\APACrefnote{Accessed: 2025-2-5}
\PrintBackRefs{\CurrentBib}

\bibitem [\protect \citeauthoryear {%
{Department for Science, Innovation and Technology}%
, Donelan%
\BCBL {}\ \BBA {} Sunak%
}{%
{Department for Science, Innovation and Technology}%
\ \protect \BOthers {.}}{%
{\protect \APACyear {2024}}%
}]{%
DepartmentForScienceInnovationAndTechnology2024v}
\APACinsertmetastar {%
DepartmentForScienceInnovationAndTechnology2024v}%
\begin{APACrefauthors}%
{Department for Science, Innovation and Technology}%
, Donelan, M.%
\BCBL {}\ \BBA {} Sunak, R.%
\end{APACrefauthors}%
\unskip\
\newblock
\APACrefYearMonthDay{2024}{}{}.
\newblock
\APACrefbtitle {Global leaders agree to launch first international network of {AI} Safety Institutes to boost cooperation of {AI}.} {Global leaders agree to launch first international network of {AI} safety institutes to boost cooperation of {AI}.}
\newblock
\begin{APACrefURL} \url{https://www.gov.uk/government/news/global-leaders-agree-to-launch-first-international-network-of-ai-safety-institutes-to-boost-understanding-of-ai} \end{APACrefURL}
\newblock
\APACrefnote{Accessed: 2025-2-5}
\PrintBackRefs{\CurrentBib}

\bibitem [\protect \citeauthoryear {%
{Department for Science, Innovation and Technology}%
, {Foreign, Commonwealth and Development Office}%
\BCBL {}\ \BBA {} {Prime Minister's Office, 10 Downing Street}%
}{%
{Department for Science, Innovation and Technology}%
, {Foreign, Commonwealth and Development Office}%
\BCBL {}\ \BBA {} {Prime Minister's Office, 10 Downing Street}%
}{%
{\protect \APACyear {2023}}%
}]{%
DepartmentForScienceInnovationAndTechnology2023v}
\APACinsertmetastar {%
DepartmentForScienceInnovationAndTechnology2023v}%
\begin{APACrefauthors}%
{Department for Science, Innovation and Technology}%
, {Foreign, Commonwealth and Development Office}%
\BCBL {}\ \BBA {} {Prime Minister's Office, 10 Downing Street}.%
\end{APACrefauthors}%
\unskip\
\newblock
\APACrefYearMonthDay{2023}{}{}.
\newblock
\APACrefbtitle {The Bletchley Declaration by Countries Attending the {AI} Safety Summit, 1-2 November 2023.} {The bletchley declaration by countries attending the {AI} safety summit, 1-2 november 2023.}
\newblock
\begin{APACrefURL} \url{https://www.gov.uk/government/publications/ai-safety-summit-2023-the-bletchley-declaration/the-bletchley-declaration-by-countries-attending-the-ai-safety-summit-1-2-november-2023} \end{APACrefURL}
\PrintBackRefs{\CurrentBib}

\bibitem [\protect \citeauthoryear {%
{Department for Science, Innovation and Technology}%
\ \BBA {} Kyle%
}{%
{Department for Science, Innovation and Technology}%
\ \BBA {} Kyle%
}{%
{\protect \APACyear {2024}}%
}]{%
DepartmentForScienceInnovationAndTechnology2024k}
\APACinsertmetastar {%
DepartmentForScienceInnovationAndTechnology2024k}%
\begin{APACrefauthors}%
{Department for Science, Innovation and Technology}%
\BCBT {}\ \BBA {} Kyle, P.%
\end{APACrefauthors}%
\unskip\
\newblock
\APACrefYearMonthDay{2024}{}{}.
\newblock
\APACrefbtitle {Game-changing tech to reach the public faster as dedicated new unit launched to curb red tape.} {Game-changing tech to reach the public faster as dedicated new unit launched to curb red tape.}
\newblock
\begin{APACrefURL} \url{https://www.gov.uk/government/news/game-changing-tech-to-reach-the-public-faster-as-dedicated-new-unit-launched-to-curb-red-tape} \end{APACrefURL}
\newblock
\APACrefnote{Accessed: 2025-2-6}
\PrintBackRefs{\CurrentBib}

\bibitem [\protect \citeauthoryear {%
{Department for Science, Innovation and Technology}%
\ \BBA {} {Office for Artificial Intelligence}%
}{%
{Department for Science, Innovation and Technology}%
\ \BBA {} {Office for Artificial Intelligence}%
}{%
{\protect \APACyear {2023}}%
}]{%
DepartmentForScienceInnovationAndTechnology2023j}
\APACinsertmetastar {%
DepartmentForScienceInnovationAndTechnology2023j}%
\begin{APACrefauthors}%
{Department for Science, Innovation and Technology}%
\BCBT {}\ \BBA {} {Office for Artificial Intelligence}.%
\end{APACrefauthors}%
\unskip\
\newblock
\APACrefYearMonthDay{2023}{}{}.
\newblock
\APACrefbtitle {A pro-innovation approach to {AI} regulation.} {A pro-innovation approach to {AI} regulation.}
\newblock
\begin{APACrefURL} \url{https://www.gov.uk/government/publications/ai-regulation-a-pro-innovation-approach/white-paper} \end{APACrefURL}
\newblock
\APACrefnote{Accessed: 2024-7-12}
\PrintBackRefs{\CurrentBib}

\bibitem [\protect \citeauthoryear {%
{Department for Science, Innovation and Technology}%
, {Prime Minister's Office}%
, {Michelle Donelan}%
\BCBL {}\ \BBA {} {Rishi Sunak}%
}{%
{Department for Science, Innovation and Technology}%
, {Prime Minister's Office}%
\BCBL {}\ \protect \BOthers {.}}{%
{\protect \APACyear {2023}}%
}]{%
DepartmentForScienceInnovationAndTechnology2023s}
\APACinsertmetastar {%
DepartmentForScienceInnovationAndTechnology2023s}%
\begin{APACrefauthors}%
{Department for Science, Innovation and Technology}%
, {Prime Minister's Office}%
, {Michelle Donelan}%
\BCBL {}\ \BBA {} {Rishi Sunak}.%
\end{APACrefauthors}%
\unskip\
\newblock
\APACrefYearMonthDay{2023}{}{}.
\newblock
\APACrefbtitle {Initial \textsterling{}100 million for expert taskforce to help {UK} build and adopt next generation of safe {AI}.} {Initial \textsterling{}100 million for expert taskforce to help {UK} build and adopt next generation of safe {AI}.}
\newblock
\begin{APACrefURL} \url{https://www.gov.uk/government/news/initial-100-million-for-expert-taskforce-to-help-uk-build-and-adopt-next-generation-of-safe-ai} \end{APACrefURL}
\newblock
\APACrefnote{Accessed: 2025-2-5}
\PrintBackRefs{\CurrentBib}

\bibitem [\protect \citeauthoryear {%
{Department for Science, Innovation and Technology}%
, {Prime Minister's Office, 10 Downing Street}%
, Kyle%
, Starmer%
\BCBL {}\ \BBA {} Reeves%
}{%
{Department for Science, Innovation and Technology}%
\ \protect \BOthers {.}}{%
{\protect \APACyear {2025}}%
}]{%
DepartmentForScienceInnovationAndTechnology2025v}
\APACinsertmetastar {%
DepartmentForScienceInnovationAndTechnology2025v}%
\begin{APACrefauthors}%
{Department for Science, Innovation and Technology}%
, {Prime Minister's Office, 10 Downing Street}%
, Kyle, P.%
, Starmer, K.%
\BCBL {}\ \BBA {} Reeves, R.%
\end{APACrefauthors}%
\unskip\
\newblock
\APACrefYearMonthDay{2025}{}{}.
\newblock
\APACrefbtitle {Prime Minister sets out blueprint to turbocharge {AI}.} {Prime minister sets out blueprint to turbocharge {AI}.}
\newblock
\begin{APACrefURL} \url{https://www.gov.uk/government/news/prime-minister-sets-out-blueprint-to-turbocharge-ai} \end{APACrefURL}
\newblock
\APACrefnote{Accessed: 2025-2-6}
\PrintBackRefs{\CurrentBib}

\bibitem [\protect \citeauthoryear {%
{Digital Regulation Cooperation Forum}%
}{%
{Digital Regulation Cooperation Forum}%
}{%
{\protect \APACyear {2024}}%
}]{%
DigitalRegulationCooperationForum2024c}
\APACinsertmetastar {%
DigitalRegulationCooperationForum2024c}%
\begin{APACrefauthors}%
{Digital Regulation Cooperation Forum}.%
\end{APACrefauthors}%
\unskip\
\newblock
\APACrefYearMonthDay{2024}{}{}.
\newblock
\APACrefbtitle {About the {DRCF}.} {About the {DRCF}.}
\newblock
\begin{APACrefURL} \url{https://www.drcf.org.uk/about-us/} \end{APACrefURL}
\newblock
\APACrefnote{Accessed: 2025-2-5}
\PrintBackRefs{\CurrentBib}

\bibitem [\protect \citeauthoryear {%
Dolgin%
}{%
Dolgin%
}{%
{\protect \APACyear {2023}}%
}]{%
Dolgin2023u}
\APACinsertmetastar {%
Dolgin2023u}%
\begin{APACrefauthors}%
Dolgin, E.%
\end{APACrefauthors}%
\unskip\
\newblock
\APACrefYearMonthDay{2023}{}{}.
\newblock
{\BBOQ}\APACrefatitle {'{R}emarkable' {AI} tool designs {mRNA} vaccines that are more potent and stable} {'{R}emarkable' {AI} tool designs {mRNA} vaccines that are more potent and stable}.{\BBCQ}
\newblock
\APACjournalVolNumPages{Nature}{}{}{}.
\newblock
\begin{APACrefURL} \url{http://dx.doi.org/10.1038/d41586-023-01487-y} \end{APACrefURL}
\PrintBackRefs{\CurrentBib}

\bibitem [\protect \citeauthoryear {%
Drechsler%
, Savov%
, Schnabl%
\BCBL {}\ \BBA {} Wang%
}{%
Drechsler%
\ \protect \BOthers {.}}{%
{\protect \APACyear {2023}}%
}]{%
Drechsler2023p}
\APACinsertmetastar {%
Drechsler2023p}%
\begin{APACrefauthors}%
Drechsler, I.%
, Savov, A.%
, Schnabl, P.%
\BCBL {}\ \BBA {} Wang, O.%
\end{APACrefauthors}%
\unskip\
\newblock
\APACrefYearMonthDay{2023}{}{}.
\newblock
\APACrefbtitle {Deposit Franchise Runs} {Deposit franchise runs}\ \APACbVolEdTR{}{\BTR{}\ \BNUM\ w31138}.
\newblock
\APACaddressInstitution{Cambridge, MA}{National Bureau of Economic Research}.
\newblock
\begin{APACrefURL} \url{https://www.nber.org/system/files/working_papers/w31138/w31138.pdf} \end{APACrefURL}
\PrintBackRefs{\CurrentBib}

\bibitem [\protect \citeauthoryear {%
Epoch%
}{%
Epoch%
}{%
{\protect \APACyear {2024}}%
}]{%
Epoch2024y}
\APACinsertmetastar {%
Epoch2024y}%
\begin{APACrefauthors}%
Epoch, A\BPBI I.%
\end{APACrefauthors}%
\unskip\
\newblock
\APACrefYearMonthDay{2024}{}{}.
\newblock
\APACrefbtitle {Notable {AI} Models.} {Notable {AI} models.}
\newblock
\begin{APACrefURL} \url{https://epoch.ai/data/notable-ai-models} \end{APACrefURL}
\newblock
\APACrefnote{Accessed: 2025-6-28}
\PrintBackRefs{\CurrentBib}

\bibitem [\protect \citeauthoryear {%
{European Commission}%
}{%
{European Commission}%
}{%
{\protect \APACyear {2018}}%
{\protect \APACexlab {{\protect \BCnt {1}}}}}]{%
EuropeanCommission2018v}
\APACinsertmetastar {%
EuropeanCommission2018v}%
\begin{APACrefauthors}%
{European Commission}.%
\end{APACrefauthors}%
\unskip\
\newblock
\APACrefYearMonthDay{2018{\protect \BCnt {1}}}{}{}.
\newblock
\APACrefbtitle {Annex to the Communication from the Commission to the European Parliament, the European Council, the Council, the European Economic and Social Committee and the Committee of the Regions.} {Annex to the communication from the commission to the european parliament, the european council, the council, the european economic and social committee and the committee of the regions.}
\newblock
\begin{APACrefURL} \url{https://eur-lex.europa.eu/resource.html?uri=cellar:22ee84bb-fa04-11e8-a96d-01aa75ed71a1.0002.02/DOC_2\&format=PDF} \end{APACrefURL}
\PrintBackRefs{\CurrentBib}

\bibitem [\protect \citeauthoryear {%
{European Commission}%
}{%
{European Commission}%
}{%
{\protect \APACyear {2018}}%
{\protect \APACexlab {{\protect \BCnt {2}}}}}]{%
EuropeanCommission2018q}
\APACinsertmetastar {%
EuropeanCommission2018q}%
\begin{APACrefauthors}%
{European Commission}.%
\end{APACrefauthors}%
\unskip\
\newblock
\APACrefYearMonthDay{2018{\protect \BCnt {2}}}{}{}.
\newblock
\APACrefbtitle {{COMMUNICATION} {FROM} {THE} {COMMISSION} {TO} {THE} {EUROPEAN} {PARLIAMENT}, {THE} {EUROPEAN} {COUNCIL}, {THE} {COUNCIL}, {THE} {EUROPEAN} {ECONOMIC} {AND} {SOCIAL} {COMMITTEE} {AND} {THE} {COMMITTEE} {OF} {THE} {REGIONS} Artificial Intelligence for Europe {COM}/2018/237 final.} {{COMMUNICATION} {FROM} {THE} {COMMISSION} {TO} {THE} {EUROPEAN} {PARLIAMENT}, {THE} {EUROPEAN} {COUNCIL}, {THE} {COUNCIL}, {THE} {EUROPEAN} {ECONOMIC} {AND} {SOCIAL} {COMMITTEE} {AND} {THE} {COMMITTEE} {OF} {THE} {REGIONS} artificial intelligence for europe {COM}/2018/237 final.}
\newblock
\begin{APACrefURL} \url{https://eur-lex.europa.eu/legal-content/EN/TXT/HTML/?uri=CELEX:52018SC0137} \end{APACrefURL}
\PrintBackRefs{\CurrentBib}

\bibitem [\protect \citeauthoryear {%
{European Commission}%
}{%
{European Commission}%
}{%
{\protect \APACyear {2018}}%
{\protect \APACexlab {{\protect \BCnt {3}}}}}]{%
EuropeanCommission2018i}
\APACinsertmetastar {%
EuropeanCommission2018i}%
\begin{APACrefauthors}%
{European Commission}.%
\end{APACrefauthors}%
\unskip\
\newblock
\APACrefYearMonthDay{2018{\protect \BCnt {3}}}{}{}.
\newblock
\APACrefbtitle {Concept Note on The High-Level Expert Group on Artificial Intelligence.} {Concept note on the high-level expert group on artificial intelligence.}
\newblock
\begin{APACrefURL} \url{https://ec.europa.eu/futurium/en/system/files/ged/concept_note_on_the_ai_hlg_0.pdf} \end{APACrefURL}
\PrintBackRefs{\CurrentBib}

\bibitem [\protect \citeauthoryear {%
{European Commission}%
}{%
{European Commission}%
}{%
{\protect \APACyear {2020}}%
}]{%
EuropeanCommission2020t}
\APACinsertmetastar {%
EuropeanCommission2020t}%
\begin{APACrefauthors}%
{European Commission}.%
\end{APACrefauthors}%
\unskip\
\newblock
\APACrefYearMonthDay{2020}{}{}.
\newblock
\APACrefbtitle {{WHITE} {PAPER} On Artificial Intelligence - A European approach to excellence and trust {COM}/2020/65 final} {{WHITE} {PAPER} on artificial intelligence - a european approach to excellence and trust {COM}/2020/65 final}\ \APACbVolEdTR{}{\BTR{}}.
\newblock
\APACaddressInstitution{}{European Union}.
\newblock
\begin{APACrefURL} \url{https://eur-lex.europa.eu/legal-content/EN/TXT/?uri=CELEX:52020DC0065} \end{APACrefURL}
\PrintBackRefs{\CurrentBib}

\bibitem [\protect \citeauthoryear {%
{European Parliament}%
}{%
{European Parliament}%
}{%
{\protect \APACyear {2023}}%
}]{%
EuropeanParliament2023m}
\APACinsertmetastar {%
EuropeanParliament2023m}%
\begin{APACrefauthors}%
{European Parliament}.%
\end{APACrefauthors}%
\unskip\
\newblock
\APACrefYearMonthDay{2023}{}{}.
\newblock
\APACrefbtitle {Artificial Intelligence Act: Amendments adopted by the European Parliament on 14 June 2023 on the proposal for a regulation of the European Parliament and of the Council on laying down harmonised rules on artificial intelligence (Artificial Intelligence Act) and amending certain Union legislative acts ({COM}(2021)0206 -- {C9}-0146/2021 -- 2021/0106({COD})).} {Artificial intelligence act: Amendments adopted by the european parliament on 14 june 2023 on the proposal for a regulation of the european parliament and of the council on laying down harmonised rules on artificial intelligence (artificial intelligence act) and amending certain union legislative acts ({COM}(2021)0206 -- {C9}-0146/2021 -- 2021/0106({COD})).}
\newblock
\begin{APACrefURL} \url{https://www.europarl.europa.eu/doceo/document/TA-9-2023-0236_EN.pdf} \end{APACrefURL}
\PrintBackRefs{\CurrentBib}

\bibitem [\protect \citeauthoryear {%
{Exposure Labs}%
}{%
{Exposure Labs}%
}{%
{\protect \APACyear {2020}}%
}]{%
ExposureLabs2020r}
\APACinsertmetastar {%
ExposureLabs2020r}%
\begin{APACrefauthors}%
{Exposure Labs}.%
\end{APACrefauthors}%
\unskip\
\newblock
\APACrefYearMonthDay{2020}{}{}.
\newblock
\APACrefbtitle {The Social Dilemma.} {The social dilemma.}
\newblock
\begin{APACrefURL} \url{https://thesocialdilemma.com/} \end{APACrefURL}
\newblock
\APACrefnote{Accessed: 2025-2-4}
\PrintBackRefs{\CurrentBib}

\bibitem [\protect \citeauthoryear {%
Fang%
, Bindu%
, Gupta%
, Zhan%
\BCBL {}\ \BBA {} Kang%
}{%
Fang%
, Bindu%
\BCBL {}\ \protect \BOthers {.}}{%
{\protect \APACyear {2024}}%
}]{%
Fang2024a}
\APACinsertmetastar {%
Fang2024a}%
\begin{APACrefauthors}%
Fang, R.%
, Bindu, R.%
, Gupta, A.%
, Zhan, Q.%
\BCBL {}\ \BBA {} Kang, D.%
\end{APACrefauthors}%
\unskip\
\newblock
\APACrefYearMonthDay{2024}{}{}.
\newblock
\APACrefbtitle {Teams of {LLM} agents can exploit zero-day vulnerabilities.} {Teams of {LLM} agents can exploit zero-day vulnerabilities.}
\newblock
\begin{APACrefURL} \url{http://arxiv.org/abs/2406.01637} \end{APACrefURL}
\PrintBackRefs{\CurrentBib}

\bibitem [\protect \citeauthoryear {%
Fang%
, Bowman%
\BCBL {}\ \BBA {} Kang%
}{%
Fang%
, Bowman%
\BCBL {}\ \BBA {} Kang%
}{%
{\protect \APACyear {2024}}%
}]{%
Fang2024d}
\APACinsertmetastar {%
Fang2024d}%
\begin{APACrefauthors}%
Fang, R.%
, Bowman, D.%
\BCBL {}\ \BBA {} Kang, D.%
\end{APACrefauthors}%
\unskip\
\newblock
\APACrefYearMonthDay{2024}{}{}.
\newblock
\APACrefbtitle {Voice-Enabled {AI} Agents can Perform Common Scams.} {Voice-enabled {AI} agents can perform common scams.}
\newblock
\begin{APACrefURL} \url{http://arxiv.org/abs/2410.15650} \end{APACrefURL}
\PrintBackRefs{\CurrentBib}

\bibitem [\protect \citeauthoryear {%
Frei%
}{%
Frei%
}{%
{\protect \APACyear {2023}}%
}]{%
Frei2023m}
\APACinsertmetastar {%
Frei2023m}%
\begin{APACrefauthors}%
Frei, M.%
\end{APACrefauthors}%
\unskip\
\newblock
\APACrefYearMonthDay{2023}{}{}.
\newblock
\APACrefbtitle {`{I} don't believe [China] should be at the {AI} summit', says Iain Duncan Smith.} {`{I} don't believe [china] should be at the {AI} summit', says iain duncan smith.}
\newblock
\APACaddressPublisher{}{4 News}.
\newblock
\begin{APACrefURL} \url{https://www.channel4.com/news/i-dont-believe-china-should-be-at-the-ai-summit-says-iain-duncan-smith} \end{APACrefURL}
\PrintBackRefs{\CurrentBib}

\bibitem [\protect \citeauthoryear {%
{Future of Life Institute}%
}{%
{Future of Life Institute}%
}{%
{\protect \APACyear {2023}}%
}]{%
FutureOfLifeInstitute2023h}
\APACinsertmetastar {%
FutureOfLifeInstitute2023h}%
\begin{APACrefauthors}%
{Future of Life Institute}.%
\end{APACrefauthors}%
\unskip\
\newblock
\APACrefYearMonthDay{2023}{}{}.
\newblock
\APACrefbtitle {Pause Giant {AI} Experiments: An Open Letter.} {Pause giant {AI} experiments: An open letter.}
\newblock
\begin{APACrefURL} \url{https://futureoflife.org/open-letter/pause-giant-ai-experiments/} \end{APACrefURL}
\newblock
\APACrefnote{Accessed: 2025-2-5}
\PrintBackRefs{\CurrentBib}

\bibitem [\protect \citeauthoryear {%
{G7 Hiroshima Summit}%
}{%
{G7 Hiroshima Summit}%
}{%
{\protect \APACyear {2023}}%
}]{%
G7HiroshimaSummit2023z}
\APACinsertmetastar {%
G7HiroshimaSummit2023z}%
\begin{APACrefauthors}%
{G7 Hiroshima Summit}.%
\end{APACrefauthors}%
\unskip\
\newblock
\APACrefYearMonthDay{2023}{}{}.
\newblock
\APACrefbtitle {Hiroshima Process International Guiding Principles for Organizations Developing Advanced {AI} Systems.} {Hiroshima process international guiding principles for organizations developing advanced {AI} systems.}
\newblock
\begin{APACrefURL} \url{https://www.mofa.go.jp/files/100573471.pdf} \end{APACrefURL}
\PrintBackRefs{\CurrentBib}

\bibitem [\protect \citeauthoryear {%
Gabriel%
\ \protect \BOthers {.}}{%
Gabriel%
\ \protect \BOthers {.}}{%
{\protect \APACyear {2024}}%
}]{%
Gabriel2024c}
\APACinsertmetastar {%
Gabriel2024c}%
\begin{APACrefauthors}%
Gabriel, I.%
, Manzini, A.%
, Keeling, G.%
, Hendricks, L\BPBI A.%
, Rieser, V.%
, Iqbal, H.%
\BDBL {}Manyika, J.%
\end{APACrefauthors}%
\unskip\
\newblock
\APACrefYearMonthDay{2024}{}{}.
\newblock
\APACrefbtitle {The Ethics of Advanced {AI} Assistants} {The ethics of advanced {AI} assistants}\ \APACbVolEdTR{}{\BTR{}}.
\newblock
\APACaddressInstitution{}{Google DeepMind}.
\newblock
\begin{APACrefURL} \url{http://arxiv.org/abs/2404.16244} \end{APACrefURL}
\PrintBackRefs{\CurrentBib}

\bibitem [\protect \citeauthoryear {%
Gambacorta%
, Qiu%
, Shan%
\BCBL {}\ \BBA {} Rees%
}{%
Gambacorta%
\ \protect \BOthers {.}}{%
{\protect \APACyear {2024}}%
}]{%
Gambacorta2024z}
\APACinsertmetastar {%
Gambacorta2024z}%
\begin{APACrefauthors}%
Gambacorta, L.%
, Qiu, H.%
, Shan, S.%
\BCBL {}\ \BBA {} Rees, D.%
\end{APACrefauthors}%
\unskip\
\newblock
\APACrefYearMonthDay{2024}{}{}.
\newblock
\APACrefbtitle {Generative {AI} and labour productivity: a field experiment on coding} {Generative {AI} and labour productivity: a field experiment on coding}\ \APACbVolEdTR{}{\BTR{}\ \BNUM\ 1208}.
\newblock
\APACaddressInstitution{}{Bank for International Settlements}.
\newblock
\begin{APACrefURL} \url{https://www.bis.org/publ/work1208.htm} \end{APACrefURL}
\PrintBackRefs{\CurrentBib}

\bibitem [\protect \citeauthoryear {%
{Google DeepMind}%
}{%
{Google DeepMind}%
}{%
{\protect \APACyear {2024}}%
{\protect \APACexlab {{\protect \BCnt {1}}}}}]{%
GoogleDeepMind2024y}
\APACinsertmetastar {%
GoogleDeepMind2024y}%
\begin{APACrefauthors}%
{Google DeepMind}.%
\end{APACrefauthors}%
\unskip\
\newblock
\APACrefYearMonthDay{2024{\protect \BCnt {1}}}{}{}.
\newblock
\APACrefbtitle {Frontier Safety Framework Version 1.0.} {Frontier safety framework version 1.0.}
\newblock
\begin{APACrefURL} \url{https://storage.googleapis.com/deepmind-media/DeepMind.com/Blog/introducing-the-frontier-safety-framework/fsf-technical-report.pdf} \end{APACrefURL}
\PrintBackRefs{\CurrentBib}

\bibitem [\protect \citeauthoryear {%
{Google DeepMind}%
}{%
{Google DeepMind}%
}{%
{\protect \APACyear {2024}}%
{\protect \APACexlab {{\protect \BCnt {2}}}}}]{%
GoogleDeepMind2024s}
\APACinsertmetastar {%
GoogleDeepMind2024s}%
\begin{APACrefauthors}%
{Google DeepMind}.%
\end{APACrefauthors}%
\unskip\
\newblock
\APACrefYearMonthDay{2024{\protect \BCnt {2}}}{}{}.
\newblock
\APACrefbtitle {Project Astra.} {Project astra.}
\newblock
\begin{APACrefURL} \url{https://deepmind.google/technologies/project-astra/} \end{APACrefURL}
\newblock
\APACrefnote{Accessed: 2025-2-6}
\PrintBackRefs{\CurrentBib}

\bibitem [\protect \citeauthoryear {%
{Google DeepMind AlphaFold team}%
\ \BBA {} {Isomorphic Labs}%
}{%
{Google DeepMind AlphaFold team}%
\ \BBA {} {Isomorphic Labs}%
}{%
{\protect \APACyear {2024}}%
}]{%
GoogleDeepMindAlphaFoldTeam2024o}
\APACinsertmetastar {%
GoogleDeepMindAlphaFoldTeam2024o}%
\begin{APACrefauthors}%
{Google DeepMind AlphaFold team}%
\BCBT {}\ \BBA {} {Isomorphic Labs}.%
\end{APACrefauthors}%
\unskip\
\newblock
\APACrefYearMonthDay{2024}{}{}.
\newblock
\APACrefbtitle {{AlphaFold} 3 predicts the structure and interactions of all of life's molecules.} {{AlphaFold} 3 predicts the structure and interactions of all of life's molecules.}
\newblock
\begin{APACrefURL} \url{https://blog.google/technology/ai/google-deepmind-isomorphic-alphafold-3-ai-model/} \end{APACrefURL}
\newblock
\APACrefnote{Accessed: 2025-2-6}
\PrintBackRefs{\CurrentBib}

\bibitem [\protect \citeauthoryear {%
{GOV.UK}%
}{%
{GOV.UK}%
}{%
{\protect \APACyear {2023}}%
}]{%
GOVUK2023b}
\APACinsertmetastar {%
GOVUK2023b}%
\begin{APACrefauthors}%
{GOV.UK}.%
\end{APACrefauthors}%
\unskip\
\newblock
\APACrefYearMonthDay{2023}{}{}.
\newblock
\APACrefbtitle {About the {AI} Safety Summit 2023.} {About the {AI} safety summit 2023.}
\newblock
\begin{APACrefURL} \url{https://www.gov.uk/government/topical-events/ai-safety-summit-2023/about} \end{APACrefURL}
\newblock
\APACrefnote{Accessed: 2025-2-5}
\PrintBackRefs{\CurrentBib}

\bibitem [\protect \citeauthoryear {%
Halstead%
}{%
Halstead%
}{%
{\protect \APACyear {2024}}%
}]{%
Halstead2024j}
\APACinsertmetastar {%
Halstead2024j}%
\begin{APACrefauthors}%
Halstead, J.%
\end{APACrefauthors}%
\unskip\
\newblock
\APACrefYearMonthDay{2024}{}{}.
\newblock
\APACrefbtitle {Managing Risks from {AI}-Enabled Biological Tools.} {Managing risks from {AI}-enabled biological tools.}
\newblock
\begin{APACrefURL} \url{https://www.governance.ai/post/managing-risks-from-ai-enabled-biological-tools} \end{APACrefURL}
\newblock
\APACrefnote{Accessed: 2025-2-6}
\PrintBackRefs{\CurrentBib}

\bibitem [\protect \citeauthoryear {%
Heim%
\ \BBA {} Koessler%
}{%
Heim%
\ \BBA {} Koessler%
}{%
{\protect \APACyear {2024}}%
}]{%
Heim2024w}
\APACinsertmetastar {%
Heim2024w}%
\begin{APACrefauthors}%
Heim, L.%
\BCBT {}\ \BBA {} Koessler, L.%
\end{APACrefauthors}%
\unskip\
\newblock
\APACrefYearMonthDay{2024}{}{}.
\newblock
\APACrefbtitle {Training compute thresholds: Features and functions in {AI} regulation.} {Training compute thresholds: Features and functions in {AI} regulation.}
\newblock
\begin{APACrefURL} \url{http://arxiv.org/abs/2405.10799} \end{APACrefURL}
\PrintBackRefs{\CurrentBib}

\bibitem [\protect \citeauthoryear {%
Helfrich%
}{%
Helfrich%
}{%
{\protect \APACyear {2024}}%
}]{%
Helfrich2024z}
\APACinsertmetastar {%
Helfrich2024z}%
\begin{APACrefauthors}%
Helfrich, G.%
\end{APACrefauthors}%
\unskip\
\newblock
\APACrefYearMonthDay{2024}{}{}.
\newblock
{\BBOQ}\APACrefatitle {The harms of terminology: why we should reject so-called ``frontier {AI}''} {The harms of terminology: why we should reject so-called ``frontier {AI}''}.{\BBCQ}
\newblock
\APACjournalVolNumPages{AI and Ethics}{}{}{}.
\newblock
\begin{APACrefURL} \url{https://doi.org/10.1007/s43681-024-00438-1} \end{APACrefURL}
\PrintBackRefs{\CurrentBib}

\bibitem [\protect \citeauthoryear {%
{Hertzberg, Robert}%
}{%
{Hertzberg, Robert}%
}{%
{\protect \APACyear {2018}}%
}]{%
HertzbergRobert2018b}
\APACinsertmetastar {%
HertzbergRobert2018b}%
\begin{APACrefauthors}%
{Hertzberg, Robert}.%
\end{APACrefauthors}%
\unskip\
\newblock
\APACrefYearMonthDay{2018}{}{}.
\newblock
\APACrefbtitle {Senate Bill No. 1001: An act to add Chapter 6 (commencing with Section 17940) to Part 3 of Division 7 of the Business and Professions Code, relating to bots.} {Senate bill no. 1001: An act to add chapter 6 (commencing with section 17940) to part 3 of division 7 of the business and professions code, relating to bots.}
\newblock
\begin{APACrefURL} \url{https://leginfo.legislature.ca.gov/faces/billTextClient.xhtml?bill_id=201720180SB1001} \end{APACrefURL}
\PrintBackRefs{\CurrentBib}

\bibitem [\protect \citeauthoryear {%
Hogarth%
}{%
Hogarth%
}{%
{\protect \APACyear {2023}}%
}]{%
Hogarth2023d}
\APACinsertmetastar {%
Hogarth2023d}%
\begin{APACrefauthors}%
Hogarth, I.%
\end{APACrefauthors}%
\unskip\
\newblock
\APACrefYearMonthDay{2023}{}{}.
\newblock
{\BBOQ}\APACrefatitle {We must slow down the race to God-like {AI}} {We must slow down the race to god-like {AI}}.{\BBCQ}
\newblock
\APACjournalVolNumPages{Financial Times}{}{}{}.
\newblock
\begin{APACrefURL} \url{https://www.ft.com/content/03895dc4-a3b7-481e-95cc-336a524f2ac2} \end{APACrefURL}
\PrintBackRefs{\CurrentBib}

\bibitem [\protect \citeauthoryear {%
Hooker%
}{%
Hooker%
}{%
{\protect \APACyear {2024}}%
}]{%
Hooker2024h}
\APACinsertmetastar {%
Hooker2024h}%
\begin{APACrefauthors}%
Hooker, S.%
\end{APACrefauthors}%
\unskip\
\newblock
\APACrefYearMonthDay{2024}{}{}.
\newblock
\APACrefbtitle {On the Limitations of Compute Thresholds as a Governance Strategy.} {On the limitations of compute thresholds as a governance strategy.}
\newblock
\begin{APACrefURL} \url{http://arxiv.org/abs/2407.05694} \end{APACrefURL}
\PrintBackRefs{\CurrentBib}

\bibitem [\protect \citeauthoryear {%
{House of Commons Committee of Public Accounts}%
}{%
{House of Commons Committee of Public Accounts}%
}{%
{\protect \APACyear {2021}}%
}]{%
HouseOfCommonsCommitteeOfPublicAccounts2021r}
\APACinsertmetastar {%
HouseOfCommonsCommitteeOfPublicAccounts2021r}%
\begin{APACrefauthors}%
{House of Commons Committee of Public Accounts}.%
\end{APACrefauthors}%
\unskip\
\newblock
\APACrefYearMonthDay{2021}{}{}.
\newblock
\APACrefbtitle {Efficiency in government: Twenty-Eighth Report of Session 2021--22} {Efficiency in government: Twenty-eighth report of session 2021--22}\ \APACbVolEdTR{}{\BTR{}}.
\newblock
\APACaddressInstitution{}{House of Commons}.
\newblock
\begin{APACrefURL} \url{https://committees.parliament.uk/publications/8070/documents/82963/default/} \end{APACrefURL}
\PrintBackRefs{\CurrentBib}

\bibitem [\protect \citeauthoryear {%
{House of Commons Hansard}%
}{%
{House of Commons Hansard}%
}{%
{\protect \APACyear {2023}}%
{\protect \APACexlab {{\protect \BCnt {1}}}}}]{%
HouseOfCommonsHansard2023f}
\APACinsertmetastar {%
HouseOfCommonsHansard2023f}%
\begin{APACrefauthors}%
{House of Commons Hansard}.%
\end{APACrefauthors}%
\unskip\
\newblock
\APACrefYearMonthDay{2023{\protect \BCnt {1}}}{}{}.
\newblock
\APACrefbtitle {Advanced Artificial Intelligence.} {Advanced artificial intelligence.}
\newblock
\APAChowpublished {Intervention of Lord Clement-Jones. Volume 832: debated on Monday 24 July 2023}.
\newblock
\begin{APACrefURL} \url{https://hansard.parliament.uk/Lords/2023-07-24/debates/5432715E-F305-4EC1-8A79-B9A0C6402BFF/AdvancedArtificialIntelligence} \end{APACrefURL}
\PrintBackRefs{\CurrentBib}

\bibitem [\protect \citeauthoryear {%
{House of Commons Hansard}%
}{%
{House of Commons Hansard}%
}{%
{\protect \APACyear {2023}}%
{\protect \APACexlab {{\protect \BCnt {2}}}}}]{%
HouseOfCommonsHansard2023a}
\APACinsertmetastar {%
HouseOfCommonsHansard2023a}%
\begin{APACrefauthors}%
{House of Commons Hansard}.%
\end{APACrefauthors}%
\unskip\
\newblock
\APACrefYearMonthDay{2023{\protect \BCnt {2}}}{}{}.
\newblock
\APACrefbtitle {Data Protection and Digital Information (No. 2) Bill: Second Reading.} {Data protection and digital information (no. 2) bill: Second reading.}
\newblock
\APAChowpublished {Intervention of Damian Collins. Volume 731: debated on Monday 17 April 2023}.
\newblock
\begin{APACrefURL} \url{https://hansard.parliament.uk/Commons/2023-04-17/debates/019D4C9E-222D-4414-829C-6E8B86C1E65D/DataProtectionAndDigitalInformation(No2)Bill?highlight=pause%20ai#contribution-88D2AD71-2B99-46C9-B1BB-3F3E04175AC3} \end{APACrefURL}
\PrintBackRefs{\CurrentBib}

\bibitem [\protect \citeauthoryear {%
{House of Commons Science and Technology Committee}%
}{%
{House of Commons Science and Technology Committee}%
}{%
{\protect \APACyear {2016}}%
}]{%
HouseOfCommonsScienceAndTechnologyCommittee2016n}
\APACinsertmetastar {%
HouseOfCommonsScienceAndTechnologyCommittee2016n}%
\begin{APACrefauthors}%
{House of Commons Science and Technology Committee}.%
\end{APACrefauthors}%
\unskip\
\newblock
\APACrefYearMonthDay{2016}{}{}.
\newblock
\APACrefbtitle {Robotics and artificial intelligence: Fifth Report of Session 2016--17} {Robotics and artificial intelligence: Fifth report of session 2016--17}\ \APACbVolEdTR{}{\BTR{}}.
\newblock
\begin{APACrefURL} \url{https://dera.ioe.ac.uk/id/eprint/27621/1/145.pdf} \end{APACrefURL}
\PrintBackRefs{\CurrentBib}

\bibitem [\protect \citeauthoryear {%
Howard%
}{%
Howard%
}{%
{\protect \APACyear {2023}}%
}]{%
Howard2023r}
\APACinsertmetastar {%
Howard2023r}%
\begin{APACrefauthors}%
Howard, J.%
\end{APACrefauthors}%
\unskip\
\newblock
\APACrefYearMonthDay{2023}{}{}.
\newblock
\APACrefbtitle {{AI} Safety and the Age of Dislightenment.} {{AI} safety and the age of dislightenment.}
\newblock
\begin{APACrefURL} \url{https://www.fast.ai/posts/2023-11-07-dislightenment.html} \end{APACrefURL}
\newblock
\APACrefnote{Accessed: 2025-2-5}
\PrintBackRefs{\CurrentBib}

\bibitem [\protect \citeauthoryear {%
B.~Hu%
\ \protect \BOthers {.}}{%
B.~Hu%
\ \protect \BOthers {.}}{%
{\protect \APACyear {2024}}%
}]{%
Hu2024z}
\APACinsertmetastar {%
Hu2024z}%
\begin{APACrefauthors}%
Hu, B.%
, Zheng, L.%
, Zhu, J.%
, Ding, L.%
, Wang, Y.%
\BCBL {}\ \BBA {} Gu, X.%
\end{APACrefauthors}%
\unskip\
\newblock
\APACrefYearMonthDay{2024}{}{}.
\newblock
{\BBOQ}\APACrefatitle {Teaching plan generation and evaluation with {GPT}-4: Unleashing the potential of {LLM} in instructional design} {Teaching plan generation and evaluation with {GPT}-4: Unleashing the potential of {LLM} in instructional design}.{\BBCQ}
\newblock
\APACjournalVolNumPages{IEEE transactions on learning technologies}{17}{}{1471--1485}.
\newblock
\begin{APACrefURL} \url{https://ieeexplore.ieee.org/abstract/document/10490240} \end{APACrefURL}
\PrintBackRefs{\CurrentBib}

\bibitem [\protect \citeauthoryear {%
K.~Hu%
}{%
K.~Hu%
}{%
{\protect \APACyear {2023}}%
}]{%
Hu2023y}
\APACinsertmetastar {%
Hu2023y}%
\begin{APACrefauthors}%
Hu, K.%
\end{APACrefauthors}%
\unskip\
\newblock
\APACrefYearMonthDay{2023}{}{}.
\newblock
{\BBOQ}\APACrefatitle {{ChatGPT} sets record for fastest-growing user base - analyst note} {{ChatGPT} sets record for fastest-growing user base - analyst note}.{\BBCQ}
\newblock
\APACjournalVolNumPages{Reuters}{}{}{}.
\newblock
\begin{APACrefURL} \url{https://www.reuters.com/technology/chatgpt-sets-record-fastest-growing-user-base-analyst-note-2023-02-01/} \end{APACrefURL}
\PrintBackRefs{\CurrentBib}

\bibitem [\protect \citeauthoryear {%
{Information Technology Industry Council}%
\ \protect \BOthers {.}}{%
{Information Technology Industry Council}%
\ \protect \BOthers {.}}{%
{\protect \APACyear {2024}}%
}]{%
InformationTechnologyIndustryCouncil2024b}
\APACinsertmetastar {%
InformationTechnologyIndustryCouncil2024b}%
\begin{APACrefauthors}%
{Information Technology Industry Council}%
, {Americans for Responsible Innovation}%
, {A.Capital}%
, {Accenture}%
, {AI Policy Institute}%
, {Amazon}%
\BDBL {}{UC Berkeley Center for Human-Compatible AI}%
\end{APACrefauthors}%
\unskip\
\newblock
\APACrefYearMonthDay{2024}{}{}.
\newblock
\APACrefbtitle {Open Letter to Speaker Johnson, Majority Leader Schumer, Minority Leader Jeffries, and Republican Leader {McConnell} Regarding {US} {AI} Safety Institute.} {Open letter to speaker johnson, majority leader schumer, minority leader jeffries, and republican leader {McConnell} regarding {US} {AI} safety institute.}
\newblock
\begin{APACrefURL} \url{https://responsibleinnovation.org/wp-content/uploads/2024/10/20241021ARI_ITIOctoberAISIHillLetter.pdf} \end{APACrefURL}
\PrintBackRefs{\CurrentBib}

\bibitem [\protect \citeauthoryear {%
{Institute for Fiscal Studies}%
}{%
{Institute for Fiscal Studies}%
}{%
{\protect \APACyear {2024}}%
}]{%
InstituteForFiscalStudies2024r}
\APACinsertmetastar {%
InstituteForFiscalStudies2024r}%
\begin{APACrefauthors}%
{Institute for Fiscal Studies}.%
\end{APACrefauthors}%
\unskip\
\newblock
\APACrefYearMonthDay{2024}{}{}.
\newblock
\APACrefbtitle {What does the government spend money on?} {What does the government spend money on?}
\newblock
\begin{APACrefURL} \url{https://ifs.org.uk/taxlab/taxlab-key-questions/what-does-government-spend-money} \end{APACrefURL}
\newblock
\APACrefnote{Accessed: 2025-2-6}
\PrintBackRefs{\CurrentBib}

\bibitem [\protect \citeauthoryear {%
{Intellectual Property Office}%
, {Department for Science, Innovation \& Technology}%
\BCBL {}\ \BBA {} {Department for Culture, Media \& Sport}%
}{%
{Intellectual Property Office}%
\ \protect \BOthers {.}}{%
{\protect \APACyear {2024}}%
}]{%
IntellectualPropertyOffice2024z}
\APACinsertmetastar {%
IntellectualPropertyOffice2024z}%
\begin{APACrefauthors}%
{Intellectual Property Office}%
, {Department for Science, Innovation \& Technology}%
\BCBL {}\ \BBA {} {Department for Culture, Media \& Sport}.%
\end{APACrefauthors}%
\unskip\
\newblock
\APACrefYearMonthDay{2024}{}{}.
\newblock
\APACrefbtitle {Copyright and Artificial Intelligence.} {Copyright and artificial intelligence.}
\newblock
\begin{APACrefURL} \url{https://www.gov.uk/government/consultations/copyright-and-artificial-intelligence/copyright-and-artificial-intelligence} \end{APACrefURL}
\newblock
\APACrefnote{Accessed: 2025-2-6}
\PrintBackRefs{\CurrentBib}

\bibitem [\protect \citeauthoryear {%
Iosad%
, Railton%
\BCBL {}\ \BBA {} Westgarth%
}{%
Iosad%
\ \protect \BOthers {.}}{%
{\protect \APACyear {2024}}%
}]{%
Iosad2024y}
\APACinsertmetastar {%
Iosad2024y}%
\begin{APACrefauthors}%
Iosad, A.%
, Railton, D.%
\BCBL {}\ \BBA {} Westgarth, T.%
\end{APACrefauthors}%
\unskip\
\newblock
\APACrefYearMonthDay{2024}{}{}.
\newblock
\APACrefbtitle {Governing in the Age of {AI}: A New Model to Transform the State} {Governing in the age of {AI}: A new model to transform the state}\ \APACbVolEdTR{}{\BTR{}}.
\newblock
\APACaddressInstitution{}{Tony Blair Institute}.
\newblock
\begin{APACrefURL} \url{https://institute.global/insights/politics-and-governance/governing-in-the-age-of-ai-a-new-model-to-transform-the-state} \end{APACrefURL}
\PrintBackRefs{\CurrentBib}

\bibitem [\protect \citeauthoryear {%
Jack%
\ \BBA {} Edwards%
}{%
Jack%
\ \BBA {} Edwards%
}{%
{\protect \APACyear {2025}}%
}]{%
Jack2025j}
\APACinsertmetastar {%
Jack2025j}%
\begin{APACrefauthors}%
Jack, S.%
\BCBT {}\ \BBA {} Edwards, C.%
\end{APACrefauthors}%
\unskip\
\newblock
\APACrefYearMonthDay{2025}{}{}.
\newblock
{\BBOQ}\APACrefatitle {Government ousts {UK} competition watchdog chair} {Government ousts {UK} competition watchdog chair}.{\BBCQ}
\newblock
\APACjournalVolNumPages{BBC News}{}{}{}.
\newblock
\begin{APACrefURL} \url{https://www.bbc.com/news/articles/c2d3e6zklxgo} \end{APACrefURL}
\PrintBackRefs{\CurrentBib}

\bibitem [\protect \citeauthoryear {%
Jacobs%
}{%
Jacobs%
}{%
{\protect \APACyear {2025}}%
}]{%
Jacobs2025e}
\APACinsertmetastar {%
Jacobs2025e}%
\begin{APACrefauthors}%
Jacobs, J.%
\end{APACrefauthors}%
\unskip\
\newblock
\APACrefYearMonthDay{2025}{}{}.
\newblock
\APACrefbtitle {Trump announces up to \$500 billion in private sector {AI} infrastructure investment.} {Trump announces up to \$500 billion in private sector {AI} infrastructure investment.}
\newblock
\begin{APACrefURL} \url{https://www.cbsnews.com/news/trump-announces-private-sector-ai-infrastructure-investment/} \end{APACrefURL}
\newblock
\APACrefnote{Accessed: 2025-2-6}
\PrintBackRefs{\CurrentBib}

\bibitem [\protect \citeauthoryear {%
Johnson%
}{%
Johnson%
}{%
{\protect \APACyear {2019}}%
}]{%
Johnson2019c}
\APACinsertmetastar {%
Johnson2019c}%
\begin{APACrefauthors}%
Johnson, B.%
\end{APACrefauthors}%
\unskip\
\newblock
\APACrefYearMonthDay{2019}{}{}.
\newblock
\APACrefbtitle {{PM} speech to the {UN} General Assembly: 24 September 2019.} {{PM} speech to the {UN} general assembly: 24 september 2019.}
\newblock
\begin{APACrefURL} \url{https://www.gov.uk/government/speeches/pm-speech-to-the-un-general-assembly-24-september-2019} \end{APACrefURL}
\newblock
\APACrefnote{Accessed: 2025-2-4}
\PrintBackRefs{\CurrentBib}

\bibitem [\protect \citeauthoryear {%
Korinek%
\ \BBA {} Suh%
}{%
Korinek%
\ \BBA {} Suh%
}{%
{\protect \APACyear {2024}}%
}]{%
Korinek2024i}
\APACinsertmetastar {%
Korinek2024i}%
\begin{APACrefauthors}%
Korinek, A.%
\BCBT {}\ \BBA {} Suh, D.%
\end{APACrefauthors}%
\unskip\
\newblock
\APACrefYearMonthDay{2024}{}{}.
\newblock
\APACrefbtitle {Scenarios for the Transition to {AGI}} {Scenarios for the transition to {AGI}}\ \APACbVolEdTR{}{\BTR{}\ \BNUM\ 32255}.
\newblock
\APACaddressInstitution{}{National Bureau of Economic Research}.
\newblock
\begin{APACrefURL} \url{https://www.nber.org/papers/w32255} \end{APACrefURL}
\PrintBackRefs{\CurrentBib}

\bibitem [\protect \citeauthoryear {%
Kwint%
}{%
Kwint%
}{%
{\protect \APACyear {2023}}%
}]{%
Kwint2023w}
\APACinsertmetastar {%
Kwint2023w}%
\begin{APACrefauthors}%
Kwint, J.%
\end{APACrefauthors}%
\unskip\
\newblock
\APACrefYearMonthDay{2023}{}{}.
\newblock
\APACrefbtitle {Artificial intelligence: 10 promising interventions for healthcare} {Artificial intelligence: 10 promising interventions for healthcare}\ \APACbVolEdTR{}{\BTR{}}.
\newblock
\APACaddressInstitution{}{National Institute for Health and Care Research}.
\newblock
\begin{APACrefURL} \url{http://dx.doi.org/10.3310/nihrevidence_59502} \end{APACrefURL}
\PrintBackRefs{\CurrentBib}

\bibitem [\protect \citeauthoryear {%
{Labour Party}%
}{%
{Labour Party}%
}{%
{\protect \APACyear {2024}}%
{\protect \APACexlab {{\protect \BCnt {1}}}}}]{%
LabourParty2024z}
\APACinsertmetastar {%
LabourParty2024z}%
\begin{APACrefauthors}%
{Labour Party}.%
\end{APACrefauthors}%
\unskip\
\newblock
\APACrefYearMonthDay{2024{\protect \BCnt {1}}}{}{}.
\newblock
\APACrefbtitle {Change: Labour Party Manifesto 2024.} {Change: Labour party manifesto 2024.}
\newblock
\begin{APACrefURL} \url{https://labour.org.uk/wp-content/uploads/2024/06/Labour-Party-manifesto-2024.pdf} \end{APACrefURL}
\PrintBackRefs{\CurrentBib}

\bibitem [\protect \citeauthoryear {%
{Labour Party}%
}{%
{Labour Party}%
}{%
{\protect \APACyear {2024}}%
{\protect \APACexlab {{\protect \BCnt {2}}}}}]{%
LabourParty2024x}
\APACinsertmetastar {%
LabourParty2024x}%
\begin{APACrefauthors}%
{Labour Party}.%
\end{APACrefauthors}%
\unskip\
\newblock
\APACrefYearMonthDay{2024{\protect \BCnt {2}}}{}{}.
\newblock
\APACrefbtitle {Labour's fiscal plan.} {Labour's fiscal plan.}
\newblock
\begin{APACrefURL} \url{https://labour.org.uk/change/labours-fiscal-plan/} \end{APACrefURL}
\newblock
\APACrefnote{Accessed: 2025-2-20}
\PrintBackRefs{\CurrentBib}

\bibitem [\protect \citeauthoryear {%
{Labour Party}%
}{%
{Labour Party}%
}{%
{\protect \APACyear {2024}}%
{\protect \APACexlab {{\protect \BCnt {3}}}}}]{%
LabourParty2024r}
\APACinsertmetastar {%
LabourParty2024r}%
\begin{APACrefauthors}%
{Labour Party}.%
\end{APACrefauthors}%
\unskip\
\newblock
\APACrefYearMonthDay{2024{\protect \BCnt {3}}}{}{}.
\newblock
\APACrefbtitle {Mission-driven government.} {Mission-driven government.}
\newblock
\begin{APACrefURL} \url{https://labour.org.uk/change/mission-driven-government/} \end{APACrefURL}
\newblock
\APACrefnote{Accessed: 2025-2-6}
\PrintBackRefs{\CurrentBib}

\bibitem [\protect \citeauthoryear {%
Lloyd%
}{%
Lloyd%
}{%
{\protect \APACyear {2023}}%
}]{%
Lloyd2023n}
\APACinsertmetastar {%
Lloyd2023n}%
\begin{APACrefauthors}%
Lloyd, N.%
\end{APACrefauthors}%
\unskip\
\newblock
\APACrefYearMonthDay{2023}{}{}.
\newblock
{\BBOQ}\APACrefatitle {Labour vows to force firms developing powerful {AI} to meet requirements} {Labour vows to force firms developing powerful {AI} to meet requirements}.{\BBCQ}
\newblock
\APACjournalVolNumPages{The Independent}{}{}{}.
\newblock
\begin{APACrefURL} \url{https://www.independent.co.uk/news/uk/politics/rishi-sunak-labour-government-prime-minister-bletchley-park-b2440275.html} \end{APACrefURL}
\PrintBackRefs{\CurrentBib}

\bibitem [\protect \citeauthoryear {%
Manancourt%
}{%
Manancourt%
}{%
{\protect \APACyear {2024}}%
}]{%
Manancourt2024z}
\APACinsertmetastar {%
Manancourt2024z}%
\begin{APACrefauthors}%
Manancourt, V.%
\end{APACrefauthors}%
\unskip\
\newblock
\APACrefYearMonthDay{2024}{}{}.
\newblock
\APACrefbtitle {Inside Britain's plan to save the world from runaway {AI}.} {Inside britain's plan to save the world from runaway {AI}.}
\newblock
\begin{APACrefURL} \url{https://www.politico.eu/article/britain-ai-silicon-valley-rishi-sunak-prime-minister-interest-cyber-attacks-national-security/} \end{APACrefURL}
\newblock
\APACrefnote{Accessed: 2025-2-5}
\PrintBackRefs{\CurrentBib}

\bibitem [\protect \citeauthoryear {%
May%
}{%
May%
}{%
{\protect \APACyear {2018}}%
}]{%
May2018l}
\APACinsertmetastar {%
May2018l}%
\begin{APACrefauthors}%
May, T.%
\end{APACrefauthors}%
\unskip\
\newblock
\APACrefYearMonthDay{2018}{}{}.
\newblock
\APACrefbtitle {{PM}'s speech at Davos 2018: 25 January.} {{PM}'s speech at davos 2018: 25 january.}
\newblock
\begin{APACrefURL} \url{https://www.gov.uk/government/speeches/pms-speech-at-davos-2018-25-january} \end{APACrefURL}
\newblock
\APACrefnote{Accessed: 2025-2-4}
\PrintBackRefs{\CurrentBib}

\bibitem [\protect \citeauthoryear {%
McAfee%
}{%
McAfee%
}{%
{\protect \APACyear {2024}}%
}]{%
McAfee2024d}
\APACinsertmetastar {%
McAfee2024d}%
\begin{APACrefauthors}%
McAfee, A.%
\end{APACrefauthors}%
\unskip\
\newblock
\APACrefYearMonthDay{2024}{}{}.
\newblock
\APACrefbtitle {Generally Faster: The Economic Impact of Generative {AI}} {Generally faster: The economic impact of generative {AI}}\ \APACbVolEdTR{}{\BTR{}}.
\newblock
\APACaddressInstitution{}{Google}.
\newblock
\begin{APACrefURL} \url{https://policycommons.net/artifacts/12281693/generally_faster_-_the_economic_impact_of_generative_ai/} \end{APACrefURL}
\PrintBackRefs{\CurrentBib}

\bibitem [\protect \citeauthoryear {%
Metz%
}{%
Metz%
}{%
{\protect \APACyear {2023}}%
}]{%
Metz2023h}
\APACinsertmetastar {%
Metz2023h}%
\begin{APACrefauthors}%
Metz, C.%
\end{APACrefauthors}%
\unskip\
\newblock
\APACrefYearMonthDay{2023}{}{}.
\newblock
{\BBOQ}\APACrefatitle {`{T}he Godfather of A.{I}.' Leaves Google and Warns of Danger Ahead} {`{T}he godfather of a.{I}.' leaves google and warns of danger ahead}.{\BBCQ}
\newblock
\APACjournalVolNumPages{The New York Times}{}{}{}.
\newblock
\begin{APACrefURL} \url{https://www.nytimes.com/2023/05/01/technology/ai-google-chatbot-engineer-quits-hinton.html} \end{APACrefURL}
\PrintBackRefs{\CurrentBib}

\bibitem [\protect \citeauthoryear {%
{Ministry of Justice}%
\ \BBA {} Davies-Jones%
}{%
{Ministry of Justice}%
\ \BBA {} Davies-Jones%
}{%
{\protect \APACyear {2025}}%
}]{%
MinistryOfJustice2025d}
\APACinsertmetastar {%
MinistryOfJustice2025d}%
\begin{APACrefauthors}%
{Ministry of Justice}%
\BCBT {}\ \BBA {} Davies-Jones, A.%
\end{APACrefauthors}%
\unskip\
\newblock
\APACrefYearMonthDay{2025}{}{}.
\newblock
\APACrefbtitle {Government crackdown on explicit deepfakes.} {Government crackdown on explicit deepfakes.}
\newblock
\begin{APACrefURL} \url{https://www.gov.uk/government/news/government-crackdown-on-explicit-deepfakes} \end{APACrefURL}
\newblock
\APACrefnote{Accessed: 2025-2-6}
\PrintBackRefs{\CurrentBib}

\bibitem [\protect \citeauthoryear {%
M{\"{o}}kander%
\ \protect \BOthers {.}}{%
M{\"{o}}kander%
\ \protect \BOthers {.}}{%
{\protect \APACyear {2024}}%
}]{%
Mokander2024e}
\APACinsertmetastar {%
Mokander2024e}%
\begin{APACrefauthors}%
M{\"{o}}kander, J.%
, Margetts, H.%
, Trager, R.%
, McBride, K.%
, Rajkumar, N.%
\BCBL {}\ \BBA {} Teo, M.%
\end{APACrefauthors}%
\unskip\
\newblock
\APACrefYearMonthDay{2024}{}{}.
\newblock
\APACrefbtitle {Getting the {UK}'s Legislative Strategy for {AI} Right.} {Getting the {UK}'s legislative strategy for {AI} right.}
\newblock
\APACaddressPublisher{}{Tony Blair Institute}.
\newblock
\begin{APACrefURL} \url{https://institute.global/insights/tech-and-digitalisation/getting-the-uks-legislative-strategy-for-ai-right} \end{APACrefURL}
\newblock
\APACrefnote{Accessed: 2025-2-6}
\PrintBackRefs{\CurrentBib}

\bibitem [\protect \citeauthoryear {%
Mollick%
\ \BBA {} Mollick%
}{%
Mollick%
\ \BBA {} Mollick%
}{%
{\protect \APACyear {2023}}%
}]{%
Mollick2023x}
\APACinsertmetastar {%
Mollick2023x}%
\begin{APACrefauthors}%
Mollick, E\BPBI R.%
\BCBT {}\ \BBA {} Mollick, L.%
\end{APACrefauthors}%
\unskip\
\newblock
\APACrefYearMonthDay{2023}{}{}.
\newblock
\APACrefbtitle {Using {AI} to implement effective teaching strategies in classrooms: Five strategies, including prompts.} {Using {AI} to implement effective teaching strategies in classrooms: Five strategies, including prompts.}
\newblock
\begin{APACrefURL} \url{https://papers.ssrn.com/abstract=4391243} \end{APACrefURL}
\PrintBackRefs{\CurrentBib}

\bibitem [\protect \citeauthoryear {%
Moulange%
, Langenkamp%
, Alexanian%
, Curtis%
\BCBL {}\ \BBA {} Livingston%
}{%
Moulange%
\ \protect \BOthers {.}}{%
{\protect \APACyear {2023}}%
}]{%
Moulange2023c}
\APACinsertmetastar {%
Moulange2023c}%
\begin{APACrefauthors}%
Moulange, R.%
, Langenkamp, M.%
, Alexanian, T.%
, Curtis, S.%
\BCBL {}\ \BBA {} Livingston, M.%
\end{APACrefauthors}%
\unskip\
\newblock
\APACrefYearMonthDay{2023}{}{}.
\newblock
\APACrefbtitle {Towards responsible governance of biological design tools.} {Towards responsible governance of biological design tools.}
\newblock
\begin{APACrefURL} \url{http://arxiv.org/abs/2311.15936} \end{APACrefURL}
\PrintBackRefs{\CurrentBib}

\bibitem [\protect \citeauthoryear {%
Mouton%
, Lucas%
\BCBL {}\ \BBA {} Guest%
}{%
Mouton%
\ \protect \BOthers {.}}{%
{\protect \APACyear {2024}}%
}]{%
Mouton2024i}
\APACinsertmetastar {%
Mouton2024i}%
\begin{APACrefauthors}%
Mouton, C\BPBI A.%
, Lucas, C.%
\BCBL {}\ \BBA {} Guest, E.%
\end{APACrefauthors}%
\unskip\
\newblock
\APACrefYearMonthDay{2024}{}{}.
\newblock
\APACrefbtitle {The Operational Risks of {AI} in Large-Scale Biological Attacks: Results of a Red-Team Study} {The operational risks of {AI} in large-scale biological attacks: Results of a red-team study}\ \APACbVolEdTR{}{\BTR{}}.
\newblock
\APACaddressInstitution{}{RAND Corporation}.
\newblock
\begin{APACrefURL} \url{https://www.rand.org/pubs/research_reports/RRA2977-2.html} \end{APACrefURL}
\PrintBackRefs{\CurrentBib}

\bibitem [\protect \citeauthoryear {%
Mozur%
}{%
Mozur%
}{%
{\protect \APACyear {2017}}%
}]{%
Mozur2017j}
\APACinsertmetastar {%
Mozur2017j}%
\begin{APACrefauthors}%
Mozur, P.%
\end{APACrefauthors}%
\unskip\
\newblock
\APACrefYearMonthDay{2017}{}{}.
\newblock
{\BBOQ}\APACrefatitle {Beijing Wants A.{I}. to Be Made in China by 2030} {Beijing wants a.{I}. to be made in china by 2030}.{\BBCQ}
\newblock
\APACjournalVolNumPages{The New York Times}{}{}{}.
\newblock
\begin{APACrefURL} \url{https://www.nytimes.com/2017/07/20/business/china-artificial-intelligence.html} \end{APACrefURL}
\PrintBackRefs{\CurrentBib}

\bibitem [\protect \citeauthoryear {%
Najjar%
}{%
Najjar%
}{%
{\protect \APACyear {2023}}%
}]{%
Najjar2023s}
\APACinsertmetastar {%
Najjar2023s}%
\begin{APACrefauthors}%
Najjar, R.%
\end{APACrefauthors}%
\unskip\
\newblock
\APACrefYearMonthDay{2023}{}{}.
\newblock
{\BBOQ}\APACrefatitle {Redefining radiology: A review of Artificial Intelligence integration in medical imaging} {Redefining radiology: A review of artificial intelligence integration in medical imaging}.{\BBCQ}
\newblock
\APACjournalVolNumPages{Diagnostics}{13}{17}{2760}.
\newblock
\begin{APACrefURL} \url{https://www.mdpi.com/2075-4418/13/17/2760} \end{APACrefURL}
\PrintBackRefs{\CurrentBib}

\bibitem [\protect \citeauthoryear {%
{National Cyber Security Centre}%
}{%
{National Cyber Security Centre}%
}{%
{\protect \APACyear {2023}}%
}]{%
NationalCyberSecurityCentre2023e}
\APACinsertmetastar {%
NationalCyberSecurityCentre2023e}%
\begin{APACrefauthors}%
{National Cyber Security Centre}.%
\end{APACrefauthors}%
\unskip\
\newblock
\APACrefYearMonthDay{2023}{}{}.
\newblock
\APACrefbtitle {Cyber security regulations and directors duties in the {UK}.} {Cyber security regulations and directors duties in the {UK}.}
\newblock
\begin{APACrefURL} \url{https://www.ncsc.gov.uk/collection/board-toolkit/cyber-security-regulation-and-directors-duties-in-the-uk} \end{APACrefURL}
\newblock
\APACrefnote{Accessed: 2025-2-6}
\PrintBackRefs{\CurrentBib}

\bibitem [\protect \citeauthoryear {%
{National Security Commission on Emerging Biotechnology}%
}{%
{National Security Commission on Emerging Biotechnology}%
}{%
{\protect \APACyear {2024}}%
}]{%
NationalSecurityCommissionOnEmergingBiotechnology2024q}
\APACinsertmetastar {%
NationalSecurityCommissionOnEmergingBiotechnology2024q}%
\begin{APACrefauthors}%
{National Security Commission on Emerging Biotechnology}.%
\end{APACrefauthors}%
\unskip\
\newblock
\APACrefYearMonthDay{2024}{}{}.
\newblock
\APACrefbtitle {{AIxBio} White Paper 3: Risks of {AIxBio}.} {{AIxBio} white paper 3: Risks of {AIxBio}.}
\newblock
\begin{APACrefURL} \url{https://www.biotech.senate.gov/press-releases/aixbio-white-paper-risks-of-aixbio/} \end{APACrefURL}
\newblock
\APACrefnote{Accessed: 2025-2-6}
\PrintBackRefs{\CurrentBib}

\bibitem [\protect \citeauthoryear {%
Nelson%
\ \BBA {} Rose%
}{%
Nelson%
\ \BBA {} Rose%
}{%
{\protect \APACyear {2023}}%
}]{%
Nelson2023p}
\APACinsertmetastar {%
Nelson2023p}%
\begin{APACrefauthors}%
Nelson, C.%
\BCBT {}\ \BBA {} Rose, S.%
\end{APACrefauthors}%
\unskip\
\newblock
\APACrefYearMonthDay{2023}{}{}.
\newblock
\APACrefbtitle {Overcoming challenges with synthetic nucleic acid screening implementation} {Overcoming challenges with synthetic nucleic acid screening implementation}\ \APACbVolEdTR{}{\BTR{}}.
\newblock
\APACaddressInstitution{}{Centre for Long-Term Resilience}.
\newblock
\begin{APACrefURL} \url{https://www.longtermresilience.org/reports/overcoming-challenges-with-synthetic-nucleic-acid-screening-implementation-2/} \end{APACrefURL}
\PrintBackRefs{\CurrentBib}

\bibitem [\protect \citeauthoryear {%
Ng%
}{%
Ng%
}{%
{\protect \APACyear {2023}}%
}]{%
Ng2023q}
\APACinsertmetastar {%
Ng2023q}%
\begin{APACrefauthors}%
Ng, A.%
\end{APACrefauthors}%
\unskip\
\newblock
\APACrefYearMonthDay{2023}{}{}.
\newblock
{\BBOQ}\APACrefatitle {Written Statement of Andrew Ng Before the {U}.{S}. Senate {AI} Insight Forum} {Written statement of andrew ng before the {U}.{S}. senate {AI} insight forum}.{\BBCQ}
\newblock
\APACjournalVolNumPages{AI FUND}{}{}{}.
\newblock
\begin{APACrefURL} \url{https://aifund.ai/insights-written-statement-of-andrew-ng-before-the-u-s-senate-ai-insight-forum/} \end{APACrefURL}
\PrintBackRefs{\CurrentBib}

\bibitem [\protect \citeauthoryear {%
{OECD.AI Policy Observatory}%
}{%
{OECD.AI Policy Observatory}%
}{%
{\protect \APACyear {{\protect \bibnodate {}}}}%
{\protect \APACexlab {{\protect \BCntND {1}}}}}]{%
OECDAIPolicyObservatoryOtherv}
\APACinsertmetastar {%
OECDAIPolicyObservatoryOtherv}%
\begin{APACrefauthors}%
{OECD.AI Policy Observatory}.%
\end{APACrefauthors}%
\unskip\
\newblock
\APACrefYearMonthDay{{\protect \bibnodate {}}{\protect \BCntND {1}}}{}{}.
\newblock
\APACrefbtitle {List of participants in the {OECD} Expert Group on {AI} ({AIGO}).} {List of participants in the {OECD} expert group on {AI} ({AIGO}).}
\newblock
\begin{APACrefURL} \url{https://oecd.ai/en/list-of-participants-oecd-expert-group-on-ai} \end{APACrefURL}
\newblock
\APACrefnote{Accessed: 2025-2-4}
\PrintBackRefs{\CurrentBib}

\bibitem [\protect \citeauthoryear {%
{OECD.AI Policy Observatory}%
}{%
{OECD.AI Policy Observatory}%
}{%
{\protect \APACyear {{\protect \bibnodate {}}}}%
{\protect \APACexlab {{\protect \BCntND {2}}}}}]{%
OECDAIPolicyObservatoryOtherd}
\APACinsertmetastar {%
OECDAIPolicyObservatoryOtherd}%
\begin{APACrefauthors}%
{OECD.AI Policy Observatory}.%
\end{APACrefauthors}%
\unskip\
\newblock
\APACrefYearMonthDay{{\protect \bibnodate {}}{\protect \BCntND {2}}}{}{}.
\newblock
\APACrefbtitle {{OECD} {AI} Principles overview.} {{OECD} {AI} principles overview.}
\newblock
\begin{APACrefURL} \url{https://oecd.ai/en/ai-principles} \end{APACrefURL}
\newblock
\APACrefnote{Accessed: 2024-5-22}
\PrintBackRefs{\CurrentBib}

\bibitem [\protect \citeauthoryear {%
{Office for Artificial Intelligence, Department for Digital, Culture, Media \& Sport, and Department for Business, Energy \& Industrial Strategy}%
}{%
{Office for Artificial Intelligence, Department for Digital, Culture, Media \& Sport, and Department for Business, Energy \& Industrial Strategy}%
}{%
{\protect \APACyear {2021}}%
}]{%
OfficeForArtificialIntelligenceDepartmentForDigitalCultureMediaSportandDepartmentForBusinessEnergyIndustrialStrategy2021b}
\APACinsertmetastar {%
OfficeForArtificialIntelligenceDepartmentForDigitalCultureMediaSportandDepartmentForBusinessEnergyIndustrialStrategy2021b}%
\begin{APACrefauthors}%
{Office for Artificial Intelligence, Department for Digital, Culture, Media \& Sport, and Department for Business, Energy \& Industrial Strategy}.%
\end{APACrefauthors}%
\unskip\
\newblock
\APACrefYearMonthDay{2021}{}{}.
\newblock
\APACrefbtitle {National {AI} Strategy} {National {AI} strategy}\ \APACbVolEdTR{}{\BTR{}\ \BNUM\ Command Paper 525}.
\newblock
\APACaddressInstitution{}{HM Government}.
\newblock
\begin{APACrefURL} \url{https://www.gov.uk/government/publications/national-ai-strategy/national-ai-strategy-html-version} \end{APACrefURL}
\PrintBackRefs{\CurrentBib}

\bibitem [\protect \citeauthoryear {%
{OpenAI}%
}{%
{OpenAI}%
}{%
{\protect \APACyear {2015}}%
}]{%
OpenAI2015n}
\APACinsertmetastar {%
OpenAI2015n}%
\begin{APACrefauthors}%
{OpenAI}.%
\end{APACrefauthors}%
\unskip\
\newblock
\APACrefYearMonthDay{2015}{}{}.
\newblock
\APACrefbtitle {Introducing {OpenAI}.} {Introducing {OpenAI}.}
\newblock
\begin{APACrefURL} \url{https://openai.com/index/introducing-openai/} \end{APACrefURL}
\newblock
\APACrefnote{Accessed: 2025-2-4}
\PrintBackRefs{\CurrentBib}

\bibitem [\protect \citeauthoryear {%
{OpenAI}%
}{%
{OpenAI}%
}{%
{\protect \APACyear {2023}}%
{\protect \APACexlab {{\protect \BCnt {1}}}}}]{%
OpenAI2023d}
\APACinsertmetastar {%
OpenAI2023d}%
\begin{APACrefauthors}%
{OpenAI}.%
\end{APACrefauthors}%
\unskip\
\newblock
\APACrefYearMonthDay{2023{\protect \BCnt {1}}}{}{}.
\newblock
\APACrefbtitle {Frontier Model Forum.} {Frontier model forum.}
\newblock
\begin{APACrefURL} \url{https://openai.com/index/frontier-model-forum/} \end{APACrefURL}
\newblock
\APACrefnote{Accessed: 2025-2-5}
\PrintBackRefs{\CurrentBib}

\bibitem [\protect \citeauthoryear {%
{OpenAI}%
}{%
{OpenAI}%
}{%
{\protect \APACyear {2023}}%
{\protect \APACexlab {{\protect \BCnt {2}}}}}]{%
OpenAI2023m}
\APACinsertmetastar {%
OpenAI2023m}%
\begin{APACrefauthors}%
{OpenAI}.%
\end{APACrefauthors}%
\unskip\
\newblock
\APACrefYearMonthDay{2023{\protect \BCnt {2}}}{}{}.
\newblock
\APACrefbtitle {Preparedness Framework (Beta)} {Preparedness framework (beta)}\ \APACbVolEdTR{}{\BTR{}}.
\newblock
\APACaddressInstitution{}{OpenAI}.
\newblock
\begin{APACrefURL} \url{https://cdn.openai.com/openai-preparedness-framework-beta.pdf} \end{APACrefURL}
\PrintBackRefs{\CurrentBib}

\bibitem [\protect \citeauthoryear {%
{OpenAI}%
}{%
{OpenAI}%
}{%
{\protect \APACyear {2024}}%
{\protect \APACexlab {{\protect \BCnt {1}}}}}]{%
OpenAI2024k}
\APACinsertmetastar {%
OpenAI2024k}%
\begin{APACrefauthors}%
{OpenAI}.%
\end{APACrefauthors}%
\unskip\
\newblock
\APACrefYearMonthDay{2024{\protect \BCnt {1}}}{}{}.
\newblock
\APACrefbtitle {{OpenAI} {o1} System Card} {{OpenAI} {o1} system card}\ \APACbVolEdTR{}{\BTR{}}.
\newblock
\APACaddressInstitution{}{OpenAI}.
\newblock
\begin{APACrefURL} \url{https://cdn.openai.com/o1-system-card-20240917.pdf} \end{APACrefURL}
\PrintBackRefs{\CurrentBib}

\bibitem [\protect \citeauthoryear {%
{OpenAI}%
}{%
{OpenAI}%
}{%
{\protect \APACyear {2024}}%
{\protect \APACexlab {{\protect \BCnt {2}}}}}]{%
OpenAI2024t}
\APACinsertmetastar {%
OpenAI2024t}%
\begin{APACrefauthors}%
{OpenAI}.%
\end{APACrefauthors}%
\unskip\
\newblock
\APACrefYearMonthDay{2024{\protect \BCnt {2}}}{}{}.
\newblock
\APACrefbtitle {Swarm (experimental, educational).} {Swarm (experimental, educational).}
\newblock
\begin{APACrefURL} \url{https://github.com/openai/swarm} \end{APACrefURL}
\PrintBackRefs{\CurrentBib}

\bibitem [\protect \citeauthoryear {%
{OpenAI}%
}{%
{OpenAI}%
}{%
{\protect \APACyear {2025}}%
}]{%
OpenAI2025g}
\APACinsertmetastar {%
OpenAI2025g}%
\begin{APACrefauthors}%
{OpenAI}.%
\end{APACrefauthors}%
\unskip\
\newblock
\APACrefYearMonthDay{2025}{}{}.
\newblock
\APACrefbtitle {{OpenAI} {o3}-mini System Card.} {{OpenAI} {o3}-mini system card.}
\newblock
\begin{APACrefURL} \url{https://cdn.openai.com/o3-mini-system-card.pdf} \end{APACrefURL}
\newblock
\APACrefnote{Accessed: 2025-2-6}
\PrintBackRefs{\CurrentBib}

\bibitem [\protect \citeauthoryear {%
Pannu%
, Gebauer%
, McKelvey%
, Cicero%
\BCBL {}\ \BBA {} Inglesby%
}{%
Pannu%
\ \protect \BOthers {.}}{%
{\protect \APACyear {2024}}%
}]{%
Pannu2024z}
\APACinsertmetastar {%
Pannu2024z}%
\begin{APACrefauthors}%
Pannu, J.%
, Gebauer, S.%
, McKelvey, G., Jr%
, Cicero, A.%
\BCBL {}\ \BBA {} Inglesby, T.%
\end{APACrefauthors}%
\unskip\
\newblock
\APACrefYearMonthDay{2024}{}{}.
\newblock
{\BBOQ}\APACrefatitle {{AI} could pose pandemic-scale biosecurity risks. Here's how to make it safer} {{AI} could pose pandemic-scale biosecurity risks. here's how to make it safer}.{\BBCQ}
\newblock
\APACjournalVolNumPages{Nature}{635}{8040}{808--811}.
\newblock
\begin{APACrefURL} \url{http://dx.doi.org/10.1038/d41586-024-03815-2} \end{APACrefURL}
\PrintBackRefs{\CurrentBib}

\bibitem [\protect \citeauthoryear {%
Parker%
}{%
Parker%
}{%
{\protect \APACyear {2023}}%
}]{%
Parker2023d}
\APACinsertmetastar {%
Parker2023d}%
\begin{APACrefauthors}%
Parker, G.%
\end{APACrefauthors}%
\unskip\
\newblock
\APACrefYearMonthDay{2023}{}{}.
\newblock
{\BBOQ}\APACrefatitle {Britain to host first global {AI} regulation summit in autumn} {Britain to host first global {AI} regulation summit in autumn}.{\BBCQ}
\newblock
\APACjournalVolNumPages{Financial Times}{}{}{}.
\newblock
\begin{APACrefURL} \url{https://www.ft.com/content/3929908e-0f6a-4223-9c1c-5cd68d82a828} \end{APACrefURL}
\PrintBackRefs{\CurrentBib}

\bibitem [\protect \citeauthoryear {%
Patwardhan%
\ \protect \BOthers {.}}{%
Patwardhan%
\ \protect \BOthers {.}}{%
{\protect \APACyear {2024}}%
}]{%
Patwardhan2024g}
\APACinsertmetastar {%
Patwardhan2024g}%
\begin{APACrefauthors}%
Patwardhan, T.%
, Liu, K.%
, Markov, T.%
, Chowdhury, N.%
, Leet, D.%
, Cone, N.%
\BDBL {}Madry, A.%
\end{APACrefauthors}%
\unskip\
\newblock
\APACrefYearMonthDay{2024}{}{}.
\newblock
\APACrefbtitle {Building an early warning system for {LLM}-aided biological threat creation} {Building an early warning system for {LLM}-aided biological threat creation}\ \APACbVolEdTR{}{\BTR{}}.
\newblock
\APACaddressInstitution{}{OpenAI}.
\newblock
\begin{APACrefURL} \url{https://openai.com/research/building-an-early-warning-system-for-llm-aided-biological-threat-creation} \end{APACrefURL}
\PrintBackRefs{\CurrentBib}

\bibitem [\protect \citeauthoryear {%
{Prime Minister's Office, 10 Downing Street}%
\ \protect \BOthers {.}}{%
{Prime Minister's Office, 10 Downing Street}%
\ \protect \BOthers {.}}{%
{\protect \APACyear {2023}}%
}]{%
PrimeMinisterSOffice10DowningStreet2023y}
\APACinsertmetastar {%
PrimeMinisterSOffice10DowningStreet2023y}%
\begin{APACrefauthors}%
{Prime Minister's Office, 10 Downing Street}%
, {Department for Science, Innovation and Technology}%
, {Foreign, Commonwealth \& Development Office}%
, {Rishi Sunak}%
, {Michelle Donelan}%
\BCBL {}\ \BBA {} {James Cleverly}.%
\end{APACrefauthors}%
\unskip\
\newblock
\APACrefYearMonthDay{2023}{}{}.
\newblock
\APACrefbtitle {Countries agree to safe and responsible development of frontier {AI} in landmark Bletchley Declaration.} {Countries agree to safe and responsible development of frontier {AI} in landmark bletchley declaration.}
\newblock
\begin{APACrefURL} \url{https://www.gov.uk/government/news/countries-agree-to-safe-and-responsible-development-of-frontier-ai-in-landmark-bletchley-declaration} \end{APACrefURL}
\newblock
\APACrefnote{Accessed: 2025-1-18}
\PrintBackRefs{\CurrentBib}

\bibitem [\protect \citeauthoryear {%
{Prime Minister's Office, 10 Downing Street and King Charles III}%
}{%
{Prime Minister's Office, 10 Downing Street and King Charles III}%
}{%
{\protect \APACyear {2024}}%
}]{%
PrimeMinisterSOffice10DowningStreetAndKingCharlesIII2024h}
\APACinsertmetastar {%
PrimeMinisterSOffice10DowningStreetAndKingCharlesIII2024h}%
\begin{APACrefauthors}%
{Prime Minister's Office, 10 Downing Street and King Charles III}.%
\end{APACrefauthors}%
\unskip\
\newblock
\APACrefYearMonthDay{2024}{}{}.
\newblock
\APACrefbtitle {Oral Statement to Parliament: The King's Speech 2024. Speech to Both Houses of Parliament.} {Oral statement to parliament: The king's speech 2024. speech to both houses of parliament.}
\newblock
\begin{APACrefURL} \url{https://www.gov.uk/government/speeches/the-kings-speech-2024} \end{APACrefURL}
\PrintBackRefs{\CurrentBib}

\bibitem [\protect \citeauthoryear {%
Quirk%
}{%
Quirk%
}{%
{\protect \APACyear {2023}}%
}]{%
Quirk2023n}
\APACinsertmetastar {%
Quirk2023n}%
\begin{APACrefauthors}%
Quirk, C.%
\end{APACrefauthors}%
\unskip\
\newblock
\APACrefYearMonthDay{2023}{}{}.
\newblock
{\BBOQ}\APACrefatitle {The High Stakes of Deepfakes: The Growing Necessity of Federal Legislation to Regulate This Rapidly Evolving Technology} {The high stakes of deepfakes: The growing necessity of federal legislation to regulate this rapidly evolving technology}.{\BBCQ}
\newblock
\APACjournalVolNumPages{The Princeton Legal Journal}{}{}{}.
\newblock
\begin{APACrefURL} \url{https://legaljournal.princeton.edu/the-high-stakes-of-deepfakes-the-growing-necessity-of-federal-legislation-to-regulate-this-rapidly-evolving-technology/} \end{APACrefURL}
\PrintBackRefs{\CurrentBib}

\bibitem [\protect \citeauthoryear {%
Rawlinson%
}{%
Rawlinson%
}{%
{\protect \APACyear {2015}}%
}]{%
Rawlinson2015a}
\APACinsertmetastar {%
Rawlinson2015a}%
\begin{APACrefauthors}%
Rawlinson, K.%
\end{APACrefauthors}%
\unskip\
\newblock
\APACrefYearMonthDay{2015}{}{}.
\newblock
{\BBOQ}\APACrefatitle {Microsoft's Bill Gates insists {AI} is a threat} {Microsoft's bill gates insists {AI} is a threat}.{\BBCQ}
\newblock
\APACjournalVolNumPages{BBC News}{}{}{}.
\newblock
\begin{APACrefURL} \url{https://www.bbc.com/news/31047780} \end{APACrefURL}
\PrintBackRefs{\CurrentBib}

\bibitem [\protect \citeauthoryear {%
Rose%
\ \BBA {} Nelson%
}{%
Rose%
\ \BBA {} Nelson%
}{%
{\protect \APACyear {2023}}%
}]{%
Rose2023n}
\APACinsertmetastar {%
Rose2023n}%
\begin{APACrefauthors}%
Rose, S.%
\BCBT {}\ \BBA {} Nelson, C.%
\end{APACrefauthors}%
\unskip\
\newblock
\APACrefYearMonthDay{2023}{}{}.
\newblock
\APACrefbtitle {Understanding {AI}-Facilitated Biological Weapon Development} {Understanding {AI}-facilitated biological weapon development}\ \APACbVolEdTR{}{\BTR{}}.
\newblock
\APACaddressInstitution{}{Centre for Long-Term Resilience}.
\newblock
\begin{APACrefURL} \url{https://www.longtermresilience.org/post/report-launch-examining-risks-at-the-intersection-of-ai-and-bio} \end{APACrefURL}
\PrintBackRefs{\CurrentBib}

\bibitem [\protect \citeauthoryear {%
Scharre%
}{%
Scharre%
}{%
{\protect \APACyear {2024}}%
}]{%
Scharre2024e}
\APACinsertmetastar {%
Scharre2024e}%
\begin{APACrefauthors}%
Scharre, P.%
\end{APACrefauthors}%
\unskip\
\newblock
\APACrefYearMonthDay{2024}{}{}.
\newblock
\APACrefbtitle {Future-Proofing Frontier {AI} Regulation: Projecting Future Compute for Frontier {AI} Models} {Future-proofing frontier {AI} regulation: Projecting future compute for frontier {AI} models}\ \APACbVolEdTR{}{\BTR{}}.
\newblock
\APACaddressInstitution{}{Center for a New American Security}.
\newblock
\begin{APACrefURL} \url{https://www.cnas.org/publications/reports/future-proofing-frontier-ai-regulation} \end{APACrefURL}
\PrintBackRefs{\CurrentBib}

\bibitem [\protect \citeauthoryear {%
Schuett%
, Anderljung%
, Carlier%
, Koessler%
\BCBL {}\ \BBA {} Garfinkel%
}{%
Schuett%
\ \protect \BOthers {.}}{%
{\protect \APACyear {2024}}%
}]{%
Schuett2024i}
\APACinsertmetastar {%
Schuett2024i}%
\begin{APACrefauthors}%
Schuett, J.%
, Anderljung, M.%
, Carlier, A.%
, Koessler, L.%
\BCBL {}\ \BBA {} Garfinkel, B.%
\end{APACrefauthors}%
\unskip\
\newblock
\APACrefYearMonthDay{2024}{}{}.
\newblock
\APACrefbtitle {From Principles to Rules: A Regulatory Approach for Frontier {AI}.} {From principles to rules: A regulatory approach for frontier {AI}.}
\newblock
\begin{APACrefURL} \url{http://arxiv.org/abs/2407.07300} \end{APACrefURL}
\PrintBackRefs{\CurrentBib}

\bibitem [\protect \citeauthoryear {%
Schuett%
\ \protect \BOthers {.}}{%
Schuett%
\ \protect \BOthers {.}}{%
{\protect \APACyear {2023}}%
}]{%
Schuett2023i}
\APACinsertmetastar {%
Schuett2023i}%
\begin{APACrefauthors}%
Schuett, J.%
, Dreksler, N.%
, Anderljung, M.%
, McCaffary, D.%
, Heim, L.%
, Bluemke, E.%
\BCBL {}\ \BBA {} Garfinkel, B.%
\end{APACrefauthors}%
\unskip\
\newblock
\APACrefYearMonthDay{2023}{}{}.
\newblock
\APACrefbtitle {Towards best practices in {AGI} safety and governance: A survey of expert opinion.} {Towards best practices in {AGI} safety and governance: A survey of expert opinion.}
\newblock
\begin{APACrefURL} \url{http://arxiv.org/abs/2305.07153} \end{APACrefURL}
\PrintBackRefs{\CurrentBib}

\bibitem [\protect \citeauthoryear {%
{Secretary of State for Digital, Culture, Media \& Sport}%
\ \BBA {} {Secretary of State for the Home Department}%
}{%
{Secretary of State for Digital, Culture, Media \& Sport}%
\ \BBA {} {Secretary of State for the Home Department}%
}{%
{\protect \APACyear {2019}}%
}]{%
SecretaryOfStateForDigitalCultureMediaSport2019r}
\APACinsertmetastar {%
SecretaryOfStateForDigitalCultureMediaSport2019r}%
\begin{APACrefauthors}%
{Secretary of State for Digital, Culture, Media \& Sport}%
\BCBT {}\ \BBA {} {Secretary of State for the Home Department}.%
\end{APACrefauthors}%
\unskip\
\newblock
\APACrefYearMonthDay{2019}{}{}.
\newblock
\APACrefbtitle {Online Harms White Paper} {Online harms white paper}\ \APACbVolEdTR{}{\BTR{}}.
\newblock
\APACaddressInstitution{}{UK Government}.
\newblock
\begin{APACrefURL} \url{https://assets.publishing.service.gov.uk/media/605e60c6e90e07750810b439/Online_Harms_White_Paper_V2.pdf} \end{APACrefURL}
\PrintBackRefs{\CurrentBib}

\bibitem [\protect \citeauthoryear {%
{Select Committee on Artificial Intelligence}%
}{%
{Select Committee on Artificial Intelligence}%
}{%
{\protect \APACyear {2018}}%
}]{%
SelectCommitteeOnArtificialIntelligence2018r}
\APACinsertmetastar {%
SelectCommitteeOnArtificialIntelligence2018r}%
\begin{APACrefauthors}%
{Select Committee on Artificial Intelligence}.%
\end{APACrefauthors}%
\unskip\
\newblock
\APACrefYearMonthDay{2018}{}{}.
\newblock
\APACrefbtitle {{AI} in the {UK}: ready, willing and able?} {{AI} in the {UK}: ready, willing and able?}\ \APACbVolEdTR{}{\BTR{}\ \BNUM\ HL Paper 100}.
\newblock
\APACaddressInstitution{}{House of Lords}.
\newblock
\begin{APACrefURL} \url{https://publications.parliament.uk/pa/ld201719/ldselect/ldai/100/100.pdf} \end{APACrefURL}
\PrintBackRefs{\CurrentBib}

\bibitem [\protect \citeauthoryear {%
Sevilla%
\ \BBA {} Rold\'{a}n%
}{%
Sevilla%
\ \BBA {} Rold\'{a}n%
}{%
{\protect \APACyear {2024}}%
}]{%
Sevilla2024q}
\APACinsertmetastar {%
Sevilla2024q}%
\begin{APACrefauthors}%
Sevilla, J.%
\BCBT {}\ \BBA {} Rold\'{a}n, E.%
\end{APACrefauthors}%
\unskip\
\newblock
\APACrefYearMonthDay{2024}{}{}.
\newblock
\APACrefbtitle {Training Compute of Frontier {AI} Models Grows by 4-{5x} per Year} {Training compute of frontier {AI} models grows by 4-{5x} per year}\ \APACbVolEdTR{}{\BTR{}}.
\newblock
\APACaddressInstitution{}{Epoch AI}.
\newblock
\begin{APACrefURL} \url{https://epoch.ai/blog/training-compute-of-frontier-ai-models-grows-by-4-5x-per-year} \end{APACrefURL}
\PrintBackRefs{\CurrentBib}

\bibitem [\protect \citeauthoryear {%
Shevlane%
\ \protect \BOthers {.}}{%
Shevlane%
\ \protect \BOthers {.}}{%
{\protect \APACyear {2023}}%
}]{%
Shevlane2023k}
\APACinsertmetastar {%
Shevlane2023k}%
\begin{APACrefauthors}%
Shevlane, T.%
, Farquhar, S.%
, Garfinkel, B.%
, Phuong, M.%
, Whittlestone, J.%
, Leung, J.%
\BDBL {}Dafoe, A.%
\end{APACrefauthors}%
\unskip\
\newblock
\APACrefYearMonthDay{2023}{}{}.
\newblock
\APACrefbtitle {Model evaluation for extreme risks} {Model evaluation for extreme risks}\ \APACbVolEdTR{}{\BTR{}}.
\newblock
\APACaddressInstitution{}{Google DeepMind}.
\newblock
\begin{APACrefURL} \url{http://arxiv.org/abs/2305.15324} \end{APACrefURL}
\PrintBackRefs{\CurrentBib}

\bibitem [\protect \citeauthoryear {%
Siegmann%
\ \BBA {} Anderljung%
}{%
Siegmann%
\ \BBA {} Anderljung%
}{%
{\protect \APACyear {2022}}%
}]{%
Siegmann2022w}
\APACinsertmetastar {%
Siegmann2022w}%
\begin{APACrefauthors}%
Siegmann, C.%
\BCBT {}\ \BBA {} Anderljung, M.%
\end{APACrefauthors}%
\unskip\
\newblock
\APACrefYearMonthDay{2022}{}{}.
\newblock
\APACrefbtitle {The Brussels Effect and Artificial Intelligence: How {EU} regulation will impact the global {AI} market} {The brussels effect and artificial intelligence: How {EU} regulation will impact the global {AI} market}\ \APACbVolEdTR{}{\BTR{}}.
\newblock
\APACaddressInstitution{}{Centre for the Governance of AI}.
\newblock
\begin{APACrefURL} \url{http://arxiv.org/abs/2208.12645} \end{APACrefURL}
\PrintBackRefs{\CurrentBib}

\bibitem [\protect \citeauthoryear {%
Siffert%
}{%
Siffert%
}{%
{\protect \APACyear {2017}}%
}]{%
Siffert2017x}
\APACinsertmetastar {%
Siffert2017x}%
\begin{APACrefauthors}%
Siffert, A.%
\end{APACrefauthors}%
\unskip\
\newblock
\APACrefYearMonthDay{2017}{}{}.
\newblock
\APACrefbtitle {Ada Lovelace and the first computer programme in the world.} {Ada lovelace and the first computer programme in the world.}
\newblock
\begin{APACrefURL} \url{https://www.mpg.de/female-pioneers-of-science/Ada-Lovelace} \end{APACrefURL}
\newblock
\APACrefnote{Accessed: 2025-2-4}
\PrintBackRefs{\CurrentBib}

\bibitem [\protect \citeauthoryear {%
Smith%
, Rose%
, Moulange%
\BCBL {}\ \BBA {} Nelson%
}{%
Smith%
\ \protect \BOthers {.}}{%
{\protect \APACyear {2024}}%
}]{%
Smith2024a}
\APACinsertmetastar {%
Smith2024a}%
\begin{APACrefauthors}%
Smith, J.%
, Rose, S.%
, Moulange, R.%
\BCBL {}\ \BBA {} Nelson, C.%
\end{APACrefauthors}%
\unskip\
\newblock
\APACrefYearMonthDay{2024}{}{}.
\newblock
\APACrefbtitle {How the {UK} Government should address the misuse risk from {AI}-enabled biological tools} {How the {UK} government should address the misuse risk from {AI}-enabled biological tools}\ \APACbVolEdTR{}{\BTR{}}.
\newblock
\APACaddressInstitution{}{Centre for Long-Term Resilience}.
\newblock
\begin{APACrefURL} \url{https://www.longtermresilience.org/wp-content/uploads/2024/07/How-the-UK-Government-should-address-the-misuse-risk-from-AI-Enabled-biological-tools-BTs-Website-Copy.pdf} \end{APACrefURL}
\PrintBackRefs{\CurrentBib}

\bibitem [\protect \citeauthoryear {%
Solon%
}{%
Solon%
}{%
{\protect \APACyear {2018}}%
}]{%
Solon2018o}
\APACinsertmetastar {%
Solon2018o}%
\begin{APACrefauthors}%
Solon, O.%
\end{APACrefauthors}%
\unskip\
\newblock
\APACrefYearMonthDay{2018}{}{}.
\newblock
{\BBOQ}\APACrefatitle {Former Facebook and Google workers launch campaign to fight tech addiction} {Former facebook and google workers launch campaign to fight tech addiction}.{\BBCQ}
\newblock
\APACjournalVolNumPages{The Guardian}{}{}{}.
\newblock
\begin{APACrefURL} \url{https://www.theguardian.com/technology/2018/feb/05/tech-addiction-former-facebook-google-employees-campaign} \end{APACrefURL}
\PrintBackRefs{\CurrentBib}

\bibitem [\protect \citeauthoryear {%
South%
\ \protect \BOthers {.}}{%
South%
\ \protect \BOthers {.}}{%
{\protect \APACyear {2025}}%
}]{%
South2025d}
\APACinsertmetastar {%
South2025d}%
\begin{APACrefauthors}%
South, T.%
, Marro, S.%
, Hardjono, T.%
, Mahari, R.%
, Whitney, C\BPBI D.%
, Greenwood, D.%
\BDBL {}Pentland, A.%
\end{APACrefauthors}%
\unskip\
\newblock
\APACrefYearMonthDay{2025}{}{}.
\newblock
\APACrefbtitle {Authenticated delegation and authorized {AI} agents.} {Authenticated delegation and authorized {AI} agents.}
\newblock
\begin{APACrefURL} \url{http://arxiv.org/abs/2501.09674} \end{APACrefURL}
\PrintBackRefs{\CurrentBib}

\bibitem [\protect \citeauthoryear {%
Starmer%
}{%
Starmer%
}{%
{\protect \APACyear {2025}}%
}]{%
Starmer2025d}
\APACinsertmetastar {%
Starmer2025d}%
\begin{APACrefauthors}%
Starmer, K.%
\end{APACrefauthors}%
\unskip\
\newblock
\APACrefYearMonthDay{2025}{}{}.
\newblock
{\BBOQ}\APACrefatitle {Britain doesn't need to walk a {US} or {EU} path on {AI}} {Britain doesn't need to walk a {US} or {EU} path on {AI}}.{\BBCQ}
\newblock
\APACjournalVolNumPages{Financial Times}{}{}{}.
\newblock
\begin{APACrefURL} \url{https://www.ft.com/content/4d448059-5a3f-405c-9343-84cf7e5b90c0} \end{APACrefURL}
\PrintBackRefs{\CurrentBib}

\bibitem [\protect \citeauthoryear {%
Sunak%
}{%
Sunak%
}{%
{\protect \APACyear {2023}}%
{\protect \APACexlab {{\protect \BCnt {1}}}}}]{%
Sunak2023e}
\APACinsertmetastar {%
Sunak2023e}%
\begin{APACrefauthors}%
Sunak, R.%
\end{APACrefauthors}%
\unskip\
\newblock
\APACrefYearMonthDay{2023{\protect \BCnt {1}}}{}{}.
\newblock
\APACrefbtitle {{PM} London Tech Week speech: 12 June 2023.} {{PM} london tech week speech: 12 june 2023.}
\newblock
\begin{APACrefURL} \url{https://www.gov.uk/government/speeches/pm-london-tech-week-speech-12-june-2023} \end{APACrefURL}
\newblock
\APACrefnote{Accessed: 2025-2-5}
\PrintBackRefs{\CurrentBib}

\bibitem [\protect \citeauthoryear {%
Sunak%
}{%
Sunak%
}{%
{\protect \APACyear {2023}}%
{\protect \APACexlab {{\protect \BCnt {2}}}}}]{%
Sunak2023x}
\APACinsertmetastar {%
Sunak2023x}%
\begin{APACrefauthors}%
Sunak, R.%
\end{APACrefauthors}%
\unskip\
\newblock
\APACrefYearMonthDay{2023{\protect \BCnt {2}}}{}{}.
\newblock
\APACrefbtitle {Prime Minister's speech on {AI}: 26 October 2023.} {Prime minister's speech on {AI}: 26 october 2023.}
\newblock
\begin{APACrefURL} \url{https://www.gov.uk/government/speeches/prime-ministers-speech-on-ai-26-october-2023} \end{APACrefURL}
\newblock
\APACrefnote{Accessed: 2025-2-5}
\PrintBackRefs{\CurrentBib}

\bibitem [\protect \citeauthoryear {%
Sunstein%
}{%
Sunstein%
}{%
{\protect \APACyear {2024}}%
}]{%
Sunstein2024b}
\APACinsertmetastar {%
Sunstein2024b}%
\begin{APACrefauthors}%
Sunstein, C\BPBI R.%
\end{APACrefauthors}%
\unskip\
\newblock
\APACrefYearMonthDay{2024}{}{}.
\newblock
\APACrefbtitle {Brave New World? Human Welfare and Paternalistic {AI}.} {Brave new world? human welfare and paternalistic {AI}.}
\newblock
\begin{APACrefURL} \url{https://papers.ssrn.com/abstract=4908836} \end{APACrefURL}
\PrintBackRefs{\CurrentBib}

\bibitem [\protect \citeauthoryear {%
Susskind%
}{%
Susskind%
}{%
{\protect \APACyear {2017}}%
}]{%
Susskind2017j}
\APACinsertmetastar {%
Susskind2017j}%
\begin{APACrefauthors}%
Susskind, D.%
\end{APACrefauthors}%
\unskip\
\newblock
\APACrefYearMonthDay{2017}{}{}.
\newblock
\APACrefbtitle {A model of technological unemployment.} {A model of technological unemployment.}
\newblock
\begin{APACrefURL} \url{https://static1.squarespace.com/static/57d002e01b631bc215df193b/t/595e5d34e3df28e874d5dd40/1499356473362/SUSSKIND%2C+A+Model+of+Technological+Unemployment+6+July+2017.pdf} \end{APACrefURL}
\PrintBackRefs{\CurrentBib}

\bibitem [\protect \citeauthoryear {%
Telle%
\ \BBA {} Votruba%
}{%
Telle%
\ \BBA {} Votruba%
}{%
{\protect \APACyear {2011}}%
}]{%
Telle2011u}
\APACinsertmetastar {%
Telle2011u}%
\begin{APACrefauthors}%
Telle, K.%
\BCBT {}\ \BBA {} Votruba, M.%
\end{APACrefauthors}%
\unskip\
\newblock
\APACrefYearMonthDay{2011}{}{}.
\newblock
{\BBOQ}\APACrefatitle {Parental Job Loss and Children's School Performance} {Parental job loss and children's school performance}.{\BBCQ}
\newblock
\APACjournalVolNumPages{Review of Economic Studies}{78}{}{1462--1489}.
\newblock
\begin{APACrefURL} \url{http://dx.doi.org/10.2307/41407068} \end{APACrefURL}
\PrintBackRefs{\CurrentBib}

\bibitem [\protect \citeauthoryear {%
{The Alan Turing Institute}%
}{%
{The Alan Turing Institute}%
}{%
{\protect \APACyear {2023}}%
}]{%
TheAlanTuringInstitute2023k}
\APACinsertmetastar {%
TheAlanTuringInstitute2023k}%
\begin{APACrefauthors}%
{The Alan Turing Institute}.%
\end{APACrefauthors}%
\unskip\
\newblock
\APACrefYearMonthDay{2023}{}{}.
\newblock
\APACrefbtitle {Our strategy.} {Our strategy.}
\newblock
\begin{APACrefURL} \url{https://www.turing.ac.uk/about-us/our-strategy} \end{APACrefURL}
\newblock
\APACrefnote{Accessed: 2025-2-4}
\PrintBackRefs{\CurrentBib}

\bibitem [\protect \citeauthoryear {%
{The Electoral Commission}%
}{%
{The Electoral Commission}%
}{%
{\protect \APACyear {2024}}%
}]{%
TheElectoralCommission2024q}
\APACinsertmetastar {%
TheElectoralCommission2024q}%
\begin{APACrefauthors}%
{The Electoral Commission}.%
\end{APACrefauthors}%
\unskip\
\newblock
\APACrefYearMonthDay{2024}{}{}.
\newblock
\APACrefbtitle {New advice for voters on disinformation, and for campaigners using generative {AI}.} {New advice for voters on disinformation, and for campaigners using generative {AI}.}
\newblock
\begin{APACrefURL} \url{https://www.electoralcommission.org.uk/media-centre/new-advice-voters-disinformation-and-campaigners-using-generative-ai} \end{APACrefURL}
\newblock
\APACrefnote{Accessed: 2025-2-6}
\PrintBackRefs{\CurrentBib}

\bibitem [\protect \citeauthoryear {%
{The Royal Society}%
}{%
{The Royal Society}%
}{%
{\protect \APACyear {2017}}%
}]{%
TheRoyalSociety2017l}
\APACinsertmetastar {%
TheRoyalSociety2017l}%
\begin{APACrefauthors}%
{The Royal Society}.%
\end{APACrefauthors}%
\unskip\
\newblock
\APACrefYearMonthDay{2017}{}{}.
\newblock
\APACrefbtitle {Machine learning: the power and promise of computers that learn by example} {Machine learning: the power and promise of computers that learn by example}\ \APACbVolEdTR{}{\BTR{}\ \BNUM\ DES4702}.
\newblock
\begin{APACrefURL} \url{https://royalsociety.org/-/media/policy/projects/machine-learning/publications/machine-learning-report.pdf} \end{APACrefURL}
\PrintBackRefs{\CurrentBib}

\bibitem [\protect \citeauthoryear {%
Toner%
\ \BBA {} Fist%
}{%
Toner%
\ \BBA {} Fist%
}{%
{\protect \APACyear {2023}}%
}]{%
Toner2023f}
\APACinsertmetastar {%
Toner2023f}%
\begin{APACrefauthors}%
Toner, H.%
\BCBT {}\ \BBA {} Fist, T.%
\end{APACrefauthors}%
\unskip\
\newblock
\APACrefYearMonthDay{2023}{}{}.
\newblock
\APACrefbtitle {Regulating the {AI} Frontier: Design Choices and Constraints} {Regulating the {AI} frontier: Design choices and constraints}\ \APACbVolEdTR{}{\BTR{}}.
\newblock
\APACaddressInstitution{}{Center for Security and Emerging Technology}.
\newblock
\begin{APACrefURL} \url{https://cset.georgetown.edu/article/regulating-the-ai-frontier-design-choices-and-constraints/} \end{APACrefURL}
\PrintBackRefs{\CurrentBib}

\bibitem [\protect \citeauthoryear {%
Torenberg%
}{%
Torenberg%
}{%
{\protect \APACyear {2024}}%
}]{%
Torenberg2024o}
\APACinsertmetastar {%
Torenberg2024o}%
\begin{APACrefauthors}%
Torenberg, E.%
\end{APACrefauthors}%
\unskip\
\newblock
\APACrefYearMonthDay{2024}{}{}.
\newblock
\APACrefbtitle {David Sacks' Intellectual Journey.} {David sacks' intellectual journey.}
\newblock
\begin{APACrefURL} \url{https://eriktorenberg.substack.com/p/david-sacks-intellectual-journey} \end{APACrefURL}
\newblock
\APACrefnote{Accessed: 2025-2-6}
\PrintBackRefs{\CurrentBib}

\bibitem [\protect \citeauthoryear {%
{Trades Union Congress}%
, Allen%
\BCBL {}\ \BBA {} Masters%
}{%
{Trades Union Congress}%
\ \protect \BOthers {.}}{%
{\protect \APACyear {2024}}%
}]{%
TradesUnionCongress2024q}
\APACinsertmetastar {%
TradesUnionCongress2024q}%
\begin{APACrefauthors}%
{Trades Union Congress}%
, Allen, R.%
\BCBL {}\ \BBA {} Masters, D.%
\end{APACrefauthors}%
\unskip\
\newblock
\APACrefYearMonthDay{2024}{}{}.
\newblock
\APACrefbtitle {Artificial Intelligence (Regulation and Employment Rights) Bill.} {Artificial intelligence (regulation and employment rights) bill.}
\newblock
\begin{APACrefURL} \url{https://www.tuc.org.uk/research-analysis/reports/artificial-intelligence-regulation-and-employment-rights-bill} \end{APACrefURL}
\newblock
\APACrefnote{Accessed: 2025-2-6}
\PrintBackRefs{\CurrentBib}

\bibitem [\protect \citeauthoryear {%
{Tunbridge Wells Labour Party}%
}{%
{Tunbridge Wells Labour Party}%
}{%
{\protect \APACyear {2024}}%
}]{%
TunbridgeWellsLabourParty2024k}
\APACinsertmetastar {%
TunbridgeWellsLabourParty2024k}%
\begin{APACrefauthors}%
{Tunbridge Wells Labour Party}.%
\end{APACrefauthors}%
\unskip\
\newblock
\APACrefYearMonthDay{2024}{}{}.
\newblock
\APACrefbtitle {Labour's Five Missions for Britain.} {Labour's five missions for britain.}
\newblock
\begin{APACrefURL} \url{https://tunbridgewells.laboursites.org/national-labour-2/} \end{APACrefURL}
\newblock
\APACrefnote{Accessed: 2025-2-6}
\PrintBackRefs{\CurrentBib}

\bibitem [\protect \citeauthoryear {%
Turing%
}{%
Turing%
}{%
{\protect \APACyear {1950}}%
}]{%
Turing1950v}
\APACinsertmetastar {%
Turing1950v}%
\begin{APACrefauthors}%
Turing, A.%
\end{APACrefauthors}%
\unskip\
\newblock
\APACrefYearMonthDay{1950}{}{}.
\newblock
{\BBOQ}\APACrefatitle {Computing Machinery and Intelligence} {Computing machinery and intelligence}.{\BBCQ}
\newblock
\APACjournalVolNumPages{Mind}{59}{236}{433--460}.
\newblock
\begin{APACrefURL} \url{http://dx.doi.org/10.1093/mind/lix.236.433} \end{APACrefURL}
\PrintBackRefs{\CurrentBib}

\bibitem [\protect \citeauthoryear {%
{UK AI Safety Institute}%
}{%
{UK AI Safety Institute}%
}{%
{\protect \APACyear {{\protect \bibnodate {}}}}%
}]{%
UKAISafetyInstituteOtherk}
\APACinsertmetastar {%
UKAISafetyInstituteOtherk}%
\begin{APACrefauthors}%
{UK AI Safety Institute}.%
\end{APACrefauthors}%
\unskip\
\newblock
\APACrefYearMonthDay{{\protect \bibnodate {}}}{}{}.
\newblock
\APACrefbtitle {Conference on frontier {AI} safety frameworks.} {Conference on frontier {AI} safety frameworks.}
\newblock
\begin{APACrefURL} \url{https://www.aisi.gov.uk/work/conference-on-frontier-ai-safety-frameworks} \end{APACrefURL}
\newblock
\APACrefnote{Accessed: 2025-2-6}
\PrintBackRefs{\CurrentBib}

\bibitem [\protect \citeauthoryear {%
{UK Office for Budget Responsibility}%
}{%
{UK Office for Budget Responsibility}%
}{%
{\protect \APACyear {2023}}%
}]{%
UKOfficeForBudgetResponsibility2023l}
\APACinsertmetastar {%
UKOfficeForBudgetResponsibility2023l}%
\begin{APACrefauthors}%
{UK Office for Budget Responsibility}.%
\end{APACrefauthors}%
\unskip\
\newblock
\APACrefYearMonthDay{2023}{}{}.
\newblock
\APACrefbtitle {Economic and fiscal outlook} {Economic and fiscal outlook}\ \APACbVolEdTR{}{\BTR{}}.
\newblock
\APACaddressInstitution{}{HM Government}.
\newblock
\begin{APACrefURL} \url{https://obr.uk/docs/dlm_uploads/OBR-EFO-March-2023_Web_Accessible.pdf} \end{APACrefURL}
\PrintBackRefs{\CurrentBib}

\bibitem [\protect \citeauthoryear {%
{UK Office for National Statistics}%
}{%
{UK Office for National Statistics}%
}{%
{\protect \APACyear {2025}}%
}]{%
UKOfficeForNationalStatistics2025y}
\APACinsertmetastar {%
UKOfficeForNationalStatistics2025y}%
\begin{APACrefauthors}%
{UK Office for National Statistics}.%
\end{APACrefauthors}%
\unskip\
\newblock
\APACrefYearMonthDay{2025}{}{}.
\newblock
\APACrefbtitle {Public sector finances, {UK}: December 2024.} {Public sector finances, {UK}: December 2024.}
\newblock
\APACaddressPublisher{}{Office for National Statistics}.
\newblock
\begin{APACrefURL} \url{https://www.ons.gov.uk/economy/governmentpublicsectorandtaxes/publicsectorfinance/bulletins/publicsectorfinances/december2024} \end{APACrefURL}
\newblock
\APACrefnote{Accessed: 2025-2-20}
\PrintBackRefs{\CurrentBib}

\bibitem [\protect \citeauthoryear {%
{UK Secretary of State for Science, Innovation and Technology}%
}{%
{UK Secretary of State for Science, Innovation and Technology}%
}{%
{\protect \APACyear {2025}}%
}]{%
UKSecretaryOfStateForScienceInnovationAndTechnology2025d}
\APACinsertmetastar {%
UKSecretaryOfStateForScienceInnovationAndTechnology2025d}%
\begin{APACrefauthors}%
{UK Secretary of State for Science, Innovation and Technology}.%
\end{APACrefauthors}%
\unskip\
\newblock
\APACrefYearMonthDay{2025}{}{}.
\newblock
\APACrefbtitle {{AI} Opportunities Action Plan.} {{AI} opportunities action plan.}
\newblock
\begin{APACrefURL} \url{https://www.gov.uk/government/publications/ai-opportunities-action-plan/ai-opportunities-action-plan} \end{APACrefURL}
\newblock
\APACrefnote{Accessed: 2025-2-4}
\PrintBackRefs{\CurrentBib}

\bibitem [\protect \citeauthoryear {%
{U.S. Department of Commerce}%
}{%
{U.S. Department of Commerce}%
}{%
{\protect \APACyear {2025}}%
}]{%
USDepartmentOfCommerce2025c}
\APACinsertmetastar {%
USDepartmentOfCommerce2025c}%
\begin{APACrefauthors}%
{U.S. Department of Commerce}.%
\end{APACrefauthors}%
\unskip\
\newblock
\APACrefYearMonthDay{2025}{}{}.
\newblock
\APACrefbtitle {Statement from {U}.{S}. Secretary of Commerce Howard Lutnick on Transforming the {U}.{S}. {AI} Safety Institute into the Pro-Innovation, Pro-Science {U}.{S}. Center for {AI} Standards and Innovation.} {Statement from {U}.{S}. secretary of commerce howard lutnick on transforming the {U}.{S}. {AI} safety institute into the pro-innovation, pro-science {U}.{S}. center for {AI} standards and innovation.}
\newblock
\begin{APACrefURL} \url{https://www.commerce.gov/news/press-releases/2025/06/statement-us-secretary-commerce-howard-lutnick-transforming-us-ai} \end{APACrefURL}
\newblock
\APACrefnote{Accessed: 2025-6-16}
\PrintBackRefs{\CurrentBib}

\bibitem [\protect \citeauthoryear {%
{US National Institute of Standards and Technology}%
}{%
{US National Institute of Standards and Technology}%
}{%
{\protect \APACyear {2022}}%
}]{%
USNationalInstituteOfStandardsAndTechnology2022f}
\APACinsertmetastar {%
USNationalInstituteOfStandardsAndTechnology2022f}%
\begin{APACrefauthors}%
{US National Institute of Standards and Technology}.%
\end{APACrefauthors}%
\unskip\
\newblock
\APACrefYearMonthDay{2022}{}{}.
\newblock
\APACrefbtitle {Chips for America.} {Chips for america.}
\newblock
\begin{APACrefURL} \url{https://www.nist.gov/chips} \end{APACrefURL}
\newblock
\APACrefnote{Accessed: 2025-2-5}
\PrintBackRefs{\CurrentBib}

\bibitem [\protect \citeauthoryear {%
{US National Security Commission on Artificial Intelligence}%
}{%
{US National Security Commission on Artificial Intelligence}%
}{%
{\protect \APACyear {2021}}%
}]{%
USNationalSecurityCommissionOnArtificialIntelligence2021r}
\APACinsertmetastar {%
USNationalSecurityCommissionOnArtificialIntelligence2021r}%
\begin{APACrefauthors}%
{US National Security Commission on Artificial Intelligence}.%
\end{APACrefauthors}%
\unskip\
\newblock
\APACrefYearMonthDay{2021}{}{}.
\newblock
\APACrefbtitle {Final Report} {Final report}\ \APACbVolEdTR{}{\BTR{}}.
\newblock
\APACaddressInstitution{}{NSCAI}.
\newblock
\begin{APACrefURL} \url{https://reports.nscai.gov/final-report/} \end{APACrefURL}
\PrintBackRefs{\CurrentBib}

\bibitem [\protect \citeauthoryear {%
Van~Loo%
}{%
Van~Loo%
}{%
{\protect \APACyear {2019}}%
}]{%
VanLoo2019n}
\APACinsertmetastar {%
VanLoo2019n}%
\begin{APACrefauthors}%
Van~Loo, R.%
\end{APACrefauthors}%
\unskip\
\newblock
\APACrefYearMonthDay{2019}{}{}.
\newblock
{\BBOQ}\APACrefatitle {Digital market perfection} {Digital market perfection}.{\BBCQ}
\newblock
\APACjournalVolNumPages{Michigan law review}{117}{117.5}{815}.
\newblock
\begin{APACrefURL} \url{https://repository.law.umich.edu/mlr/vol117/iss5/2} \end{APACrefURL}
\PrintBackRefs{\CurrentBib}

\bibitem [\protect \citeauthoryear {%
Vincent%
}{%
Vincent%
}{%
{\protect \APACyear {2019}}%
}]{%
Vincent2019b}
\APACinsertmetastar {%
Vincent2019b}%
\begin{APACrefauthors}%
Vincent, J.%
\end{APACrefauthors}%
\unskip\
\newblock
\APACrefYearMonthDay{2019}{}{}.
\newblock
\APACrefbtitle {`{G}odfathers of {AI}' honored with Turing Award, the Nobel Prize of computing.} {`{G}odfathers of {AI}' honored with turing award, the nobel prize of computing.}
\newblock
\begin{APACrefURL} \url{https://www.theverge.com/2019/3/27/18280665/ai-godfathers-turing-award-2018-yoshua-bengio-geoffrey-hinton-yann-lecun} \end{APACrefURL}
\newblock
\APACrefnote{Accessed: 2025-2-18}
\PrintBackRefs{\CurrentBib}

\bibitem [\protect \citeauthoryear {%
Volpicelli%
}{%
Volpicelli%
}{%
{\protect \APACyear {2023}}%
}]{%
Volpicelli2023j}
\APACinsertmetastar {%
Volpicelli2023j}%
\begin{APACrefauthors}%
Volpicelli, G.%
\end{APACrefauthors}%
\unskip\
\newblock
\APACrefYearMonthDay{2023}{}{}.
\newblock
\APACrefbtitle {Power grab by France, Germany and Italy threatens to kill {EU}'s {AI} bill.} {Power grab by france, germany and italy threatens to kill {EU}'s {AI} bill.}
\newblock
\begin{APACrefURL} \url{https://www.politico.eu/article/france-germany-power-grab-kill-eu-blockbuster-ai-artificial-intelligence-bill/} \end{APACrefURL}
\newblock
\APACrefnote{Accessed: 2025-2-5}
\PrintBackRefs{\CurrentBib}

\bibitem [\protect \citeauthoryear {%
Whittlestone%
, Shane%
\BCBL {}\ \BBA {} Robinson%
}{%
Whittlestone%
\ \protect \BOthers {.}}{%
{\protect \APACyear {2024}}%
}]{%
Whittlestone2024x}
\APACinsertmetastar {%
Whittlestone2024x}%
\begin{APACrefauthors}%
Whittlestone, J.%
, Shane, T\BPBI S.%
\BCBL {}\ \BBA {} Robinson, B.%
\end{APACrefauthors}%
\unskip\
\newblock
\APACrefYearMonthDay{2024}{}{}.
\newblock
\APACrefbtitle {The {UK} is heading in the right direction on {AI} regulation, but must move faster} {The {UK} is heading in the right direction on {AI} regulation, but must move faster}\ \APACbVolEdTR{}{\BTR{}}.
\newblock
\APACaddressInstitution{}{The Centre for Long-Term Resilience}.
\newblock
\begin{APACrefURL} \url{https://www.longtermresilience.org/the-uk-is-heading-in-the-right-direction-on-ai-regulation-but-must-move-faster/} \end{APACrefURL}
\PrintBackRefs{\CurrentBib}

\bibitem [\protect \citeauthoryear {%
Wills%
}{%
Wills%
}{%
{\protect \APACyear {2024}}%
}]{%
Wills2024h}
\APACinsertmetastar {%
Wills2024h}%
\begin{APACrefauthors}%
Wills, P.%
\end{APACrefauthors}%
\unskip\
\newblock
\APACrefYearMonthDay{2024}{}{}.
\newblock
\APACrefbtitle {Care for Chatbots.} {Care for chatbots.}
\newblock
\begin{APACrefURL} \url{https://papers.ssrn.com/abstract=4814272} \end{APACrefURL}
\PrintBackRefs{\CurrentBib}

\bibitem [\protect \citeauthoryear {%
Wilson%
}{%
Wilson%
}{%
{\protect \APACyear {2024}}%
}]{%
Wilson2024e}
\APACinsertmetastar {%
Wilson2024e}%
\begin{APACrefauthors}%
Wilson, C.%
\end{APACrefauthors}%
\unskip\
\newblock
\APACrefYearMonthDay{2024}{}{}.
\newblock
\APACrefbtitle {The {US} Has Committed to Spend Far Less Than Peers on {AI} Safety.} {The {US} has committed to spend far less than peers on {AI} safety.}
\newblock
\begin{APACrefURL} \url{https://www.centeraipolicy.org/work/the-us-has-committed-to-spend-far-less-than-peers-on-ai-safety} \end{APACrefURL}
\newblock
\APACrefnote{Accessed: 2025-2-5}
\PrintBackRefs{\CurrentBib}

\bibitem [\protect \citeauthoryear {%
Wong%
, Frank%
, Nobles%
\BCBL {}\ \BBA {} Brown-Kaiser%
}{%
Wong%
\ \protect \BOthers {.}}{%
{\protect \APACyear {2023}}%
}]{%
Wong2023x}
\APACinsertmetastar {%
Wong2023x}%
\begin{APACrefauthors}%
Wong, S.%
, Frank, T\BPBI V.%
, Nobles, R.%
\BCBL {}\ \BBA {} Brown-Kaiser, L.%
\end{APACrefauthors}%
\unskip\
\newblock
\APACrefYearMonthDay{2023}{}{}.
\newblock
\APACrefbtitle {Elon Musk warns of 'civilizational risk' posed by {AI} in meeting with tech {CEOs} and senators.} {Elon musk warns of 'civilizational risk' posed by {AI} in meeting with tech {CEOs} and senators.}
\newblock
\begin{APACrefURL} \url{https://www.nbcnews.com/politics/congress/big-tech-ceos-ai-meeting-senators-musk-zuckerberg-rcna104738} \end{APACrefURL}
\newblock
\APACrefnote{Accessed: 2025-2-6}
\PrintBackRefs{\CurrentBib}

\bibitem [\protect \citeauthoryear {%
Wright%
\ \BBA {} Sellman%
}{%
Wright%
\ \BBA {} Sellman%
}{%
{\protect \APACyear {2024}}%
}]{%
Wright2024b}
\APACinsertmetastar {%
Wright2024b}%
\begin{APACrefauthors}%
Wright, O.%
\BCBT {}\ \BBA {} Sellman, M.%
\end{APACrefauthors}%
\unskip\
\newblock
\APACrefYearMonthDay{2024}{}{}.
\newblock
\APACrefbtitle {Britain must treat tech giants like nation states, minister warns.} {Britain must treat tech giants like nation states, minister warns.}
\newblock
\begin{APACrefURL} \url{https://www.thetimes.com/uk/politics/article/britain-must-treat-tech-giants-like-nation-states-minister-warns-ktmm5vmc9} \end{APACrefURL}
\newblock
\APACrefnote{Accessed: 2025-2-6}
\PrintBackRefs{\CurrentBib}

\end{thebibliography}

\end{document}